PUCRS

ESCOLA POLITÉCNICA
PROGRAMA DE PÓS-GRADUAÇÃO EM CIÊNCIA DA COMPUTAÇÃO MESTRADO EM CIÊNCIA DA COMPUTAÇÃO

ANDREW RAFAEL FRITSCH

# OVERLAPPING ERROR CORRECTION CODES ON TWO-DIMENSIONAL STRUCTURES

Porto Alegre
2025

PÓS-GRADUAÇÃO - *STRICTO SENSU*

Pontifícia Universidade Católica do Rio Grande do Sul

PONTIFICAL CATHOLIC UNIVERSITY OF RIO GRANDE DO SUL
SCHOOL OF TECHNOLOGY
COMPUTER SCIENCE GRADUATE PROGRAM

# OVERLAPPING ERROR CORRECTION CODES ON TWO-DIMENSIONAL STRUCTURES

ANDREW RAFAEL FRITSCH

Master Thesis submitted to the Pontifical Catholic University of Rio Grande do Sul in partial fulfillment of the requirements for the degree of master's in computer science.

Advisor: Prof. Dr. César Augusto Missio Marcon

# Ficha Catalográfica

ANDREW RAFAEL FRITSCH

# OVERLAPPING ERROR CORRECTION CODES ON TWO-DIMENSIONAL STRUCTURES

This Master Thesis has been submitted in partial fulfillment of the requirements for the degree of master's in computer science, of the Computer Science Graduate Program, School of Technology of the Pontifical Catholic University of Rio Grande do Sul

Sanctioned on March 2025.

## COMMITTEE MEMBERS

Prof. Dr. César Marcon (PPGCC/PUCRS – Advisor)

Prof. Dr. Avelino Zorzo (PPGCC/PUCRS)

Prof. Dr. Eduardo Bezerra (PPGEEL/UFSC)

# AGRADECIMENTOS

Agradeço ao professor e orientador César Marcon pela atenção e revisões que foram imprescindíveis para o desenvolvimento deste trabalho.

Agradeço aos meus familiares pelo apoio durante essa jornada de dois anos de mestrado.

Agradeço também ao Conselho Nacional de Desenvolvimento Científico e Tecnológico (CNPq), o Ministério da Ciência, Tecnologia, Inovações e Comunicações (MCTIC) e a Pontifícia Universidade Católica do Rio Grande do Sul (PUCRS) pelo apoio financeiro, estrutural e tecnológico disponibilizados.

# Sobreposição de Códigos de Correção de Erros em Estruturas Bidimensionais

## RESUMO

A crescente demanda por sistemas de comunicação altamente confiáveis impulsiona a pesquisa e o desenvolvimento de algoritmos capazes de identificar e corrigir erros que ocorrem durante a transmissão e o armazenamento de dados. Essa necessidade torna-se ainda mais crítica em sistemas de difícil acesso ou de natureza sensível, como os utilizados em aplicações espaciais, no transporte de passageiros e no setor financeiro. Nesse contexto, os Códigos de Correção de Erros (Error Correction Codes – ECCs) são ferramentas essenciais para garantir um certo nível de confiabilidade. Este trabalho propõe uma técnica para aumentar a capacidade de correção dos ECCs por meio da sobreposição de regiões de dados. A abordagem consiste em proteger uma mesma área de dados com múltiplos ECCs organizados em uma estrutura bidimensional, permitindo inferências lógicas que correlacionam os códigos e ampliam sua capacidade de detecção e correção de erros. Mais especificamente, a sobreposição é caracterizada pela organização de múltiplos ECCs cuja interseção abrange exclusivamente toda a região de dados. Para avaliar a proposta, diferentes organizações de ECCs sobrepostos foram analisadas em termos de capacidade de detecção e correção de erros, escalabilidade e confiabilidade. Os resultados experimentais comprovam a eficácia da técnica e demonstram que a mesma possui alto potencial de escalabilidade, reduzindo a necessidade de bits de redundância em relação ao número de bits de dados. Além disso, comparações com abordagens do estado da arte em ECC indicam a aplicabilidade da técnica em sistemas críticos que exigem alta confiabilidade.

# Overlapping Error Correction Codes on Two-Dimensional Structures

## ABSTRACT

The growing demand for highly reliable communication systems drives research and development of algorithms capable of identifying and correcting errors that occur during data transmission and storage. This need becomes even more critical in hard-to-access or sensitive systems, such as those used in space applications, passenger transportation, and the financial sector. In this context, Error Correction Codes (ECCs) are essential tools for ensuring a certain level of reliability. This work proposes a technique to enhance ECC error correction capability through the overlapping of data regions. The approach consists of protecting the same data area with multiple ECCs organized in a two-dimensional structure, enabling logical inferences that correlate the codes and improve their error detection and correction capabilities. More specifically, the overlapping is characterized by the organization of multiple ECCs, whose intersection exclusively covers the entire data region. To evaluate the proposal, different configurations of overlapping ECCs were analyzed in terms of error detection and correction capability, scalability, and reliability. Experimental results confirm the effectiveness of the technique and demonstrate its high scalability potential, reducing the need for redundancy bits relative to the number of data bits. Furthermore, comparisons with state-of-the-art ECC approaches indicate the applicability of the technique in critical systems that require high reliability.

## LIST OF ABBREVIATIONS

| | |
|---|---|
| 1D | One-Dimensional |
| 1D-ECC | One-Dimensional ECC |
| 2D | Two-Dimensional |
| 2D-ECC | Two-Dimensional ECC |
| BCH | Bose-Chaudhuri-Hocquenghem |
| DMC | Decimal Matrix Code |
| ECC | Error Correction Code |
| EDC | Error Detection Code |
| EMPC-SA | Modified Product Code for Space Applications |
| eMRSC | Extended Matrix Region Selection Code |
| EPC | Extended Product Code |
| HVDB | Horizontal-Vertical-Diagonal-Block |
| HVDD | Horizontal-Vertical-Double-Bit Diagonal |
| IC | Integrated Circuit |
| LPC | Line Product Code |
| MBU | Multiple Bit Upset |
| MC | Mixed Code |
| MCU | Multiple Cell Upset |
| MRSC | Matrix Region Selection Code |
| MTTF | Mean Time To Failure |
| PC | Product Code |
| PCoSA | Product Code for Space Applications |
| RM | Reed-Muller |
| RTL | Register-Transfer Level |
| S2E | Straightforward 2D-ECC |
| SBU | Single Bit Upset |
| SEC | Single Error Correction |
| SECDED | Single Error Correction - Double Error Detection |
| SEE | Single Event Effect |
| SEFI | Single Event Functional Interruption |
| SEL | Single Event Latch-up |
| SET | Single Event Transient |
| SEU | Single Event Upset |
| SPC | Single Parity Check |
| SRAM | Static Random-Access Memories |
| TLC | Triple-Level Cell |
| TMR | Triple Modular Redundancy |
| UBER | Uncorrectable Bit Error Rate |
| VHDL | VHSIC Hardware Description Language |
| VHSIC | Very High-Speed Integrated Circuits |

# LIST OF TABLES

# LIST OF FIGURES

# SUMMARY

# 1. INTRODUCTION

Error correction and detection coding are methods used to handle errors in data transmitted through communication channels. Together, they form error control coding, crucial for properly functioning communication and storage systems. These techniques are essential in the telecommunications revolution, the Internet, digital recording, and space exploration. They are widely employed in devices such as compact discs, DVDs, hard drives, memory systems, and cell phones. Error control coding is fundamental in the modern information-based society, ensuring data integrity across various technologies and applications [37].

In digital computing and telecommunications, information is almost always represented in binary form as sequences of bits where values are defined as 0 or 1. To transmit or store this information—which may include words, punctuation, or videos—via analog or digital signals, communication in binary form always occurs during the process. The content of this message, or encoded word, is organized into a logical format. To verify the information's correctness, it is necessary to evaluate the logical structure of the information to ensure the completeness and integrity of the received content; this verification relies on a method incorporating some redundancy [12][37].

An Error Correction Code (ECC) is commonly used to verify the integrity of stored or transmitted/received information. Error identification and analysis are enabled through coding and decoding techniques, allowing for validation whenever data is transmitted, received, or accessed during memory read/write operations [30]. ECC is a fault-tolerance technique widely used across a broad spectrum of applications, ranging from large-scale systems, such as memory controllers in high-performance servers [65], to high-criticality systems, such as satellites [66], and even in on-chip subsystems, such as transmission or storage in networks-on-chip [44][55].

Defining *efficacy* as the capability of a system to achieve its intended goal and *efficiency* as the cost of designing and operating the system, error detection, and correction are key elements for assessing the efficacy of an ECC. Similarly, area consumption, power dissipation, energy consumption, and latency are the primary elements for evaluating its efficiency.

The first ECCs were independently designed, meaning that only one type of ECC was applied to a specific data region. This approach resulted in low-complexity encoding/decoding processes and limited error correction capabilities, but with higher synthesis and operational efficiency [18]. Applying multiple ECCs to the same data region enables error correction and detection information to be correlated. This correlation can enhance the capacity to identify and correct errors during data transmission, reception, or storage, thereby improving information reliability. A product code is a typical example of ECC cross-correlation, where the correction capabilities of two independent ECCs are combined through cross-verification in a Two-Dimensional (2D) organization [35][18].

Theoretically, infinite ECC correlations can be performed, leading to varying levels of encoding/decoding complexity, different efficacies in error detection and correction, and varying synthesis and operational efficiencies. Literature reviews indicate that cross-correlation is the most common approach, where two ECCs protect the same data bit (e.g., product codes). However, no studies have explored ECCs with complete overlap of the data region, which is the focus of this dissertation.

## 1.1 General Objective

The primary goal of this master’s dissertation is to research, explore, implement, and validate ECCs with overlapping data regions to improve error correction and detection efficacy while concurrently evaluating their efficiency, reliability, and scalability.

## 1.2 Specific Objectives

To achieve the main goal of this study, in addition to conducting an extensive review of related works, the following specific objectives are proposed:

- Develop a framework to simulate various error patterns and evaluate the error correction and detection efficacy of ECCs with overlapping data regions;
- Explore error patterns in ECC regions (data, redundancy, and both);
- Synthesize the selected ECCs and collect data on power dissipation, area consumption, and latency;
- Define a fault model to assess the reliability of ECCs over time;
- Mathematically evaluate the scalability of the selected ECCs using metrics such as efficacy and redundancy costs.

# 2. THEORETICAL FOUNDATION

This chapter presents the scientific foundations employed in this study, including fault, error, and failure events, their causes, and the mechanisms used to mitigate these occurrences.

## 2.1 Concepts of Fault, Error, and Failure

Avizienis et al. [2] explain that a service is correct when it implements the functionality defined for the system. A ***failure*** occurs when the service deviates from its definition. If the service is understood as a sequence of states, then a failure implies that at least one state deviates from its definition, and this deviation is called an ***error***. The cause of an error is referred to as a ***fault***, which can be internal or external to the system. The presence of an internal fault allows an external fault to affect the system, causing errors and potentially resulting in subsequent failures. In most cases, a fault first causes an error in the service state of a component that is part of the system's internal state, while the external state remains unaffected initially.

Consider a logical AND gate with two inputs as an example of a system component. A grounded input (i.e., stuck at 0) represents a latent fault. An error occurs when both inputs assume a logical value of 1, as the output value becomes 0 (due to the fault) instead of 1. A failure arises if the system's decision deviates from its specification due to the AND gate's result; otherwise, the error remains latent.

This study focuses on faults caused by physical factors [51], such as voltage variations, temperature changes, radiation, magnetic fluctuations, electrical noise, electromagnetic interference, and stress over time. Chabot et al. [5] classify faults into three types based on their duration:

I. *Permanent Fault* – Caused by a physical event that affects the entire lifetime of the system, such as a short circuit or an open circuit, which can only be corrected by replacing the hardware;
II. *Intermittent Fault* – Occurs sporadically at irregular intervals. Intermittent faults are often early indicators of potential permanent faults;
III. *Transient Fault* – Occurs randomly, primarily due to the impact of charged particles [27]. This type of fault manifests as one or more bitflips and can be corrected without hardware replacement.

In this study, we focus on addressing transient faults by employing mechanisms that detect them and restore the system to its correct state.

## 2.2 Single Event Effect (SEE)

The downscaling of Integrated Circuits (ICs) enhances the computational power of systems but also makes devices more susceptible to faults caused by radiation effects. One of the most common fault-inducing phenomena in electronic circuits is the Single Event Effect (SEE) — an electrical disturbance that alters the operation of a circuit. Charged particles passing through transistor junctions can induce SEEs. The transistor's behavior depends on the ion charge at impact. Figure 1 illustrates how a highly charged ion affects a transistor junction [3]: (a) the ion crosses the junction, generating a cylindrical track of highly charged electron-hole pairs; (b) the charge imbalance induces the creation of a temporary funnel; (c) when the funnel dissipates, the remaining ions are balanced through diffusion.

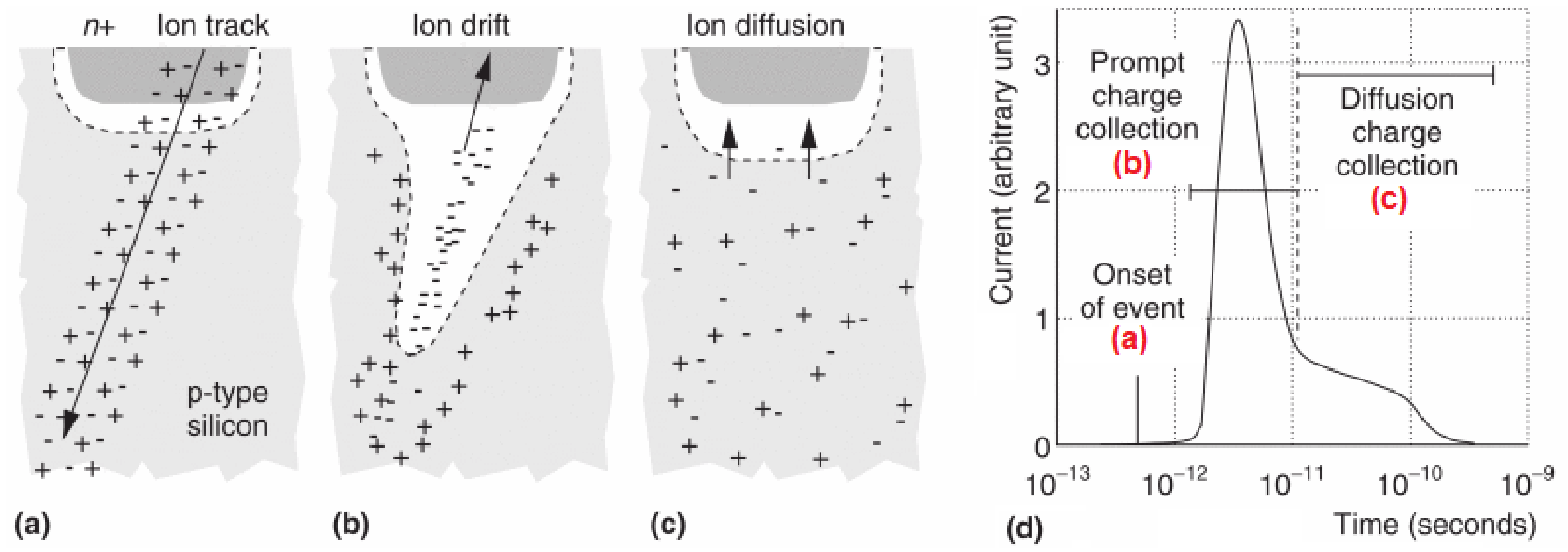

**Figure 1. Effect of a charged particle passing through a transistor junction: (a) cylindrical track of electron-hole pairs; (b) funnel extending the depletion region; (c) diffusion dominating the collection process; (d) resulting current pulse (Source: adapted from [3]).**

While SEEs are commonly observed in space applications due to solar radiation and cosmic rays, at ground level, alpha particles (more prevalent) and neutrons can modify the system state, causing transient and sometimes permanent faults [67]. Nicolaidis [42] categorizes SEEs into types of faults based on their logical and physical scope:

I. *Single Bit Upset* (SBU) – An SEE affects a single bitflip in a single memory cell, often synonymous with *Single Event Upset* (SEU);
II. *Multiple Cell Upset* (MCU) – An SEE alters two or more memory cells;
III. *Multiple Bit Upset* (MBU) – An SEE flips two or more bits within the same word;
IV. *Single Event Transient* (SET) – An SEE causes a voltage fault in a circuit;

V. *Single Event Functional Interruption* (SEFI) – An SEE disrupts functionality due to interference with registers, clocks, resets, and others;

VI. *Single Event Latch-up* (SEL) – An SEE induces an abnormally high current, requiring a power reset.

This study focuses solely on SEEs that result in transient faults such as SBU/SEU, MCU, and MBU, excluding systemic faults.

Memories are susceptible to radiation, making SEE constantly threatening systems exposed to charged particles. Designers must understand the most likely SEEs to mitigate operational issues. Memory errors have been extensively analyzed over decades; studies indicate that a significant fraction of these errors recur at the same address [62][63][64]. Furthermore, faults often cluster spatially and temporally, showing strong correlations [28].

The continuous scaling down of transistors aligns with Moore's Law [38], introducing the challenge of scaling faults [34][41]. Smaller transistors directly increase hardware sensitivity to temporary errors, leading to higher MCU occurrences in modern technologies [10]. Typically, a SEE affects a single bit (SBU); however, Figure 2 demonstrates the growing presence of MBUs as technology scales down. For example, in SRAMs below 40 nm, more than 40% of particle impacts result in MBUs due to reduced threshold voltages, lower capacitances per transistor, and the smaller volume of transistors, which increases the likelihood of a single SEE affecting multiple neighboring transistors [5].

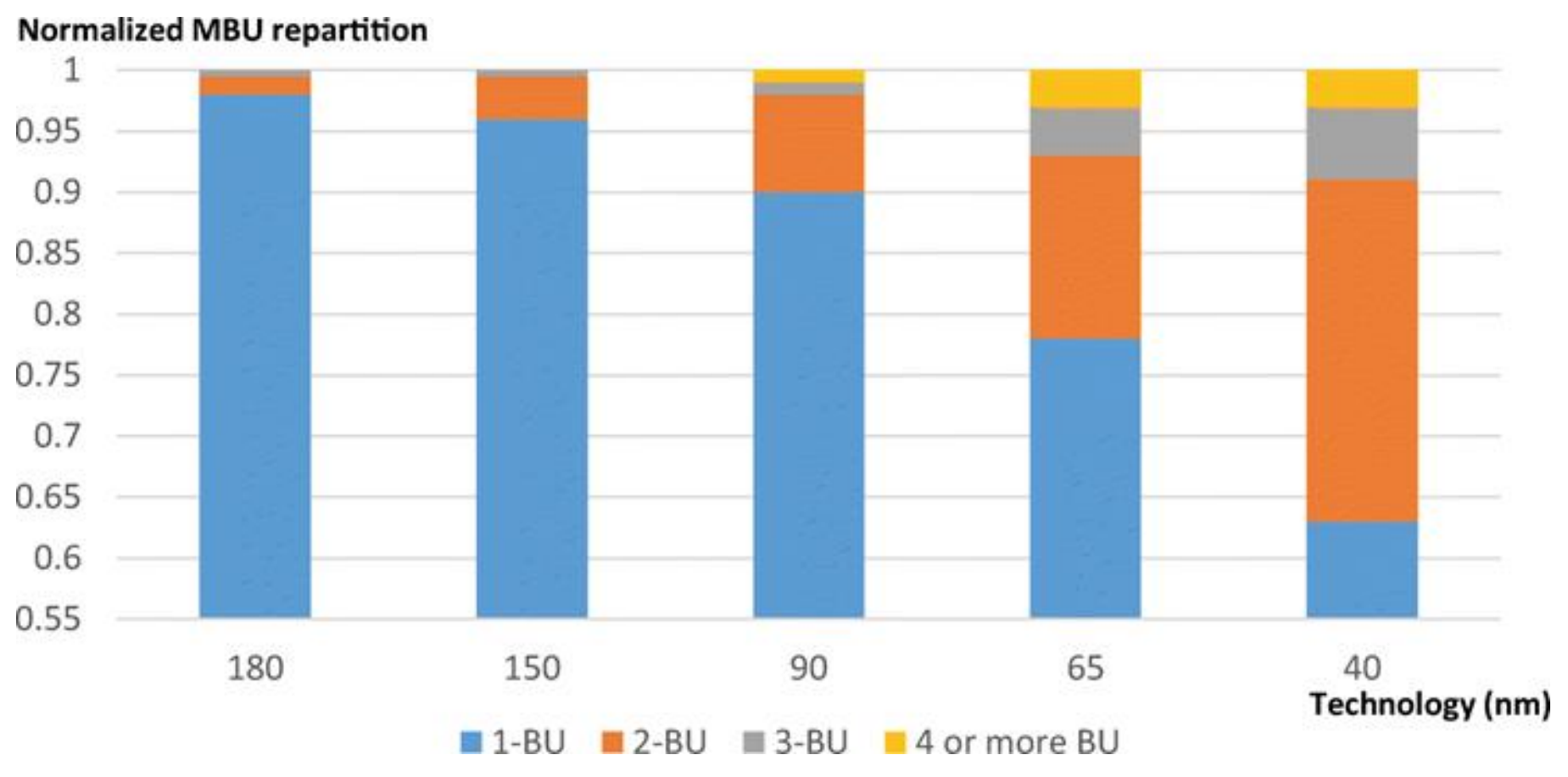

**Figure 2. Percentage of MBU occurrences by technology node (nm) for SRAMs (Source: [5]).**

Gracia-Morán et al. [24] simulated SEEs in 45 nm memories under terrestrial radiation levels. As illustrated in Figure 3, while SBUs predominate, MBUs represent nearly half of the occurrences, including many double and triple errors.

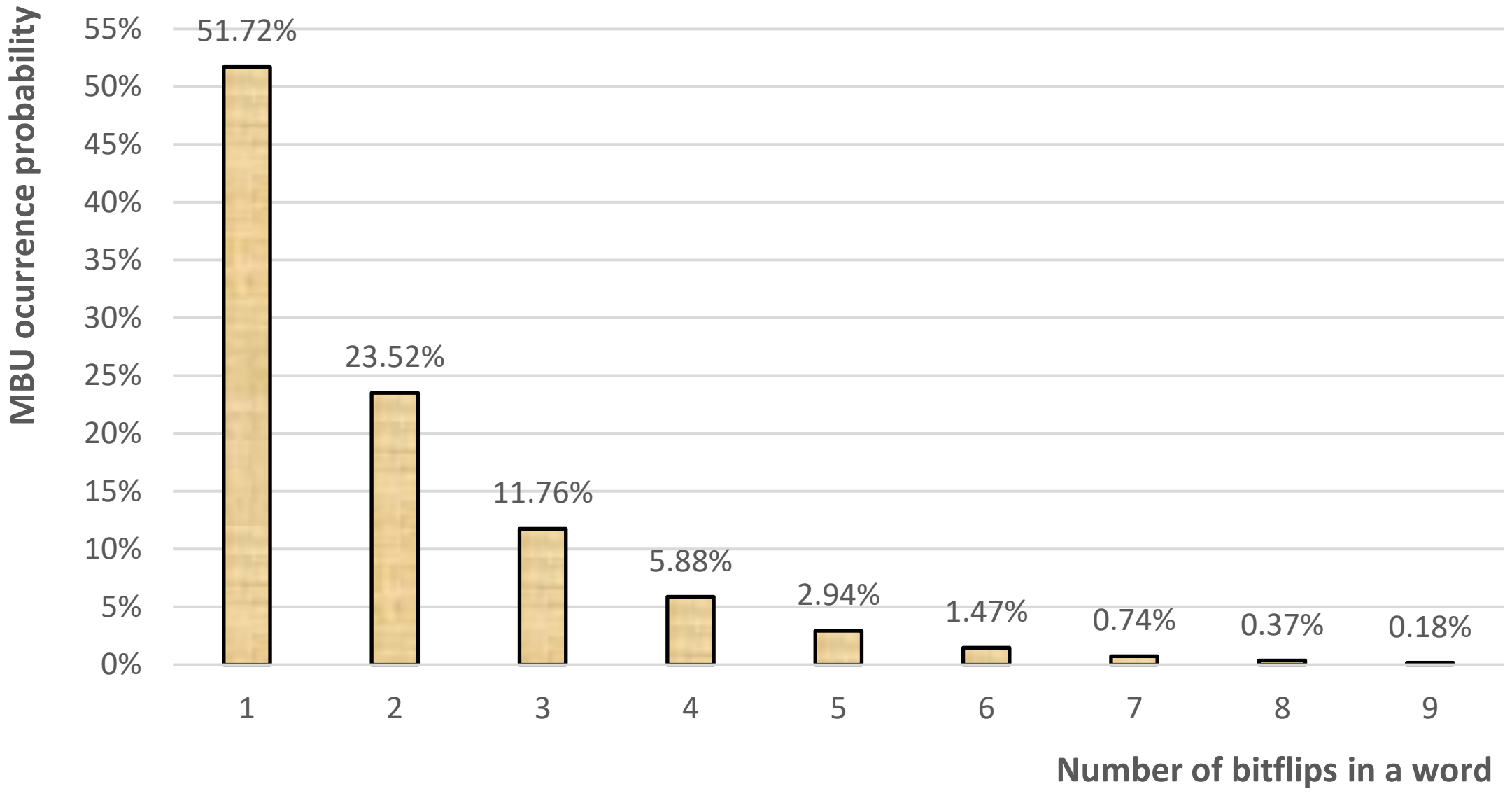

**Figure 3. Probability of an MBU in memory words caused by radiation impact (Source: adapted from [19]).**

## 2.3 Error Correction Code (ECC)

Various mechanisms exist to handle transient faults, applied at the hardware, software, or hybrid levels. One example is modular redundancy, which can be implemented in hardware and software. Triple Modular Redundancy (TMR) is a typical instance known for its cost-benefit ratio [5]. Another example is temporal redundancy, implemented in software, where the system state is periodically saved at checkpoints. In the event of an error, the system can revert to a safe checkpoint [45]. This dissertation focuses on Error Correction Code (ECC), a mechanism typically applied at the hardware level to achieve low latency, making it suitable for real-time applications. However, ECC can also be implemented in software when timing requirements are less critical.

ECC is composed of a structure that includes information and redundancy bits, along with encoding and decoding algorithms, aiming to detect and correct errors. Depending on its composition, this bit structure is called a codeword or a codestruct. While a codeword represents a word consisting of a vector of data bits and a vector of check bits, a codestruct is a collection of codewords, often organized in a matrix format. A single codeword or codestruct enables the implementation of multiple ECCs, depending on the chosen encoding and decoding methods, resulting in varying area and energy consumption, power dissipation, and latency [18].

The history of ECCs dates to the early days of computing and telecommunications. With increasing system complexity and growing demand for reliability, the need for effective error detection and correction methods became evident. In the 1940s, during the early computing era, ECCs began to be explored with the advent of the first electronic computers. Basic parity-check algorithms were used for error detection, but error correction capabilities remained limited [54].

Advancements in information theory and coding led to the development of robust ECCs, such as the Hamming code, which enabled error detection and correction in specific contexts [26]. Continuous progress in error correction research eventually produced more sophisticated codes, such as Reed-Solomon, widely used in modern storage and transmission systems [50]. Today, ECCs are extensively implemented across a variety of devices and technologies. The ongoing evolution of these codes aims to address the challenges posed by increasingly complex systems and diverse technologies [50].

## 2.4 Hamming Distance

The Hamming Distance is the number of positions at which the corresponding bits of two equal-length vectors differ. It is a metric that expresses the minimum number of errors required to transform the content of one vector into another. The Hamming Distance determines the error detection and correction limits of a code [37]. This distance is exclusively related to the organization of data and check bits (i.e., codeword or codestruct), with encoding and decoding algorithms serving as enablers to achieve the theoretical limits defined by Hamming.

Let $d$ be the Hamming Distance. Then, (1) and (2) calculate the maximum number of errors a code can correct $EC$ or detect $ED$ based on $d$ [37].

$$EC = \left\lfloor \frac{d-1}{2} \right\rfloor \tag{1}$$

$$ED = d - 1 \tag{2}$$

These equations are mutually exclusive, meaning a code can perform either error correction $EC$ or error detection $ED$, but not simultaneously. To simultaneously perform error detection and correction, (3) must replace (2). Consequently, $ED$ is reduced for applications that aim to correct and detect errors at the same time [18].

$$ED = d - EC - 1 \tag{3}$$

## 2.5 Parity Code

A parity code adds a bit to a data vector, indicating the number of 1s or 0s in the vector. For example, the parity encoder counts the number of data bits set to 1 in an even parity code. If this number is even, the encoder sets the parity bit to 0; otherwise, it sets the parity bit to 1. During data reading, the parity decoder checks whether the number of 1s matches the expected parity to determine if the data is correct or corrupted. Table 1 illustrates an example of even parity validation.

**Table 1. Example of even parity with a 9-bit codeword, consisting of a data byte (Data) and a parity bit (P) (Source: Author).**

| Codeword | | Number of 1s in the codeword | Error detected? |
|---|---|---|---|
| Data | P | | |
| 00000000 | 0 | 0 | NO |
| 10000001 | 1 | 3 | YES |
| 11100000 | 1 | 4 | NO |
| 10000000 | 1 | 2 | NO |

Parity errors are detectable only when an odd number of bits are corrupted. Consequently, an even number of bit flips can remain undetected. Fortunately, SBUs are the most common type of data corruption [52].

Parity bits are frequently used in memory modules, serial communications, and other systems where error detection is important, but error correction is not critical [30].

The decoding algorithm sums all bits set to 1 using XOR (exclusive OR) logic on the data vector. The logical result is compared to the stored parity bit. Figure 4 illustrates the read and write operations; XOR logic is applied to all bits during a write operation (encoding), and the result is stored as the parity bit. During a read operation (decoding), the stored parity bit is compared to the XOR result of the data bits.

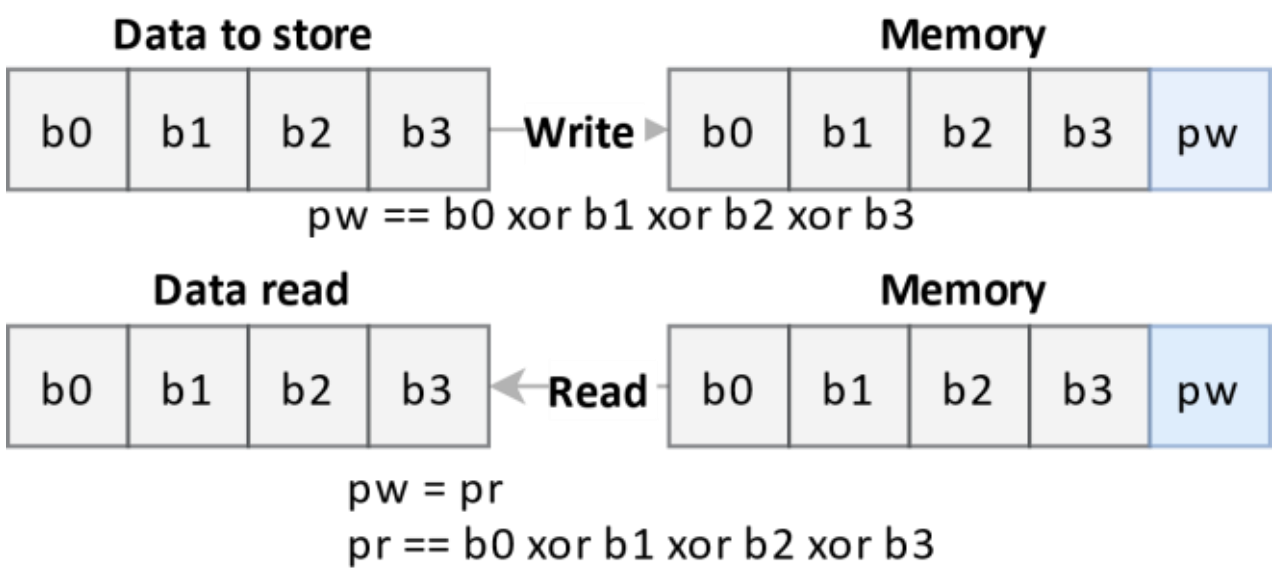

**Figure 4. Reading and writing 4 data bits (b0 to b3) in memory using the parity code stored in pw (Source: [5]).**

Parity code is not technically an ECC but an Error Detection Code (EDC), with a Hamming distance of $d = 2$. Consequently, it has $EC = 0$ (no error correction) and $ED = 1$ (detection of 1 error). The code can detect an odd number of errors but cannot identify the number of errors or their locations, making it incapable of error correction.

However, parity code is often presented as a foundational ECC, mainly because many ECCs are composed of parity combined with other codes. The correlation of parity codes enables the creation of error correction algorithms, as seen in 2D ECCs [18].

## 2.6 Hamming Code

Hamming was one of the first ECCs developed for computational systems to correct SBUs, given that $d = 3$ [18], resulting in $EC = 1$ and $ED = 2$. This attribute places Hamming in the class of ECCs called *Single Error Correction* (SEC) codes. Let $M$ be the data vector and $k$ the number of check bits; then $N$ represents the codeword, and the total number of bits in the codeword is given by (4). The Hamming code is represented as $Ham(|N|, |M|)$.

$$|N| = |M| + k \tag{4}$$

Additionally, the Hamming code must satisfy inequality (5) [37], which means that $k$ check bits generate $2^k$ values. These values allow the identification of the error's location within the vector of check bits (referenced by $k$) and data bits ($M$), as well as an error-free condition (referenced by 1). The inequality is defined as "≥" because $M$ can represent a vector with fewer bits, resulting in $2^k$ values greater than the required number of addresses, with the extra addresses discarded.

$$2^{\mathrm{k}} \geq \mathrm{k} + |\mathrm{M}| + 1 \tag{5}$$

Hamming encoding and decoding processes utilize identity, generation, and verification matrices, as illustrated in the $Ham(7{,}4)$ example below – an example of a Hamming code that employs 3 check bits to protect 4 data bits. The square identity matrix $I_{2^k-k-1}$, of order $2^k - k - 1$, and matrix $Q$, covering the data addresses 011, 101, 110, and 111 in the event of an error, are described by (6) and (7), respectively.

$$I_{2^k-k-1} = \begin{bmatrix} 1 & 0 & 0 & 0 \\ 0 & 1 & 0 & 0 \\ 0 & 0 & 1 & 0 \\ 0 & 0 & 0 & 1 \end{bmatrix} \tag{6}$$

$$Q = \begin{bmatrix} 0 & 1 & 1 \\ 1 & 0 & 1 \\ 1 & 1 & 0 \\ 1 & 1 & 1 \end{bmatrix} \tag{7}$$

The generator matrix $G$, used to initiate the Hamming encoding process, is represented by (8), while (9) illustrates $G(4,7)$.

$$\mathrm{G} = [\ \mathrm{I}_{2^{\mathrm{k}}-\mathrm{k}-1}\ \mathrm{Q}] \tag{8}$$

$$\mathrm{G}(4,7) = \begin{bmatrix} 1 & 0 & 0 & 0 & 0 & 1 & 1 \\ 0 & 1 & 0 & 0 & 1 & 0 & 1 \\ 0 & 0 & 1 & 0 & 1 & 1 & 0 \\ 0 & 0 & 0 & 1 & 1 & 1 & 1 \end{bmatrix} \tag{9}$$

Equation (10) describes the linear transformation that multiplies the data vector $M(1,4)$ by the generator matrix $G(4,7)$ to encode $Ham(7,4)$ into the codeword $N(1,7)$.

$$\mathrm{N} = \mathrm{M} \times \mathrm{G} \tag{10}$$

For example, (11) illustrates that the vector $M = [1000]$ encoded with $Ham(4,7)$ results in the codeword $N = [1000011]$. Note that the first four bits $[1000]$ represent the data, while the remaining three bits $[011]$ are check bits.

$$N = [1 \quad 0 \quad 0 \quad 0] \times \begin{bmatrix} 1 & 0 & 0 & 0 & 0 & 1 & 1 \\ 0 & 1 & 0 & 0 & 1 & 0 & 1 \\ 0 & 0 & 1 & 0 & 1 & 1 & 0 \\ 0 & 0 & 0 & 1 & 1 & 1 & 1 \end{bmatrix} = [1 \quad 0 \quad 0 \quad 0 \quad 0 \quad 1 \quad 1] \tag{11}$$

Decoding is the inverse encoding process, requiring examining whether the check bits match the values obtained during encoding. Let $Q^T$ be the transposed matrix of $H$ and $I_k$, the square identity matrix of order $k$, computed by (12) and (13), respectively. Then, (14) computes the redundancy matrix $H$.

$$\mathrm{Q}^{\mathrm{T}} = \begin{bmatrix} 0 & 1 & 1 & 1 \\ 1 & 0 & 1 & 1 \\ 1 & 1 & 0 & 1 \end{bmatrix} \tag{12}$$

$$\mathrm{I}_{\mathrm{k}} = \begin{bmatrix} 1 & 0 & 0 \\ 0 & 1 & 0 \\ 0 & 0 & 1 \end{bmatrix} \tag{13}$$

$$H = [Q^T I_k] = \begin{bmatrix} 0 & 1 & 1 & 1 & 1 & 0 & 0 \\ 1 & 0 & 1 & 1 & 0 & 1 & 0 \\ 1 & 1 & 0 & 1 & 0 & 0 & 1 \end{bmatrix} \quad (14)$$

Let $N'$ be the codeword read during the decoding process, such that $N' = N$ in the absence of errors. Then, $N'^T$ represents the transposed matrix of $N'$, as described in (15). Equation (16) multiplies $H$ by $N'^T$ to compute the syndrome vector $S[s_0\ s_1\ s_2]$. Note that this example produces $S = [000]$, as the matrices used in encoding ($N$) and decoding ($N'$) are identical, indicating no errors.

$$N'^T = \begin{bmatrix} 1 \\ 0 \\ 0 \\ 0 \\ 0 \\ 1 \\ 1 \end{bmatrix} \quad (15)$$

$$S = H \times N'^T = \begin{bmatrix} 0 & 1 & 1 & 1 & 1 & 0 & 0 \\ 1 & 0 & 1 & 1 & 0 & 1 & 0 \\ 1 & 1 & 0 & 1 & 0 & 0 & 1 \end{bmatrix} \times \begin{bmatrix} 1 \\ 0 \\ 0 \\ 0 \\ 0 \\ 1 \\ 1 \end{bmatrix} = [0 \quad 0 \quad 0] \quad (16)$$

Although the mathematical modeling of Hamming is based on matrix computation, the encoding and decoding processes can also be performed using XOR logic ($\oplus$). Equations (17) to (19) illustrate, for the $Ham(7{,}4)$ example, the calculations for the check vector $C[c_0\ c_1\ c_2]$ produced during encoding, which, along with the data $D[d_0\ d_1\ d_2\ d_3]$, can be transmitted or stored.

$$c_0 = d_1 \oplus d_2 \oplus d_3 \quad (17)$$

$$c_1 = d_0 \oplus d_2 \oplus d_3 \quad (18)$$

$$c_2 = d_0 \oplus d_1 \oplus d_3 \quad (19)$$

During decoding, a similar process is performed; the check vector is recomputed as $C'[c'_0\ c'_1\ c'_2]$, now considering the received/read data vector $D'[d'_0\ d'_1\ d'_2\ d_{\prime 3}]$, as shown in (20) to (22).

$$c'_0 = d'_1 \oplus d'_2 \oplus d'_3 \quad (20)$$

$$c'_1 = d'_0 \oplus d'_2 \oplus d'_3 \quad (21)$$

$$c'_2 = d'_0 \oplus d'_1 \oplus d'_3 \quad (22)$$

Additionally, (23) to (25) demonstrate that $S$ is computed during decoding through $\oplus$ operations between the bits of $C$ and $C'$.

$$s_0 = c_0 \oplus c'_0 \quad (23)$$

$$s_1 = c_1 \oplus c'_1 \quad (24)$$

$$s_2 = c_2 \oplus c'_2 \quad (25)$$

In Hamming codes, the codeword bits are numbered consecutively, starting with bit 1 on the far left, followed by bit 2 immediately to its right, and so on. Also, check bits occupy positions that are powers of 2 (i.e., 1, 2, and 4), while the remaining are data bits (i.e., 3, 5, 6, and 7). Thus, the codeword for $Ham(7,4)$ is represented as $N = \begin{bmatrix} c_0 & c_1 & d_0 & c_2 & d_1 & d_2 & d_3 \\ 1 & 2 & 3 & 4 & 5 & 6 & 7 \end{bmatrix}$. However, Hamming codewords are typically presented with data and check bits grouped, as shown in the first column of Table 2. This table relates the eight possible syndromes to their corresponding error addresses and identifies which bit in the codeword is erroneous. Note that $S = [000]$ indicates no errors, while all other combinations correspond to the addresses where errors occurred. The error address is derived by combining the weights of the check bits, as illustrated in (26).

$$\text{Error address} = 4 \times c_2 + 2 \times c_1 + c_0 \quad (26)$$

**Table 2. Relations between the codeword, syndrome bit vector, and error address (Source: Author).**

| Codeword | $S$ | Error address | Bit with error |
|---|---|---|---|
| $N = \begin{bmatrix} d_0 & d_1 & d_2 & d_3 & c_0 & c_1 & c_2 \\ 3 & 5 & 6 & 7 & 1 & 2 & 4 \end{bmatrix}$ *data bits* *check bits* | [000] | 0 | Ø |
| | [001] | 1 | $[\quad\quad\quad\quad\quad\quad c_2]$ |
| | [010] | 2 | $[\quad\quad\quad\quad\quad c_1 \quad]$ |
| | [011] | 3 | $[d_0 \quad\quad\quad\quad\quad\quad]$ |
| | [100] | 4 | $[\quad\quad\quad\quad c_0 \quad\quad]$ |
| | [101] | 5 | $[\quad d_1 \quad\quad\quad\quad\quad]$ |
| | [110] | 6 | $[\quad\quad d_2 \quad\quad\quad\quad]$ |
| | [111] | 7 | $[\quad\quad\quad d_3 \quad\quad\quad]$ |

For example, encoding the data vector $M = [1000]$ results in $C = [011]$ and, consequently, the codeword $N = [1000\ 011]$. Suppose an error occurs in the second data bit ($d_1$), resulting in the codeword $N' = [1100011]$. In this case, the recalculation of the check bits produces $C' = [110]$, during decoding. Applying (24) to (26) generates a syndrome vector $S = [101]$, indicating that two check bits have changed. Finally, using (17), it is determined that the error address is 5, which corresponds to $d_1$, as shown in Table 2.

Note that to protect larger data areas, the Hamming code can grow indefinitely while maintaining a growth ratio close to the logarithm base 2. In all cases, it always retains the characteristic of being a SEC code. Additionally, data areas can be protected using Hamming combinations [19], which enhances error correction efficacy but increases efficiency penalties due to the higher number of check bits.

## 2.7 Extended Hamming Code

The Extended Hamming Code is formed by adding a parity bit to the standard Hamming code, increasing the Hamming distance to $d = 3$ [18]. This enhancement allows the code to correct one error ($EC = 1$) and detect two errors ($ED = 2$). Thus, the Extended Hamming Code belongs to the class of ECCs known as S*ingle Error Correction - Double Error Detection* (SECDED) [46].

The Extended Hamming Code is represented as $ExHam(|N^*|, |M|)$, where $|N^*| = |N| + 1$, with $N$ being the codeword of the basic Hamming code and “1” referring to the additional parity bit. In addition to the syndrome vector $S$, obtained by comparing the computed and recomputed check bits, $ExHam(|N^*|, |M|)$ includes the parity syndrome $\delta$, calculated by applying XOR between the parity bit generated during encoding ($p$) and the parity bit recalculated during decoding ($p'$), as described in (27).

$$\delta = p \oplus p' \tag{27}$$

Let $\boldsymbol{s}$ be the logical OR (∨) operation applied to all bits of $S$, as described in (28). Table 3 then describes the type of error based on the combination of the Hamming and parity syndromes.

$$\boldsymbol{s} = s_0 \lor s_1 \lor s_2 \tag{28}$$

**Table 3. Error types of based on the combination of Hamming and parity syndromes (Source: Author).**

| $s$ | $\delta$ | Type of error |
|---|---|---|
| 0 | 0 | No error |
| 0 | 1 | Single error in parity |
| 1 | 0 | Double error |
| 1 | 1 | Single error |

Using an example like Section 2.6, $M = [1000]$ results in $N^* = [1000011\mathbf{1}]$, as parity is calculated by considering all data and check bits, as illustrated in (29).

$$p = d_0 \oplus d_1 \oplus d_2 \oplus d_3 \oplus c_0 \oplus c_1 \oplus c_2 \quad (29)$$

In the occurrence of two errors in $N^* = [\mathbf{01}000111]$, the parity syndrome is zero ($\delta = 0$) and $\boldsymbol{s} = 1$ ($S = [110]$). The combination of $\delta$ and $\boldsymbol{s}$ indicates a double error detection; however, the positions of the errors are unknown, and therefore, $N^*$ cannot be corrected.

## 2.8 Two-Dimensional Error Correction Code (2D-ECC)

The scaling of integrated circuits increases their susceptibility to MBUs, requiring more effective ECCs that, in turn, dissipate more power and consume more area and energy. Additionally, a *One-Dimensional ECC (1D-ECC)* demands long redundancy sequences to achieve high efficacy, especially for large data words [5]. Thus, a *Two-Dimensional Error Correction Code (2D-ECC)* emerges as an efficient solution for handling MBUs. Its format combines low-cost codes to achieve efficacy similar to 1D-ECCs without overloading the codewords with redundancy bits.

A 2D-ECC is characterized by having data and/or redundancy bits in two dimensions, typically referred to as rows and columns. This definition includes any 1D-ECC physically organized into rows and columns within the 2D-ECC class. Consequently, Freitas et al. [18] classify 2D-ECCs into four categories:

VII. *Straightforward 2D-ECC* (S2E) – A code organized in a 2D physical structure but correcting errors using 1D algorithms.

VIII. *Product Code* (PC) – An ECC treated as a product of two codes, allowing the construction of long codes from smaller ones.

IX. *Extended Product Code* (EPC) – A special case of PC that uses more than one code per row and/or column.

X. *Mixed Code* (MC) – A 2D-ECC that contains at least one data or redundancy bit whose modification impacts encoding in both dimensions but cannot be classified as PC or EPC.

Let $\alpha$ and $\beta$ be the number of columns comprising the data and redundancy areas, and $\vartheta$ and $\varepsilon$ the number of rows comprising the data and redundancy areas, respectively, such that $\gamma = \alpha + \beta$ and $\theta = \vartheta + \varepsilon$. Then, each row and column of the PC is encoded using the codes $C_1(\gamma, \alpha, d_1)$ and $C_2(\theta, \vartheta, d_2)$, respectively, forming the code $C_1 \times C_2$. Any bitflip in the data region disturbs the row and column of the corresponding bit. Figure 5(a) illustrates the basic structure of the PC. Furthermore, PC adds a region containing check bits for the check bits, increasing the Hamming distance and, consequently, the code's correction potential [37]. However, some codes implement the product code without including the region of check bits for the check bits, aiming to reduce the redundancy costs. These codes are referred to as modified PCs, as illustrated in Figure 5(b).

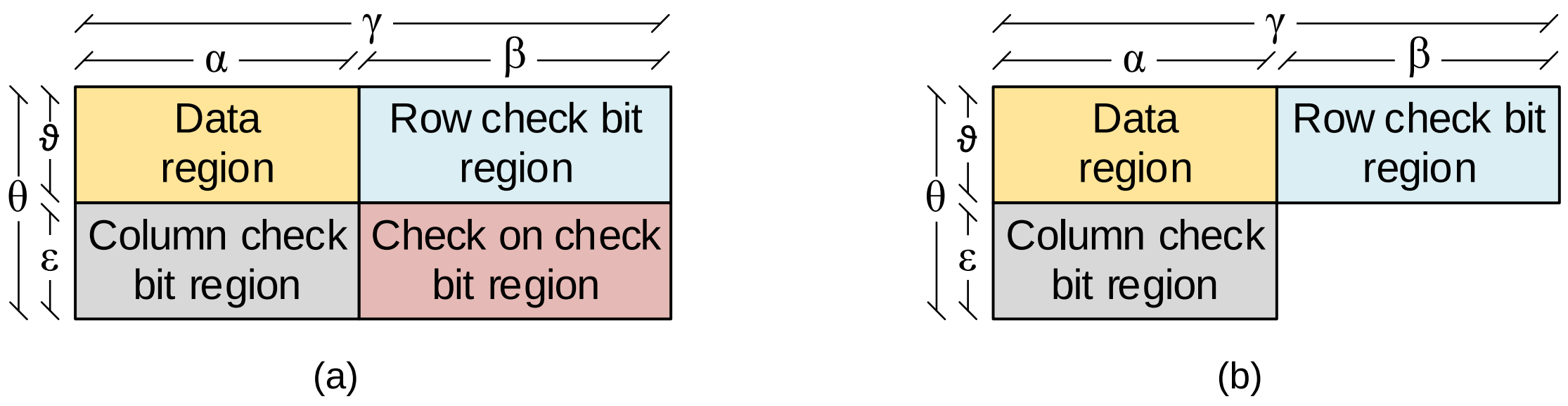

**Figure 5. Basic structures of (a) PC and (b) modified PC (Source: [18]).**

Several 2D-ECCs exist, such as Matrix [1], Decimal Matrix Code (DMC) [25], Horizontal-Vertical-Double-Bit Diagonal (HVDD) [48], Product Code for Space Applications (PCoSA) [14], Extended Matrix Region Selection Code (eMRSC) [56], and Line Product Code (LPC) [15]. These 2D-ECCs are often compositions of basic ECCs, resulting in heterogeneous structures. While they all employ cross-correlation of ECCs, none utilize the data overlapped technique proposed in this work. Additionally, the 2D organizations proposed in this study are not fully aligned with any classifications presented by Freitas et al. [18], with the closest match being the MC code class.

# 3. RELATED WORK

This chapter presents a bibliographic review focusing on the analysis of 2D-ECCs conducted using the following research platforms:

- IEEE Xplore (https://ieeexplore.ieee.org/)
- Elsevier (https://www.sciencedirect.com/)
- ACM (https://dl.acm.org/)
- MDPI (https://www.mdpi.com/)

Additionally, for greater comprehensiveness, Google search tools were also employed, allowing the inclusion of articles from other research platforms.

To conduct the research, we based our work on the foundational use of ECC as an error correction technique pioneered by Richard W. Hamming in his seminal 1950 publication, "Error Detecting and Error Correcting Codes", in the Bell System Technical Journal. Since then, hundreds of studies have explored ECC with various applications, such as providing reliability for different communication media and data storage systems. Two-dimensional codes (2D-ECC) also have a long history, with one of the earliest implementations being the Product Code, introduced by Peter Elias in 1954 in his paper "Error-Free Coding" published in the IRE Transactions on Information Theory.

To evaluate the level of innovation in our proposed approach and its relationship to existing works, we initially conducted a broad exploratory search for error correction codes that referenced concepts similar to what we term here as "***overlapping***". These included keywords such as "overlay," "crossing," and "sharing," focusing primarily on the titles and abstracts of articles without restricting the search to a specific starting date. This exploratory search did not identify prior works proposing a methodology like ours.

We emphasize that the innovation in our proposal lies in the *overlapping model*, where more than one independently encoded ECC protects the same data region. Additionally, although the codes are encoded independently, they must be pre-related so that their encodings have characteristics enabling enhanced error correction efficacy during decoding. The distinct algorithmic organization and characteristics of our method prevented it from being classified within any of the categories proposed by Freitas et al. [18]. Thus, we classified the method as ***overlapping***. Given these considerations, we conducted a detailed search focusing solely on articles written in English and published since 2010. The search for related articles was carried out using the query string:

*(“Reliability” OR “Fault Tolerance” OR “Fault-Tolerant”) AND (“ECC 2D” OR “2D-ECC” OR “Two-Dimensional Error Correction Code”)*.

A total of 97 articles were retrieved, and their abstracts and part of the text content were analyzed to identify the primary techniques employed and the methods used for algorithm validation. Notably, no studies employed the ***Overlapping Technique*** proposed in this work. Consequently, we prioritized selecting highly cited articles, techniques, or ECCs that were most comparable to the methods proposed here, as well as recent research.

## 3.1 Matrix-Based Codes for Adjacent Error Correction

Argyrides et al. [1] addressed the challenges of MCUs in memory systems caused by technology scaling and radiation effects. The authors propose the *Matrix*-based 2D-ECC that combines horizontal and vertical parity checks to correct adjacent bit errors efficiently. This solution targets applications where conventional methods like interleaving and Hamming codes fall short in handling closely spaced error patterns.

The proposed technique organizes data into a 2D matrix, applying parity bits along rows and columns, as illustrated in Figure 6. This configuration allows the correction of up to four adjacent bit errors within a word, possibly scaling to larger matrices, such as 8×8, to address more complex error patterns. The design is tailored for environments with common clustered MCUs, such as space applications and advanced SRAM memories, where interleaving is not feasible due to physical constraints or performance trade-offs.

| | | | | | | | | | |
|---|---|---|---|---|---|---|---|---|---|
| $X_1$ | $X_2$ | $X_3$ | $X_4$ | $X_5$ | $X_6$ | $X_7$ | $X_8$ | $C_1$ | $C_2$ |
| $X_9$ | $X_{10}$ | $X_{11}$ | $X_{12}$ | $X_{13}$ | $X_{14}$ | $X_{15}$ | $X_{16}$ | $C_3$ | $C_4$ |
| $X_{17}$ | $X_{18}$ | $X_{19}$ | $X_{20}$ | $X_{21}$ | $X_{22}$ | $X_{23}$ | $X_{24}$ | $C_5$ | $C_6$ |
| $X_{25}$ | $X_{26}$ | $X_{27}$ | $X_{28}$ | $X_{29}$ | $X_{30}$ | $X_{31}$ | $X_{32}$ | $C_7$ | $C_8$ |
| $D_1$ | $D_3$ | $D_5$ | $D_7$ | $D_9$ | $D_{11}$ | $D_{13}$ | $D_{15}$ | | |
| $D_2$ | $D_4$ | $D_6$ | $D_8$ | $D_{10}$ | $D_{12}$ | $D_{14}$ | $D_{16}$ | | |

**Figure 6. Structure of the Matrix code for a 4x8 data set ($X_1$ to $X_{32}$); $C_1$ to $C_8$ and $D_1$ to $D_{16}$ represent the row and column parity bits, respectively (Source: [1]).**

Experimental evaluations demonstrated the matrix-based codes’ superiority over traditional ECCs like Hamming and Reed-Muller (RM) [37] codes. For a 32-bit codestruct, the proposed method achieves a 675× improvement in memory reliability compared to Hamming codes and a 38× improvement compared to Reed-Muller codes for distance-1 errors. When expanded to distance-3 errors, the reliability gains are even more pronounced, with memory Mean Time To Failure (MTTF) significantly extended. Cost analysis revealed that while the proposed codes require more redundant bits than Hamming codes, their fault coverage and efficiency make them a compelling choice for critical systems.

## 3.2 Enhanced Memory Reliability Against Multiple Cell Upsets Using Decimal Matrix Code

Guo et al. [25] focused on addressing the increasing challenge of MCUs in memories, especially in radiation-prone environments, leading to the development of a novel Decimal Matrix Code (DMC) that employs a decimal algorithm for error detection and correction, which enhances the fault tolerance of memories while minimizing performance overhead.

Figure 7 depicts that DMC organizes data into a 2D logical matrix structure, with rows and columns protected by redundant bits generated through decimal integer addition (for rows) and binary operations (for columns). This dual-layered approach allows the correction of up to five errors per word while requiring a reasonable number of redundant bits. The study emphasizes the encoder-reuse technique, which integrates the encoding process into the decoding circuitry, significantly reducing area and circuit complexity.

DMC demonstrated a significantly higher MTTF and lower delay overhead than established codes like Hamming and Matrix codes. However, DMC requires more redundant bits than its counterparts, which is a trade-off for its superior error correction capabilities. The researchers argue that the higher correction capability justifies the additional redundancy, particularly for applications in high-reliability environments, such as aerospace and high-performance computing, where fault tolerance is critical.

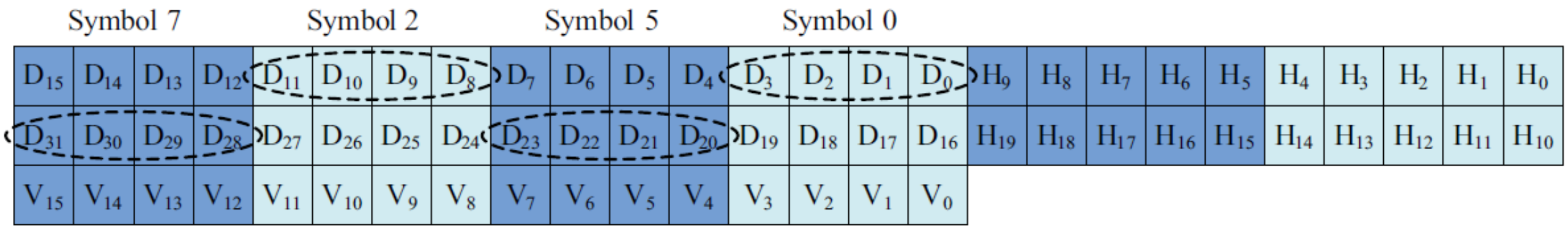

**Figure 7. 32-bit DMC organization treats each symbol as a decimal integer. Horizontal redundant bits H are generated by adding selected symbols per row, while vertical redundant bits V come from binary operations on column bits (Source: [25]).**

## 3.3 Low Redundancy Two-Dimensional Matrix-Based HVDB Code for Double Error Correction

Rahman et al. [48] proposed the Horizontal-Vertical-Double-Bit Diagonal (HVDD) parity method to mitigate soft errors in memory systems. The HVDD methodology employs a 2D-matrix structure for data and redundant bits, allowing for parallel calculation of horizontal, vertical, and diagonal parity during encoding and decoding; Figure 8 exemplifies an 8×8 HVDD. Double-bit diagonal parity enables the detection and correction of up to three-bit errors in any combination, significantly improving over traditional single and double-bit error correction codes. This approach ensures that, even in complex error patterns, most errors can be effectively identified and corrected.

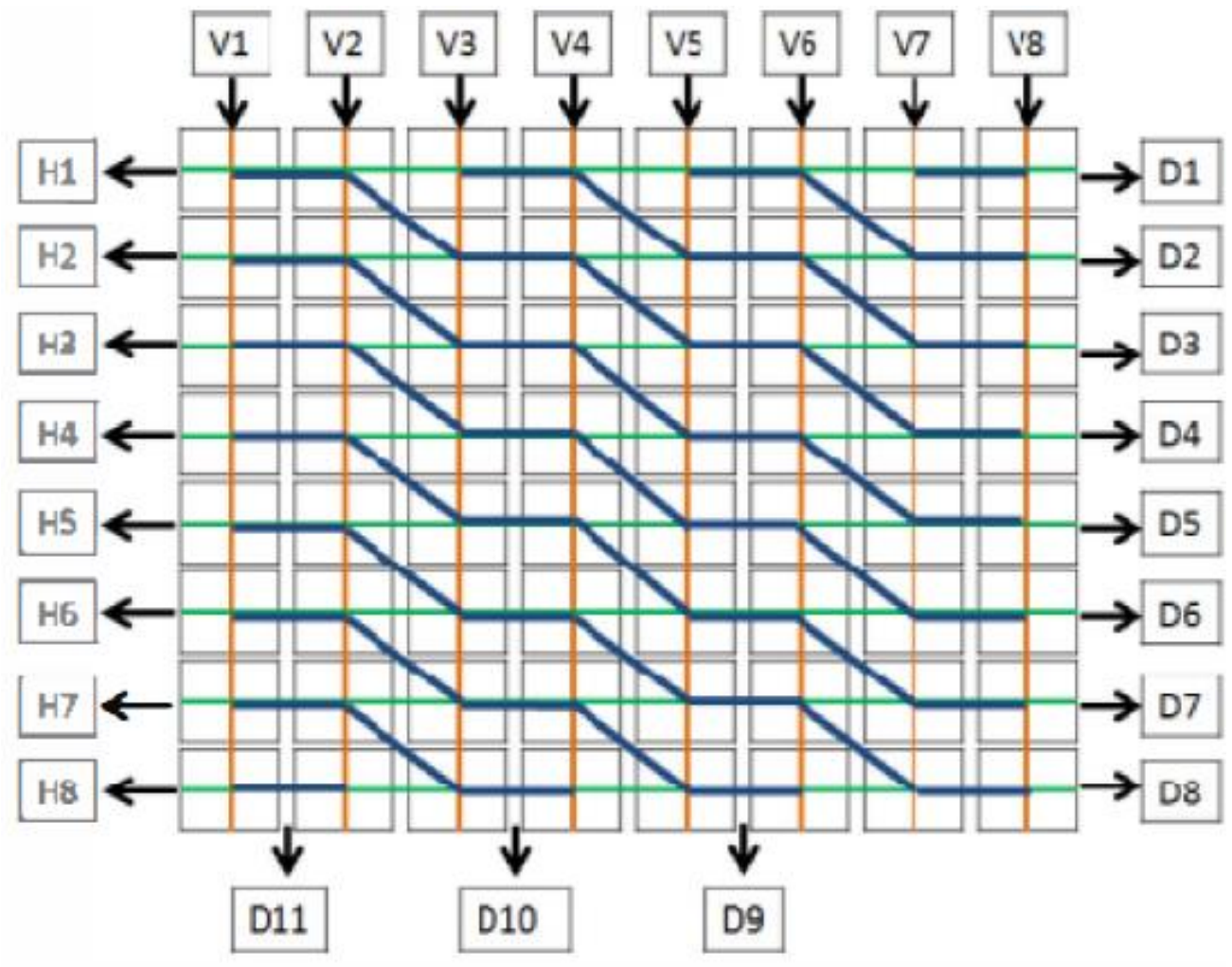

**Figure 8. HVDD structure for an 8×8 data matrix; $D_1$ to $D_{11}$, $V_1$ to $V_8$, and $H_1$ to $H_8$ are diagonal, vertical, and horizontal redundant bits (Source: [48]).**

Compared to Hamming and Bose-Chaudhuri-Hocquenghem (BCH) [36][31] codes, the HVDD technique provides a superior trade-off between error correction capability and implementation cost. Experimental evaluations validated HVDD's robustness, showing it can detect and correct all possible three-bit errors while maintaining a high code rate of 70.33%. This makes HVDD a viable solution for high-reliability applications where minimizing bit overhead is critical. The methodology is especially beneficial for real-time systems and memory architectures where error correction must be fast and efficient.

## 3.4 EG-LDPC Based 2-Dimensional Error Correcting Code for Mitigating MBUs of SRAM Memories

Erozan e Çavuş [11] proposed a 2D-ECC architecture utilizing Low-Density Parity Check codes based on Euclidean Geometry (EG-LDPC) combined with Single Parity Check (SPC) as a solution to the issue of MBUs in SRAMs. EG-LDPC codes were selected for their superior capabilities in detecting and correcting multiple errors and low-complexity decoders, making them particularly suitable for fault-tolerant memory applications.

The proposed architecture employs an EG-LDPC(15, 7, 5) code, where a 15-bit codestruct is divided into 7 data bits ($M_{ij}$) and 8 check bits ($R_{ij}$). Additionally, the 7×5 data matrix is validated by column parities ($C_{ij}$), exemplifying the application of a 2D code. This setup is organized into a two-dimensional matrix structure to reduce decoding complexity. The decoding process follows a standard matrix procedure tailored for 2D structures, as illustrated in Figure 9.

| | | | | | | | | | | | | | | |
|---|---|---|---|---|---|---|---|---|---|---|---|---|---|---|
| $M_{00}$ | $M_{01}$ | $M_{02}$ | $M_{03}$ | $M_{04}$ | $M_{05}$ | $M_{06}$ | $R_{00}$ | $R_{01}$ | $R_{02}$ | $R_{03}$ | $R_{04}$ | $R_{05}$ | $R_{06}$ | $R_{07}$ |
| $M_{10}$ | $M_{11}$ | $M_{12}$ | $M_{13}$ | $M_{14}$ | $M_{15}$ | $M_{16}$ | $R_{10}$ | $R_{11}$ | $R_{12}$ | $R_{13}$ | $R_{14}$ | $R_{15}$ | $R_{16}$ | $R_{17}$ |
| $M_{20}$ | $M_{21}$ | $M_{22}$ | $M_{23}$ | $M_{24}$ | $M_{25}$ | $M_{26}$ | $R_{20}$ | $R_{21}$ | $R_{22}$ | $R_{23}$ | $R_{24}$ | $R_{25}$ | $R_{26}$ | $R_{27}$ |
| $M_{30}$ | $M_{31}$ | $M_{32}$ | $M_{33}$ | $M_{34}$ | $M_{35}$ | $M_{36}$ | $R_{30}$ | $R_{31}$ | $R_{32}$ | $R_{33}$ | $R_{34}$ | $R_{35}$ | $R_{36}$ | $R_{37}$ |
| $M_{40}$ | $M_{41}$ | $M_{42}$ | $M_{43}$ | $M_{44}$ | $M_{45}$ | $M_{46}$ | $R_{40}$ | $R_{41}$ | $R_{42}$ | $R_{43}$ | $R_{44}$ | $R_{45}$ | $R_{46}$ | $R_{47}$ |
| $C_{00}$ | $C_{01}$ | $C_{02}$ | $C_{03}$ | $C_{04}$ | $C_{05}$ | $C_{06}$ | | | | | | | | |

**Figure 9. 2D Structure of EG-LDPC with data bits $M_{ij}$, row parity checks $R_{ij}$, and column parity checks $C_{ij}$ (Source: Adapted from [11]).**

The decoding algorithm uses pre-calculated one- and two-bit error combinations and their corresponding syndromes stored in a lookup table. At the start of decoding, syndrome values for each row and the first seven columns (data columns) are simultaneously calculated. Any non-zero syndromes are compared against the pre-saved values to identify a matching syndrome pattern. If a row syndrome is identified, it is cross-referenced with the corresponding column syndrome, representing a convergence of ECCs for error correction.

This approach achieves an error correction coverage of over 95% for up to 4 errors and 100% error detection for up to 2 errors. The authors demonstrate that their method significantly improves error correction and detection capabilities compared to other two-dimensional codes. They conclude that 2D architectures based on EG-LDPC codes offer a robust solution for mitigating MBUs in SRAM technologies, highlighting their potential for enhanced fault tolerance in modern memory systems.

## 3.5 Horizontal-Vertical Parity and Diagonal Hamming Based Soft Error Detection and Correction for Memories

Raha, Vinodhini, and Murty [47] designed the Horizontal-Vertical Parity and Diagonal Hamming (HVPDH) 2D-ECC. Figure 10 illustrates the HVPDH codestruct, which consists of 32 data bits ($D_0$-$D_{31}$) protected by 28 check bits organized into three sets: (i) 4 horizontal parity bits ($H_{z0}$-$H_{z3}$), (ii) 8 vertical parity bits ($V_0$-$V_7$), and (iii) 16 Hamming bits grouped diagonally ($H_{m11}$-$H_{m44}$).

| | | | | | | | | | | | |
|---|---|---|---|---|---|---|---|---|---|---|---|
| D31 | D27 | D20 | D13 | $H_{m41}$ | D6 | D2 | D9 | $H_{m31}$ | D16 | $H_{m21}$ | $H_{m11}$ |
| D23 | D30 | D26 | D19 | $H_{m42}$ | D12 | D5 | D1 | $H_{m32}$ | D8 | $H_{m22}$ | $H_{m12}$ |
| D15 | D22 | D29 | D25 | $H_{m43}$ | D18 | D11 | D4 | $H_{m33}$ | D0 | $H_{m23}$ | $H_{m13}$ |
| D7 | D14 | D21 | D28 | $H_{m44}$ | D24 | D17 | D10 | $H_{m34}$ | D3 | $H_{m24}$ | $H_{m14}$ |

(a)

| | | | | | | | | |
|---|---|---|---|---|---|---|---|---|
| D7 | D6 | D5 | D4 | D3 | D2 | D1 | D0 | Hz0 |
| D15 | D14 | D13 | D12 | D11 | D10 | D9 | D8 | Hz1 |
| D23 | D22 | D21 | D20 | D19 | D18 | D17 | D16 | Hz2 |
| D31 | D30 | D29 | D28 | D27 | D26 | D25 | D24 | Hz3 |
| V7 | V6 | V5 | V4 | V3 | V2 | V1 | V0 | |

(b)

**Figure 10. (a) Part of the HVPDH codestruct showing the grouping used to compute the Hamming code, e.g., $H_{m11}$ to $H_{m41}$ protect bits $D_2$, $D_6$, $D_9$, $D_{13}$, $D_{16}$, $D_{20}$, $D_{27}$, and $D_{31}$. (b) Part of the codestruct verifying horizontal parity ($H_z$) and vertical parity ($V_n$) for the data matrix. (Source: [47]).**

The research emphasizes the increasing necessity for efficient error correction codes due to the prevalence of MCUs in modern memory systems. Simulation tests, conducted using Verilog HDL, validated the HVPDH's error detection and correction capabilities. The HVPDH can detect up to 8-bit errors and correct all combinations of up to 2-bit errors and most combinations of 3- to 5-bit errors. These correction rates are competitive compared to other ECC methods, such as the Decimal Matrix Code (DMC) [25].

This design demonstrates a robust and scalable solution for fault tolerance in memory systems, particularly in environments prone to high error rates. The combination of horizontal, vertical, and diagonal error correction provides an advanced approach to mitigating the challenges MCUs pose.

## 3.6 Correction of Adjacent Errors with Low Redundant Matrix Error Correction Codes

Moran et al. [39] designed two low-redundancy 2D-ECCs to correct adjacent error patterns. Both codes exhibit the same error correction capabilities but differ in redundancy levels. The authors analyzed the impact of these low-redundancy designs on area overhead, power consumption, and delay. The proposed ECCs were compared with the Matrix code [1], as illustrated in Figure 11.

The two proposed 2D-ECCs were evaluated based on error patterns and their corresponding correction capabilities: (i) Horizontal Adjacent Errors – 100% correction was achieved for 1 to 3 errors across all proposals, with decreasing coverage for 4 to 8 errors;

(ii) Vertical Adjacent Errors – all errors up to 3 bits were corrected, but coverage dropped significantly for 6 or more errors; (iii) Square Adjacent Errors – 100% correction was achieved for 2×2 patterns, but coverage decreased for 3×3 and 4×4 patterns.

| | | | | | | | | | | | | |
|---|---|---|---|---|---|---|---|---|---|---|---|---|
| $X_0$ | $X_1$ | $X_2$ | $X_3$ | $X_4$ | $X_5$ | $X_6$ | $X_7$ | $C_0$ | $C_1$ | $C_2$ | $C_3$ | $C_4$ |
| $X_8$ | $X_9$ | $X_{10}$ | $X_{11}$ | $X_{12}$ | $X_{13}$ | $X_{14}$ | $X_{15}$ | $C_5$ | $C_6$ | $C_7$ | $C_8$ | $C_9$ |
| $X_{16}$ | $X_{17}$ | $X_{18}$ | $X_{19}$ | $X_{20}$ | $X_{21}$ | $X_{22}$ | $X_{23}$ | $C_{10}$ | $C_{11}$ | $C_{12}$ | $C_{13}$ | $C_{14}$ |
| $X_{24}$ | $X_{25}$ | $X_{26}$ | $X_{27}$ | $X_{28}$ | $X_{29}$ | $X_{30}$ | $X_{31}$ | $C_{15}$ | $C_{16}$ | $C_{17}$ | $C_{18}$ | $C_{19}$ |
| $P_0$ | $P_1$ | $P_2$ | $P_3$ | $P_4$ | $P_5$ | $P_6$ | $P_7$ | | | | | |

**Figure 11. 2D codestruct with 31 data bits ($X_0$-$X_{31}$), 20 Hamming check bits ($C_0$-$C_{19}$), and eight parity bits ($P_0$-$P_7$) (Source: [39]).**

Proposal 1, shown in Figure 12(a), provides robust coverage for horizontal and square error patterns with less redundancy than the Matrix code [1]. Its Hamming-based verification bits (C) protect more data bits (X) per check bit, reducing redundancy and offering a more efficient balance between performance and overhead. In contrast, Proposal 2, illustrated in Figure 12(b), uses 16 check bits (C) to cover the same data area (X). This results in a higher redundancy level compared to Proposal 1 but requires simpler encoding circuitry, leading to lower area, power, and delay overheads.

| | | | | | | | |
|---|---|---|---|---|---|---|---|
| $C_0$ | $C_1$ | $C_2$ | $C_3$ | $C_4$ | $C_5$ | $C_6$ | $C_7$ |
| $X_0$ | $X_1$ | $X_2$ | $X_3$ | $X_4$ | $X_5$ | $X_6$ | $X_7$ |
| $X_8$ | $X_9$ | $X_{10}$ | $X_{11}$ | $X_{12}$ | $X_{13}$ | $X_{14}$ | $X_{15}$ |
| $X_{16}$ | $X_{17}$ | $X_{18}$ | $X_{19}$ | $X_{20}$ | $X_{21}$ | $X_{22}$ | $X_{23}$ |
| $X_{24}$ | $X_{25}$ | $X_{26}$ | $X_{27}$ | $X_{28}$ | $X_{29}$ | $X_{30}$ | $X_{31}$ |

(a)

| | | | | | | | |
|---|---|---|---|---|---|---|---|
| $C_0$ | $C_1$ | $C_2$ | $C_3$ | $C_4$ | $C_5$ | $C_6$ | $C_7$ |
| $C_8$ | $C_9$ | $C_{10}$ | $C_{11}$ | $C_{12}$ | $C_{13}$ | $C_{14}$ | $C_{15}$ |
| $X_0$ | $X_1$ | $X_2$ | $X_3$ | $X_4$ | $X_5$ | $X_6$ | $X_7$ |
| $X_8$ | $X_9$ | $X_{10}$ | $X_{11}$ | $X_{12}$ | $X_{13}$ | $X_{14}$ | $X_{15}$ |
| $X_{16}$ | $X_{17}$ | $X_{18}$ | $X_{19}$ | $X_{20}$ | $X_{21}$ | $X_{22}$ | $X_{23}$ |
| $X_{24}$ | $X_{25}$ | $X_{26}$ | $X_{27}$ | $X_{28}$ | $X_{29}$ | $X_{30}$ | $X_{31}$ |

(b)

**Figure 12. Codestruct of (a) Proposal 1 and (b) Proposal 2 (Source: [39]).**

The redundancy levels across the codes were quantified as follows: (i) the Matrix code exhibits 87.50% redundancy with 28 check bits, (ii) Proposal 2 achieves 50% redundancy with 16 check bits, and (iii) Proposal 1 achieves the lowest redundancy at 25% with 8 check bits. These percentages reflect the proportion of additional bits required to implement the ECC relative to the total number of bits. The authors concluded that Proposals 1 and 2 outperform the Matrix code in horizontal and square error scenarios, offering significantly reduced area overhead, power consumption, and latency.

Both proposals were developed using a methodology called Flexible Unequal Error Control (FUEC), which allows for the automatic generation of an efficient parity-check matrix. The proposed ECCs improve error coverage compared to the Matrix code and substantially reduce redundancy and associated overheads. The paper concludes by emphasizing the significance of the results for enhancing fault tolerance and reliability in memory systems.

## 3.7 PCoSA: A product error correction code for use in memory devices targeting space applications

Freitas et al. [14] introduced the Product Code for Space Applications (PCoSA), an ECC designed to address radiation-induced errors in memory systems used in space environments. PCoSA combines Hamming codes and parity bits applied to rows and columns, achieving a 2D-ECC structure. This design ensures high fault tolerance and error detection capabilities, which is particularly suited for environments where reliability is paramount.

The core structure of PCoSA(64, 16) encodes a 16-bit data word into 64 bits, comprising 16 data bits, 12 row-check bits, 7 row-parity bits, 21 column-check bits, and 8 column-parity bits. This configuration results in a minimum Hamming distance of 16, allowing for robust error detection and correction. The code demonstrated high correction rates for error patterns containing up to 7 bitflips in simulated memory scenarios. The combination of row and column redundancy enhances PCoSA's ability to detect and correct multi-bit errors, outperforming other ECCs like Matrix and Reed-Muller codes.

| | | | | | | | |
|---|---|---|---|---|---|---|---|
| $D_0$ | $D_1$ | $D_2$ | $D_3$ | $C1_0$ | $C1_1$ | $C1_2$ | $P1_0$ |
| $D_4$ | $D_5$ | $D_6$ | $D_7$ | $C1_3$ | $C1_4$ | $C1_5$ | $P1_1$ |
| $D_8$ | $D_9$ | $D_{10}$ | $D_{11}$ | $C1_6$ | $C1_7$ | $C1_8$ | $P1_2$ |
| $D_{12}$ | $D_{13}$ | $D_{14}$ | $D_{15}$ | $C1_9$ | $C1_{10}$ | $C1_{11}$ | $P1_3$ |
| $C2_0$ | $C2_1$ | $C2_2$ | $C2_3$ | $C2_4$ | $C2_5$ | $C2_6$ | $P1_4$ |
| $C2_7$ | $C2_8$ | $C2_9$ | $C2_{10}$ | $C2_{11}$ | $C2_{12}$ | $C2_{13}$ | $P1_5$ |
| $C2_{14}$ | $C2_{15}$ | $C2_{16}$ | $C2_{17}$ | $C2_{18}$ | $C2_{19}$ | $C2_{20}$ | $P1_6$ |
| $P2_0$ | $P2_1$ | $P2_2$ | $P2_3$ | $P2_4$ | $P2_5$ | $P2_6$ | $P2_7$ |

**Figure 13. PCoSA structure with 16 data bits; the codestruct is divided into five regions: data (D), row check bits for D (C1), column check bits for D and C1 (C2), row parity for D, C1, and C2 (P1), and overall column parity (P2) (Source: [14]).**

The scalability of PCoSA was explored by extending its configurations to protect larger memory words, such as PCoSA(256, 121) and PCoSA(1024, 676). These extended formats maintain the same detection and correction capabilities while reducing the redundancy rate from 75% to as low as 33.98%.

Simulation results highlighted superior performance for PCoSA, with a 100%

detection rate for all tested error patterns. For correction capabilities, PCoSA achieved 100% for up to three bitflips and over 80% for four bitflips, significantly outperforming other evaluated ECCs. Furthermore, the reliability analysis showed that PCoSA maintains high system dependability over time, making it a suitable choice for space applications where long-term memory integrity is critical.

Despite its advantages, the synthesis cost analysis revealed trade-offs. The PCoSA encoder and decoder modules exhibited higher area consumption and power dissipation compared to other ECCs, such as PBD. However, the increased reliability and error correction performance justify these costs in critical applications like space missions.

## 3.8 Extended Matrix Region Selection Code: An ECC for adjacent Multiple Cell Upset in memory arrays

Silva et al. [59] introduced the Extended Matrix Region Selection Code (eMRSC), a 2D-ECC designed to enhance reliability in memory systems, particularly against MCUs. The eMRSC is an evolution of the Matrix Region Selection Code (MRSC) [56], expanding its capabilities from handling 16-bit data to 32-bit data, and comes in two main configurations: eMRSC(32,3,64) and eMRSC(32,7,56). These configurations differ in the number of regions they utilize for error correction and the redundancy they require, offering trade-offs between correction capability and implementation cost. The structure of eMRSC divides data and redundancy into regions, enabling localized error detection and correction within defined spatial groupings. This is particularly effective for correcting adjacent errors, the most common MCU pattern in memory devices. Figure 14(a) and (b) illustrate the structure and the three data regions of eMRSC(32,3,64).

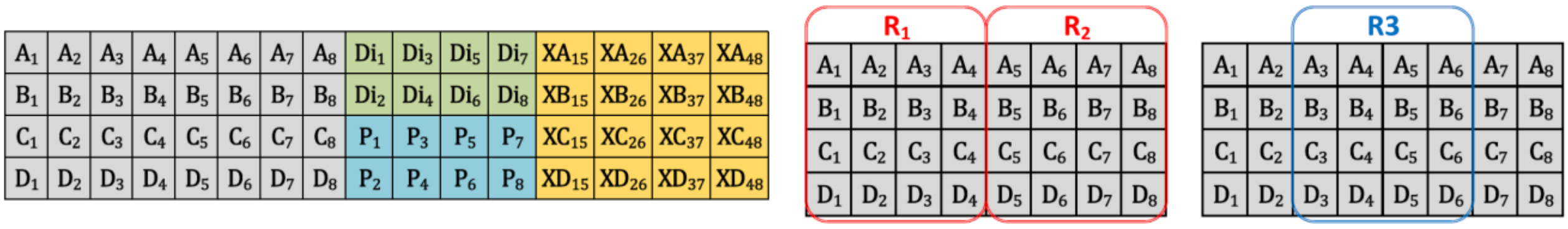

(a) (b)

**Figure 14. (a) Structure and (b) data regions of the eMRSC(32,3,64) codestruct (Source: [59]).**

Key features of eMRSC include the ability to address adjacent error patterns using a matrix-based approach to syndromes and redundancy bits, and an algorithmic process that selects error-prone regions for correction. The eMRSC(32,3,64) prioritizes robust correction capability with larger regions, while the eMRSC(32,7,56) minimizes redundancy overhead by dividing the matrix into smaller regions.

The article presented extensive experimental results comparing eMRSC to other ECCs, such as Matrix [1], Orthogonal Latin Squares (OLS) [9], and Decimal Matrix Code (DMC) [25]. eMRSC outperforms these codes in error correction capability and MTTF, particularly in scenarios involving multiple adjacent errors. While eMRSC(32,3,64) demonstrates superior error correction in more aggressive scenarios, eMRSC(32,7,56) offers efficiency for applications where lower redundancy is acceptable.

## 3.9 Error Coverage, Reliability and Cost Analysis of Fault Tolerance Techniques for 32-bit Memories used on Space Missions

Freitas et al. [13] analyzed fault tolerance techniques for 32-bit memories used in space missions, focusing on error correction rates, reliability, and implementation costs. The study evaluates seven schemes with and without interleaving, including Extended Hamming, Reed-Muller (RM) [37], and Triple Modular Redundancy (TMR) [5]. Bit interleaving, illustrated in Figure 15, is a technique that transforms an MBU into SBUs distributed across multiple codewords. This technique achieves maximum effectiveness when the physical MBU size is less than or equal to the specified interleaving scheme.

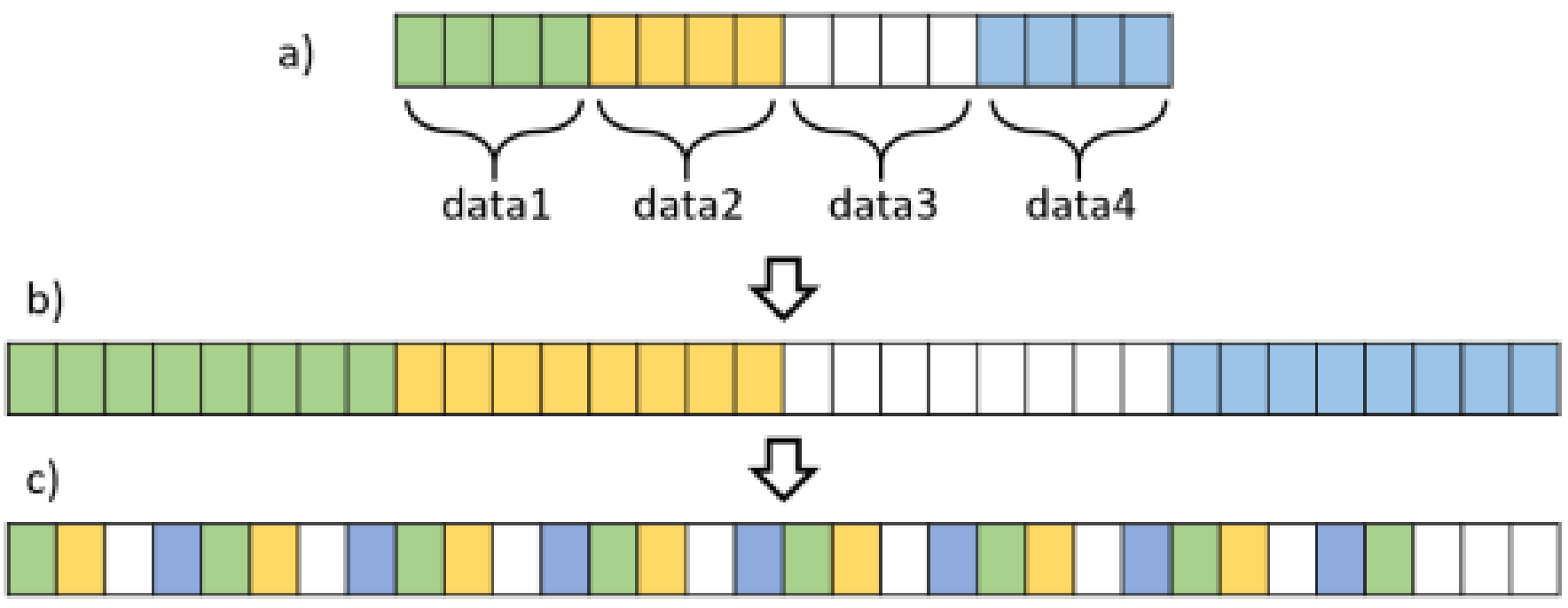

**Figure 15. Interleaving technique illustrated with (a) four 4-bit data blocks encoded into (b) four interleaved codewords organized in (c) a codestruct (Source: [13]).**

Figure 16 presents simulation results showing that the interleaving technique yields higher correction rates for adjacent errors, as these errors are distributed across other codewords through interleaving. However, when errors occur in all positions of the codestruct, the correction rates are independent of interleaved. Additionally, the results reveal that RM is the most reliable for up to 3 or 4 errors, depending on the error pattern injected during the simulation. On the other hand, TMR achieves high correction rates for up to 10 adjacent errors.

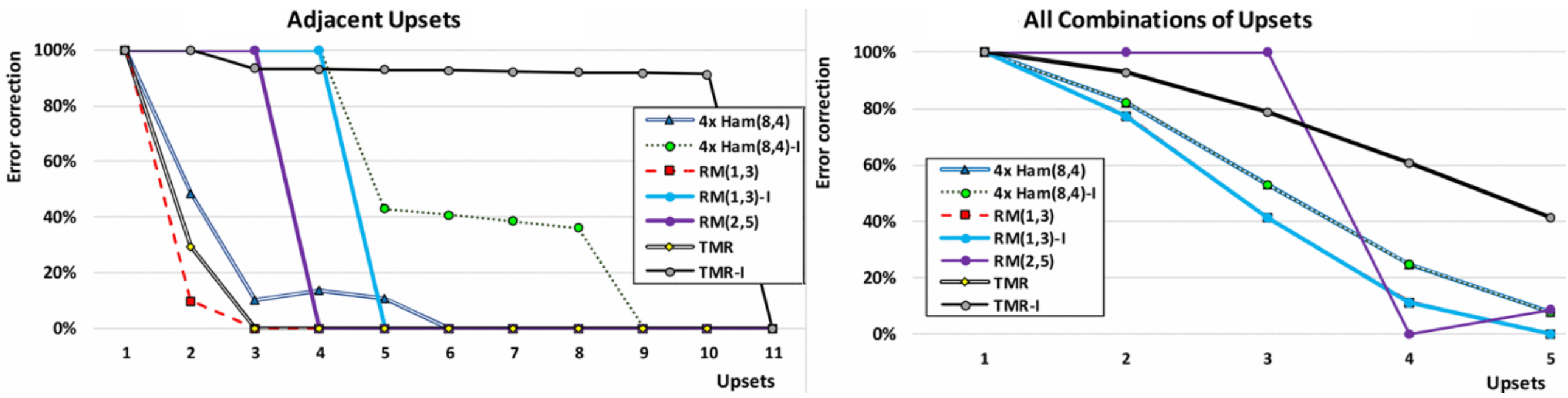

(a) (b)

**Figure 16. Simulation results showing correction rates for the proposed configurations using (a) only adjacent errors and (b) all combinations of errors (Source: [13]).**

## 3.10 LPC: An Error Correction Code for Mitigating Faults in 3D Memories

Freitas et al. [15] proposed the Line Product Code (LPC), a modified product-type ECC designed to address the reliability challenges of 3D memories, which are increasingly susceptible to faults due to transistor scaling and environmental factors like radiation and heat. LPC incorporates Hamming and parity codes across rows and columns, leveraging a matrix-based organization to enhance error correction capacity while minimizing redundancy overhead. This makes LPC particularly suitable for critical applications such as space memory systems.

Figure 17(a) illustrates the LPC codestruct, composed of 16 data bits ($D_0$ to $D_{15}$), 16 extended Hamming code bits for rows ($CR_0$ to $CR_{11}$ along with $PR_0$ to $PR_3$), and another 16 extended Hamming code bits for columns ($CC_0$ to $CC_{11}$ along with $PC_0$ to $PC_3$). Figure 17(b), however, presents the logical format used for encoding and decoding. This new figure includes row and column syndrome bits and the SEr, DEr, SEc, and DEc bits, which indicate whether single or double errors occurred in the rows or columns.

Experimental results validate LPC’s reliability and efficiency, showing its capacity to correct up to 20-bit flips within a data field under extreme error scenarios. Furthermore, LPC balances error correction rates and implementation costs with a redundancy rate of 66.7%, enabling robust performance for high-fault environments. The lightweight decoding algorithms reduce synthesis costs, making LPC a cost-effective solution for advanced memory technologies.

The study also highlights LPC’s adaptability for use in diverse memory architectures, including on-die ECC integration, enhancing memory reliability while maintaining transparency to controllers. By addressing the unique fault patterns of 3D memory stacks, LPC sets a precedent for heterogeneous ECC models tailored to varying error susceptibilities across memory layers.

| | | | | | | | | | | | | | | | |
|---|---|---|---|---|---|---|---|---|---|---|---|---|---|---|---|
| $D_0$ | $D_1$ | $D_2$ | $D_3$ | $D_4$ | $D_5$ | $D_6$ | $D_7$ | $D_8$ | $D_9$ | $D_{10}$ | $D_{11}$ | $D_{12}$ | $D_{13}$ | $D_{14}$ | $D_{15}$ |
| $CR_0$ | $CR_1$ | $CR_2$ | $CR_3$ | $CR_4$ | $CR_5$ | $CR_6$ | $CR_7$ | $CR_8$ | $CR_9$ | $CR_{10}$ | $CR_{11}$ | $PR_0$ | $PR_1$ | $PR_2$ | $PR_3$ |
| $CC_0$ | $CC_1$ | $CC_2$ | $CC_3$ | $CC_4$ | $CC_5$ | $CC_6$ | $CC_7$ | $CC_8$ | $CC_9$ | $CC_{10}$ | $CC_{11}$ | $PC_0$ | $PC_1$ | $PC_2$ | $PC_3$ |

(a)

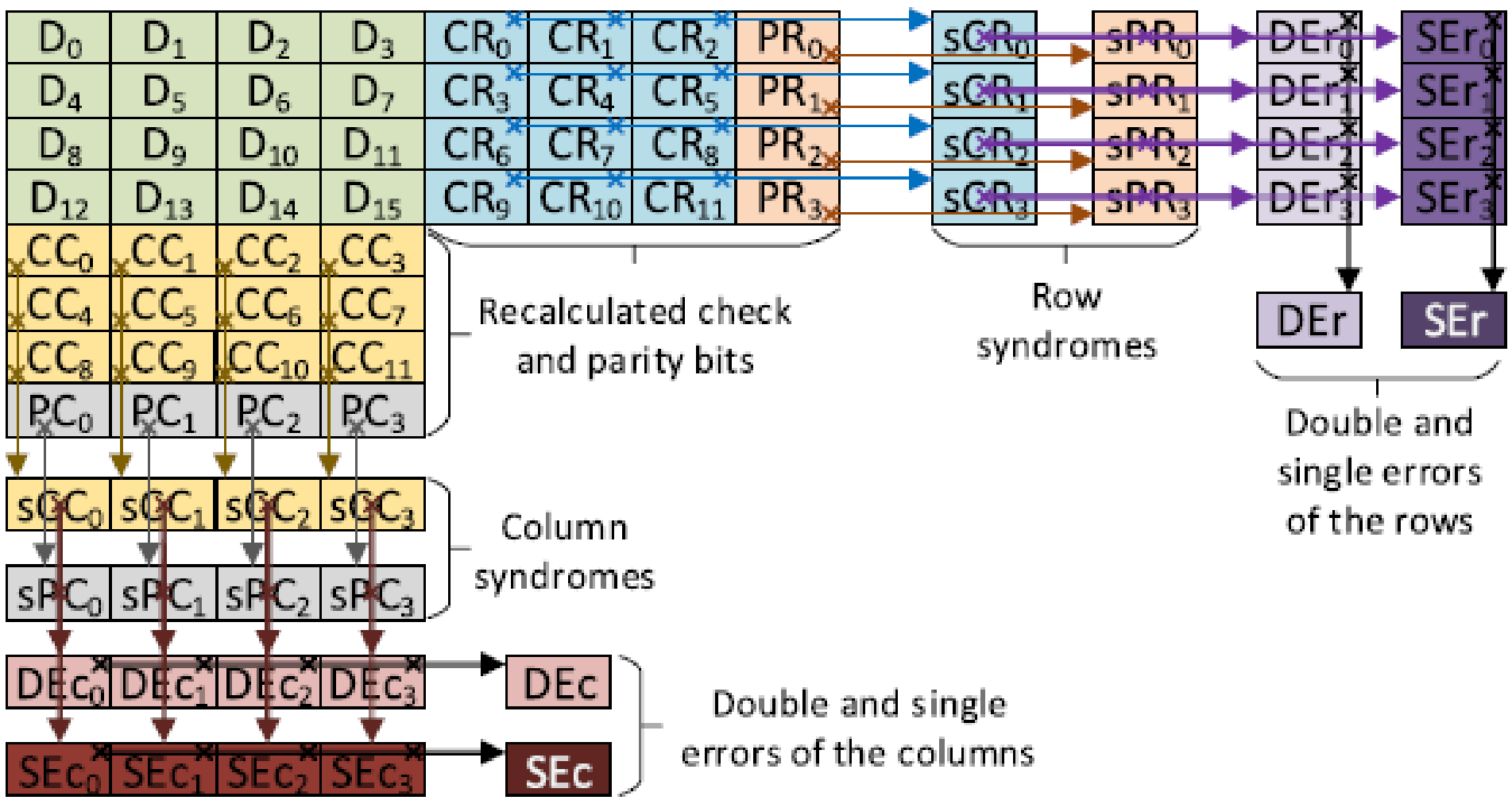

(b)

**Figure 17. (a) LPC codestruct and (b) LPC logical format used for ECC encoding and decoding (Source: [15]).**

## 3.11 A New Error Correcting Coding Technique to Tolerate Soft Errors

Sen et al. [53] proposed an enhancement to the Decimal Matrix Code (DMC) [25] to improve memory reliability by reducing processing time and information overhead. Simulations demonstrate that the proposed technique outperforms various existing ECCs regarding MCU detection and correction capabilities, making it suitable for critical applications. The proposed Enhanced DMC seeks to improve correction and detection capabilities while minimizing overhead in redundant bits, a critical factor in high-density memory applications.

The Enhanced DMC employs a 2D structure for encoding and decoding, arranging data into a matrix format. Figure 18 displays the data bits are grouped into symbols, each consisting of 4 bits, and organized into a logical $M_1 \times M_2$ symbol matrix. Redundant bits are calculated using successive XOR operations horizontally and vertically, with further decimal additions applied to generate vertical check bits. This design ensures better error correction rates than the original DMC while using fewer redundant bits. For example, while the original DMC uses 36 redundant bits for a 32-bit data word, the Enhanced DMC reduces this to 32 redundant bits, achieving improved coding efficiency.

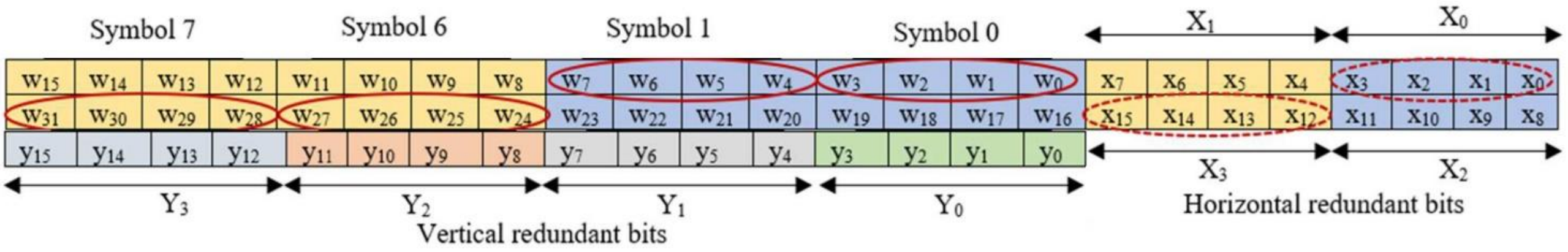

**Figure 18. Proposed structure of the Enhanced DMC (Source:[53]).**

During decoding, the Enhanced DMC identifies errors by generating syndrome bits through XOR operations on the horizontal and vertical check bits. The proposed method achieves 100% correction capability for up to five erroneous bits in a data word. The correction rate gradually decreases for more than five errors but remains superior to competing ECCs.

The article emphasizes that the Enhanced DMC also reduces the processing overhead compared to its predecessors. The method ensures faster operation by simplifying the encoding and decoding processes, which is critical in real-time systems. Furthermore, the authors highlight that the Enhanced DMC offers a competitive edge by achieving a higher error correction rate with fewer redundant bits than similar ECC methods.

## 3.12 Multiple Bit Error Correction Codes for Memories in Satellites

Tejas, Kumar, and Sunita [66] proposed a 2D-ECC called Vertical Parity and Diagonal Hamming (VPDH). This hybrid code builds upon the theoretical foundation of HVPDH [47] by reorganizing 4 parity bits. The VPDH codestruct protects 32 data bits using 16 diagonal Hamming bits, 8 vertical parity bits, and 4 diagonal parity bits.

The VPDH encoder and decoder were simulated and synthesized using the Xilinx Vivado tool with a 180nm standard cell library in implementation and testing. Automated tests evaluated the VPDH's error correction capabilities by simulating errors ranging from 1 to 5 bits. Results demonstrated that VPDH corrects approximately 30%, 24%, and 6% more 3-bit, 4-bit, and 5-bit errors, respectively, compared to HVPDH while maintaining similar bit overhead and power dissipation.

The authors conclude that VPDH is an effective solution for space applications, balancing error correction capability, area consumption, delay, and power dissipation. This makes it a promising approach for environments where reliability and efficiency are critical.

## 3.13 TECED: A Two-Dimensional Error-Correction Codes Based Energy-Efficiency SRAM Design

Chen et al. [6] proposed the Two-Dimensional Error-Correction Codes Based Energy-Efficiency SRAM Design (TECED) method to design SRAMs using 2D-ECCs. TECED focuses on balancing latency, energy consumption, and area constraints in the context of increasing chip scaling and operational frequencies while enhancing memory reliability against soft errors. By utilizing 2D-ECCs, TECED improves energy efficiency and reduces hardware costs, offering higher performance and reliability.

Compared to conventional Hamming codes, the TECED method employs horizontal and vertical memory word encoding with parity techniques, optimizing error detection unit usage and minimizing energy consumption. However, specific characteristic errors remain undetectable, as shown in Figure 19.

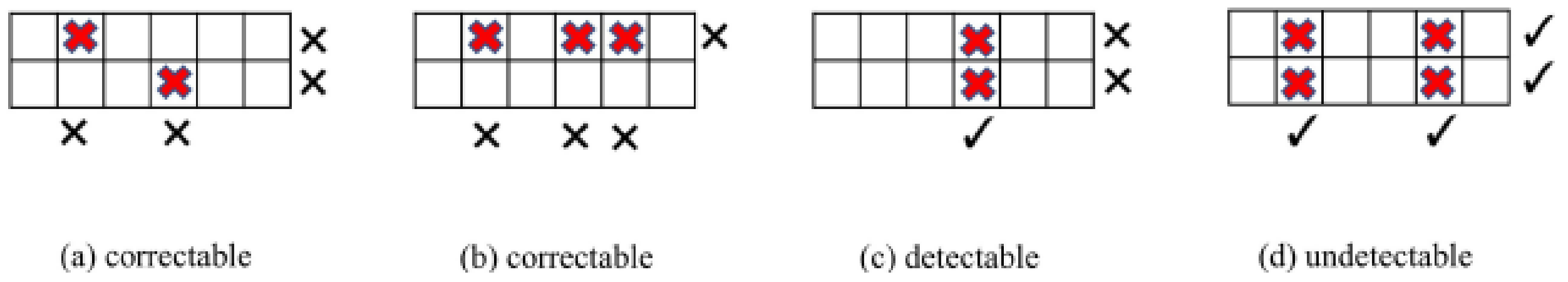

**Figure 19. Correction vs. detection capability of TECED (Source: [6]).**

The authors highlight that while the proposed method uses well-known techniques, the key contribution lies in its memory access architecture, which provides significant energy efficiency advantages and reduced impact in error-free scenarios. This makes TECED a valuable approach for modern SRAM design, where energy efficiency and reliability are critical considerations.

## 3.14 New Decoding Techniques for Modified Product Code Used in Critical Applications

Freitas et al. [16] explored advancements in error correction algorithms of the previous 2D-ECC work – LPC [15]. This new work introduces innovative decoding strategies, namely the Single Error (AlgSE) and Double Error (AlgDE) correction algorithms, which leverage LPC's unique structure to improve error correction efficiency.

The LPC organizes data and parity bits in a matrix format, enabling cross-references between rows and columns to identify and correct errors. The study employs Ham(8,4) codes for LPC's underlying structure, achieving a minimum Hamming distance of 7. This configuration allows LPC to correct up to three-bit errors and detect three simultaneously, potentially handling more errors depending on their distribution. The AlgSE, displayed in Figure 20, utilizes iterative processes and heuristics to optimize correction capabilities, such as prioritizing rows or columns with the highest error counts.

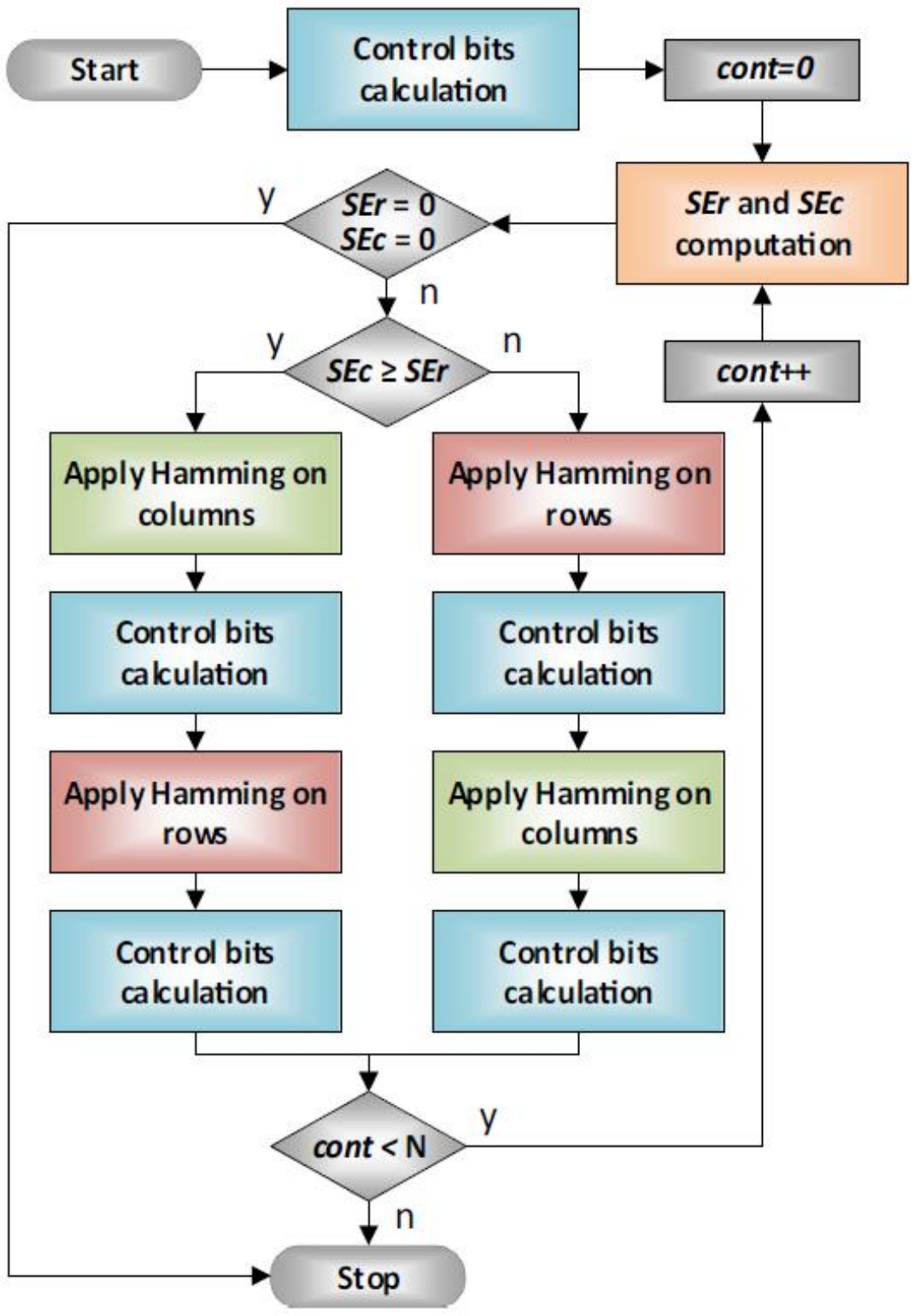

**Figure 20. Iterative AlgSE Algorithm. Compares the number of single errors found in columns (SEc) with those found in rows (SEr), applying successive iterative and interleaved rounds of Hamming correction for rows and columns (Source: [16]).**

Experimental results demonstrate that LPC achieves a high correction rate, effectively mitigating MCUs in scenarios with significant radiation exposure. Integrating AlgSE and AlgDE in decoding processes ensures a robust approach to handling single- and double-bit errors. The paper highlights trade-offs between correction efficacy and implementation costs, such as redundancy overhead and computational latency.

Applications for LPC include memory systems in space missions and other environments prone to high radiation. LPC minimizes redundancy costs without sacrificing performance by focusing on data bits for correction and recalculating redundancy from these. These findings position LPC as a practical solution for fault tolerance in critical electronic systems, balancing high error correction capacity and efficient resource utilization.

## 3.15 Low Redundancy Two-Dimensional Matrix-Based HVDB Code for Double Error Correction

Yuqi, Xi, and Tang [69] proposed the Horizontal-Vertical-Diagonal-Block (HVDB) code to protect memories from MCUs caused by radiation at nanometric levels. This 2D-ECC uses parity information in horizontal, vertical, and diagonal directions within data blocks arranged in a two-dimensional matrix to detect and correct 2-bit errors, requiring low redundancy and minimal decoding overhead.

The decoding algorithm leverages parity syndromes to detect and correct errors. The proposed matrix codes organize k data bits into $k_1 \times k_2$ matrices. For example, the authors illustrate a 64-bit data word arranged in an 8×8 matrix to apply the proposed algorithm. Other matrices with similar structures are also used for error correction. Figure 21 demonstrates the 8×8 data matrix, which is analyzed by horizontal ($h_0$-$h_7$), vertical ($v_0$-$v_7$), and diagonal ($d_0$-$d_7$) lines. Each line computes the parity of eight data bits, except for diagonals, where all but the main diagonal are computed using two lines. For instance, the parity bit $d_2$ is obtained using the XOR of bits $D_{0,6}$, $D_{1,7}$, $D_{2,0}$, $D_{3,1}$, $D_{4,2}$, $D_{5,3}$, $D_{6,4}$, and $D_{7,5}$.

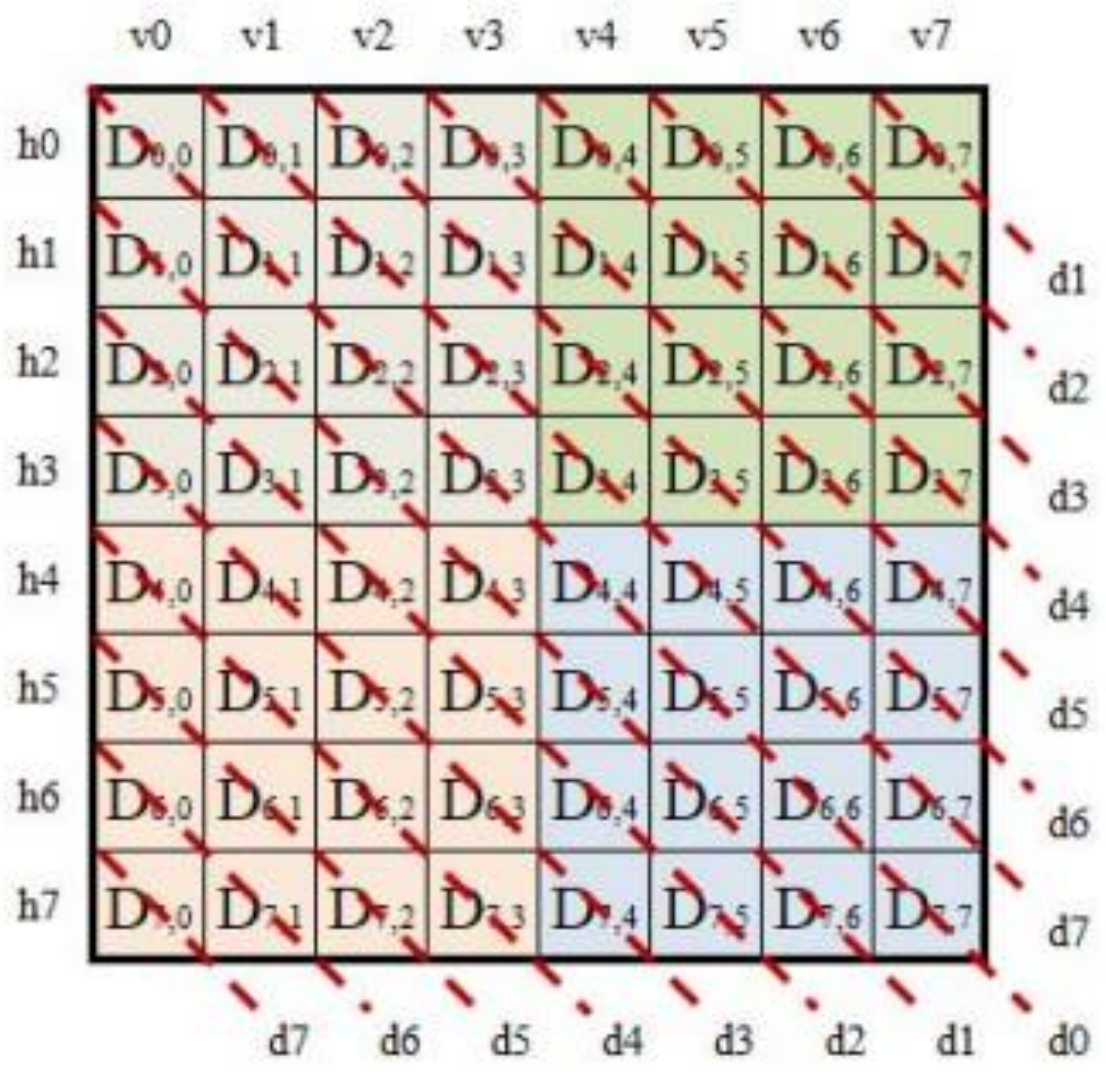

**Figure 21. Encoding structure of the 2D-ECC HVDB (Source: [69]).**

The authors suggest that matrix structures similar to those in Figure 21 can be adapted for error correction in other configurations based on specific requirements. The flexibility in matrix sizing allows adaptation to the characteristics of the target system.

The HVDB does not guarantee 100% correction for all 2-bit error cases. While the method can correct up to 2-bit errors, the exact effectiveness depends on specific conditions and error distributions. Nevertheless, the authors conclude that HVDB is an efficient 2D-ECC for memory error correction, particularly in radiation-prone environments. It offers lower redundancy and faster decoding speeds than other matrix-based methods, making it a viable choice for critical applications.

## 3.16 EMPC-SA: Error Correction Scheme using Modified Product Code for Space Applications

J. Magalhães et al. [32] designed the Modified Product Code for Space Applications (EMPC-SA) to address radiation-induced errors in critical space applications. The EMPC-SA employs 24 redundancy bits to protect 16 data bits. Experimental results demonstrate that for three-bit errors, the EMPC-SA performs competitively with the CLC code [4][57] and outperforms Matrix [1] and PBD [23] codes. Under more severe error conditions, specifically with four-bit flips, the EMPC-SA achieves a higher error correction rate than CLC-A [58][60]. However, it falls behind OPCoSA [17] and TBEC-RSC [61]. Notably, the EMPC-SA exhibits lower redundancy than OPCoSA and implements error correction only when bit flips are entirely contained within the data matrix. This limitation highlights opportunities for improvement, particularly in handling bit flips that occur at the intersections of data and parity bit regions.

The authors adopted a codestruct with fewer bits than traditional product codes, omitting the verification region for check bits. The 2D-ECC is implemented with 16 data bits, comprising 12 row check bits and 12 column check bits, as illustrated in Figure 22.

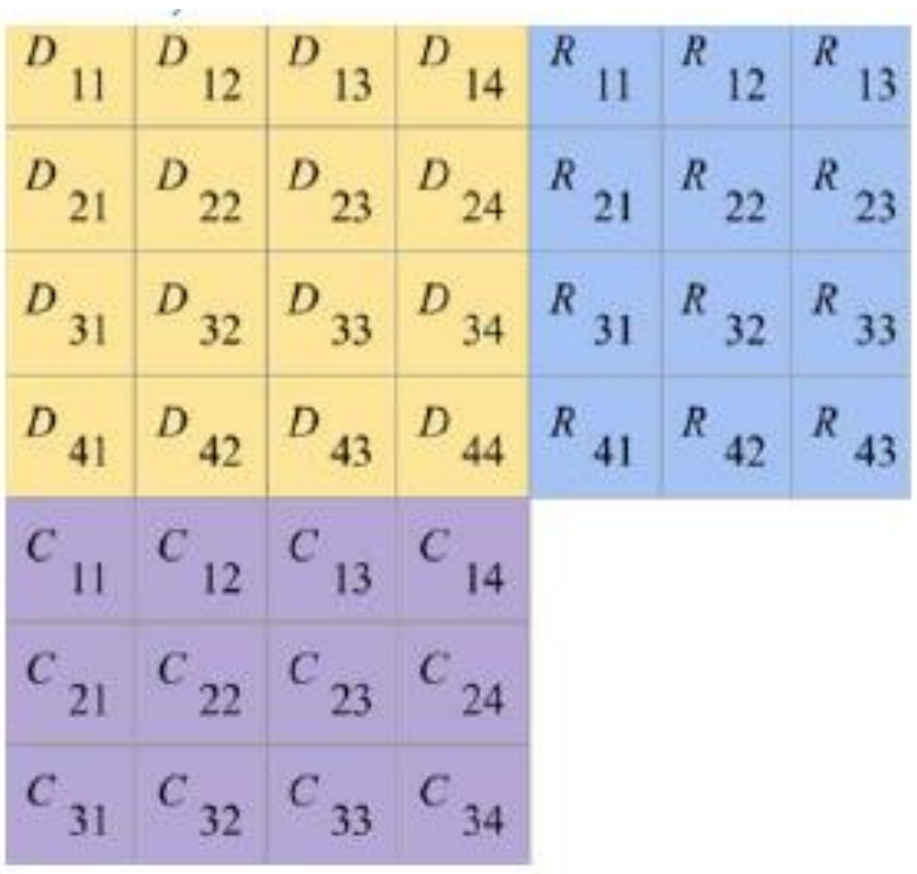

**Figure 22. Codestruct of EMPC-SA; D represents the 4×4 data matrix, R is the 4×3 row redundancy matrix using Ham(7,4), and e C is the 3×4 column redundancy matrix using Ham(7,4) (Source: [32]).**

Experimental results show that the EMPC-SA achieved 100% error correction for specific patterns of up to four errors when these patterns were entirely contained within the data matrix. Using a commercial error evaluation tool, the authors evaluated the effectiveness of EMPC-SA with 36 high-occurrence error patterns (Figure 23) obtained through neutron particle strike simulations [49].

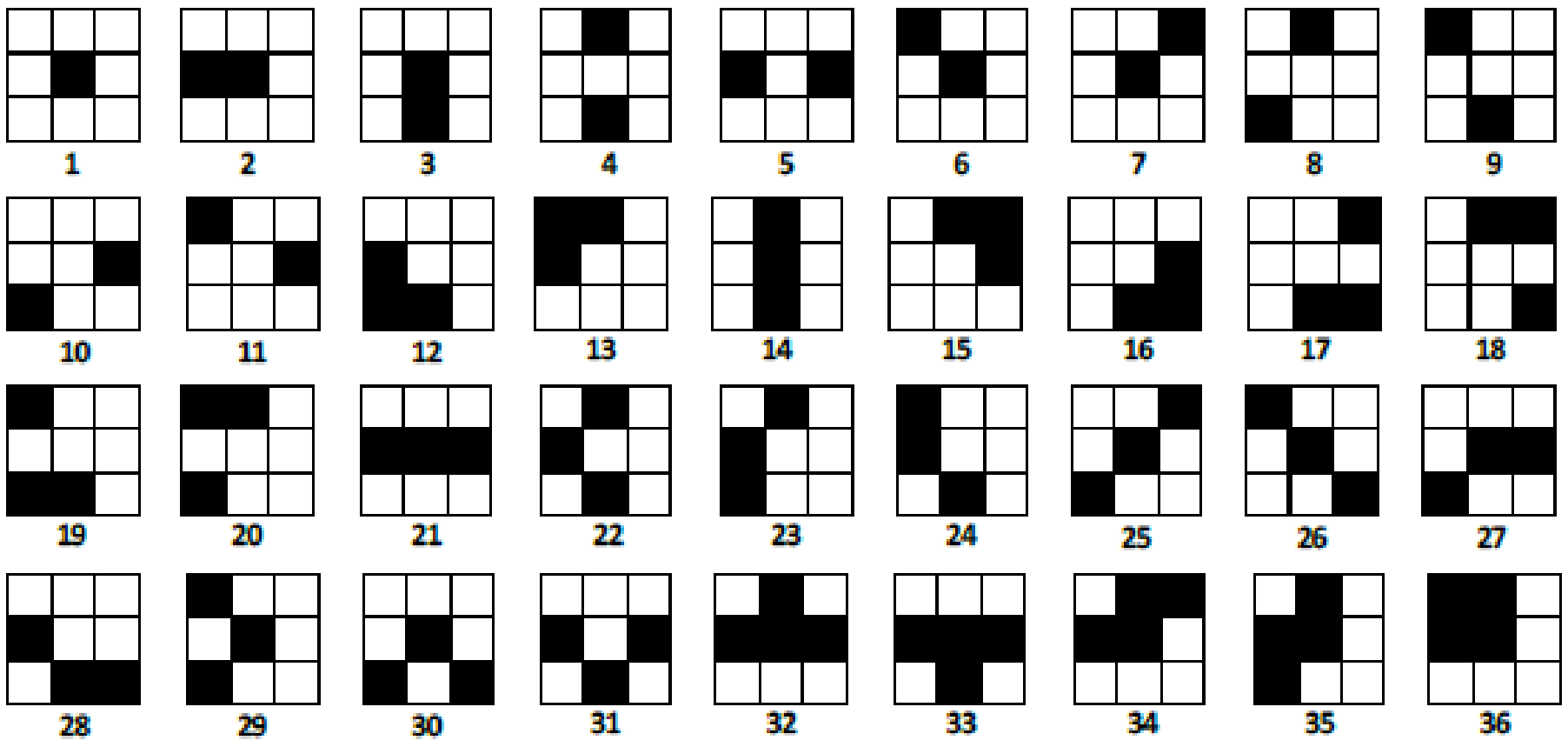

**Figure 23. High-occurrence error patterns (Source: Adapted from [32]).**

The authors placed error patterns in all feasible regions and intersections within the data matrix to determine correction rates. This iterative process tested all potential positions for the detected error bits. The results were categorized based on whether the error source was entirely within, partially, or outside the data region. This analysis allowed the evaluation of how EMPC-SA handles different error configurations and the conditions under which it achieves 100% error correction.

The testing methodology comprehensively assesses EMPC-SA's performance under varied error conditions, enabling a robust comparison of its effectiveness against other error correction codes. This makes EMPC-SA a promising candidate for improving fault tolerance in space-critical memory systems.

## 3.17 Fault tolerant micro-programmed control unit for SEU and MBU mitigation in space based digital systems

Deepanjali and Noor [8] introduced a novel ECC methodology to mitigate both SEUs and MBUs. The study minimizes hardware complexity and latency while maintaining a consistent code rate. It is particularly suitable for space-based digital systems where radiation-induced faults are a significant concern.

The proposed ECC, known as the SYMmetric Segmented Code (SYMSEG), is designed for fault mitigation in the control words of a Micro-Programmed Control Unit (MPCU). This code uses a symmetric segmentation approach, dividing control words into smaller segments. Each segment includes parity bits, while additional check bits are computed using XOR operations to enhance robustness. The design ensures that adjacent MBUs within a segment can be corrected while random errors across segments can be detected.

The encoded control words of SYMSEG are expanded by adding check and parity bits, which improve fault tolerance without significantly increasing overhead. The methodology is particularly effective for high-speed on-chip memory applications and has been implemented and validated on a LEON3 processor. Its utility is further demonstrated in space and radiation-intensive environments, where fault resilience is crucial.

The ECC exhibited superior fault mitigation to traditional ECCs, such as the Hamming and Reed-Solomon codes. Its performance metrics highlight its ability to balance error correction and detection with moderate resource use. For instance, implementing SYMSEG on FPGA hardware results in a 1.66% increase in logic usage compared to standalone systems, which is a reasonable trade-off for its enhanced fault-tolerant capabilities.

The research also addressed the limitations of Hamming codes in mitigating MBUs and avoids the high latency associated with iterative codes like Turbo and LDPC. However, its ability to correct random MBUs is limited when errors span across different segments of the control word, which remains a notable limitation. Despite this, the proposed ECC offers a promising solution for applications requiring robust fault tolerance with minimal latency, particularly in mission-critical environments such as space exploration.

## 3.18 Comparing structures of two-dimensional error correction codes

Muniz et al. [40] explored various 2D-ECC organizations and analyzed their performance in error correction and detection rates, scalability, and synthesis costs. The study investigated the increasing susceptibility of ICs to bitflips due to device miniaturization and explored the role of 2D-ECCs in ensuring fault tolerance in modern systems.

Four specific 2D-ECC organizations employ combinations of parity, and Hamming codes organized in a matrix format are evaluated —N×4p, N×ExHam, N×Ham_p, and N×Ham2_2p. The N×ExHam employs Extended Hamming codes for each row, classified as SECDED. The N×4p utilizes parities of rows, columns, and diagonals for error detection and correction. The N×Ham_p incorporates row-wise Hamming encoding combined with column parities, allowing for selective double error corrections. Lastly, N×Ham2_2p combines row and column parity checks with Hamming encoding across two rows, offering trade-offs between error correction efficacy and implementation complexity.

Figure 24 shows the configuration used to collect data on the corrected error rate and the number of undetected error scenarios. Exhaustive errors ranging from 1 to 16 were injected into the data, check, and entire codestruct regions for the four organizations.

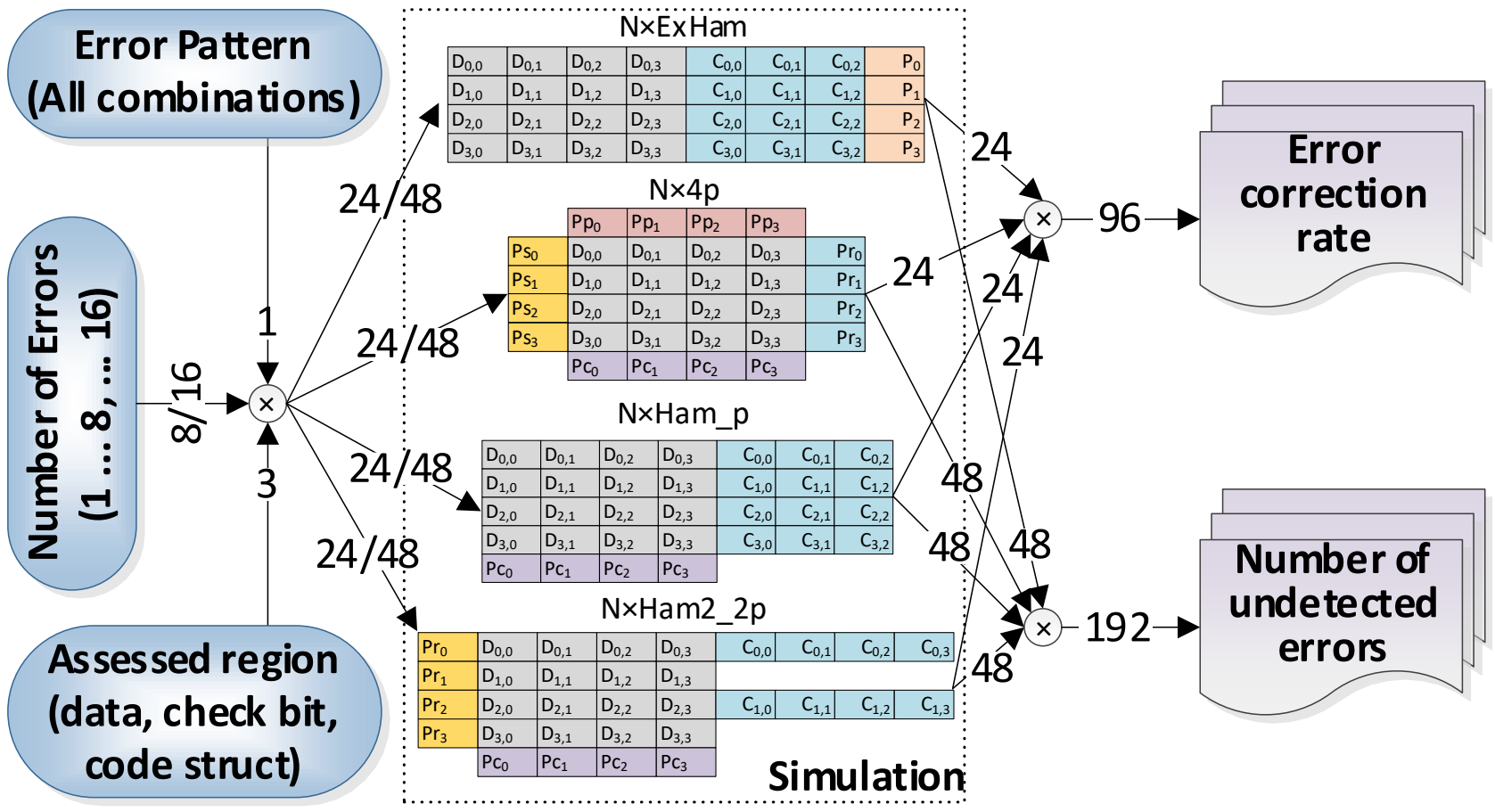

**Figure 24. Experimental setup employed to collect the error correction rates and number of undetected errors (Source: [40]).**

Experimental results highlighted differences in error correction rates depending on the error location (data or check bit regions) and the code organization. While cross-checking ECCs like N×Ham_p and N×Ham2_2p offer improved data area error correction rates, independent codes like N×ExHam demonstrate superior performance in check bit areas. Scalability metrics indicate that redundancy costs remain low as the data matrix size increases, underscoring the efficiency of these 2D-ECC designs for large-scale applications.

Finally, the paper concludes with a synthesis analysis revealing trade-offs between area consumption, power dissipation, and decoding latency among the evaluated ECCs.

## 3.19 Novel Latin square matrix code of large burst error correction for MRAM applications

Jin et al. [29] introduced the Latin Square Matrix (LSM) code, a 2D-ECC developed to address the challenges of large burst errors in Magnetic Random Access Memory (MRAM) systems. MRAM is increasingly utilized in modern computing due to its non-volatility, high speed, and low power consumption. However, it remains vulnerable to MBUs caused by radiation and other environmental factors.

The proposed LSM code leverages the structure of orthogonal Latin squares to encode and decode data. A typical implementation organizes 16 information bits into a 6-order Latin square and adds 12 parity bits to provide robust protection against up to 5-bit burst errors. Encoding involves calculating horizontal and vertical parity bits using XOR operations, while decoding uses syndromes derived from these parity bits to identify erroneous positions precisely. This design ensures unique error syndromes for each burst error pattern, enabling precise correction.

The LSM code achieves superior error correction with significantly lower hardware overhead when compared to LDPC; i.e., the area consumption and power dissipation of LSM are reduced by over 40%. Furthermore, the error correction capability of LSM codes can be scaled by increasing the order of the Latin square, allowing it to correct larger burst errors with minimal additional parity bits.

The authors synthesized LSM code using Verilog HDL and validated it through extensive fault injection simulations. The evaluation highlights the LSM code's efficient trade-off between reliability and hardware overhead, with applications extending to high-density MRAM in embedded systems and space exploration.

## 3.20 nMatrix: A New Decoding Algorithm for the Matrix ECC

Freitas et al. [20] developed the nMatrix 2D-ECC to address reliability issues in electronic memory systems, especially in critical applications such as space missions. This code is based on the existing Matrix ECC [1]. However, it introduces a more efficient decoding algorithm that improves error correction capacity and reduces implementation costs, albeit with slight decoder delay and area trade-offs.

nMatrix retains the original Matrix structure, protecting memory data with redundancy bits distributed in rows and columns. It enhances the error correction rates for multiple adjacent bitflips in specific scenarios compared to Matrix and other ECCs, such as eMRSC [59] and DMC [25]. The code supports various configurations, such as 16-bit and 32-bit data versions and incorporates techniques like interleaving to mitigate MBUs. Figure 25(a) illustrates the organization of nMatrix(16,32) – 16 data bits and 16 redundancy bits – highlighting the data areas and check bits (parity and Hamming). Figure 25(b) demonstrates the same nMatrix organization, focusing on the proposed bit interleaving approach.

Experimental results highlight that nMatrix has better error correction for more complex error patterns, higher reliability over extended operational periods, and superior energy efficiency compared to its counterparts. While its correction efficacy for large-scale adjacent errors does not surpass all alternatives, its balance between performance and hardware cost positions it as a robust choice for applications requiring enhanced fault tolerance.

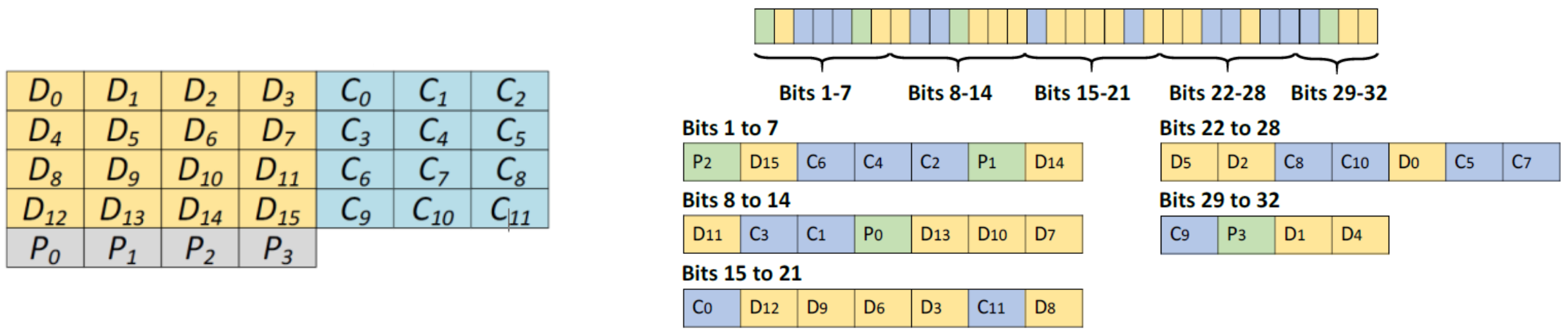

**Figure 25. (a) nMatrix(16,32) organization Matrix code structure. $D_0$ to $D_{15}$, $C_0$ to $C_{11}$, and $P_0$ to $P_3$ are data, check, and parity bits, respectively; (b) nMatrix(16,32) interleaving format (Source: [20]).**

## 3.21 An Embedded Module of Enhanced Turbo Product Code Algorithm

Luo et al. [31] proposed the Bit-Flipping Enhanced Turbo Product Code (BFE-TPC) for protecting NAND flash memory applications. Figure 26 illustrates the generic organization of a Turbo Product Code (TPC), which is characterized as a product code that includes three check areas in addition to the data area. The BFE-TPC version is optimized to avoid overloading the check areas. The authors do not specify the composition of each check region but only provide the total number of check and data bits.

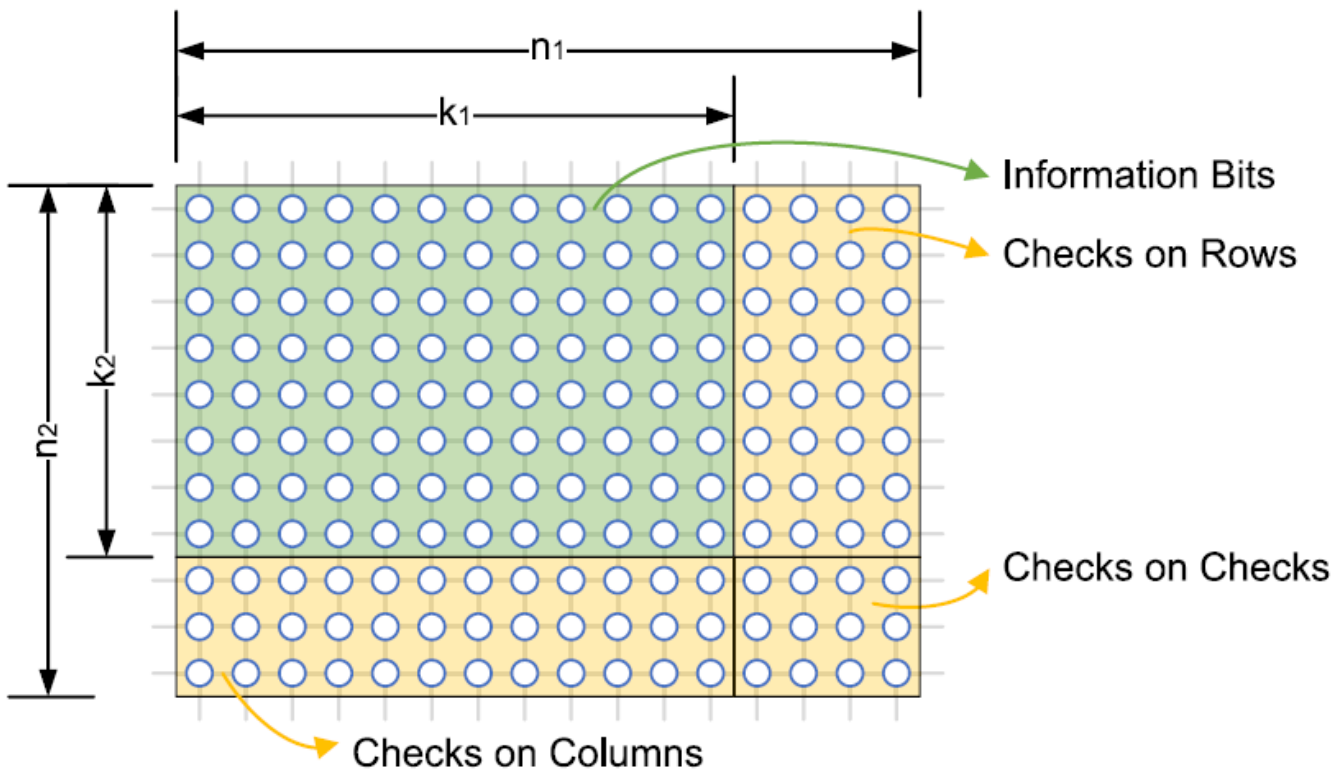

**Figure 26. Generic organization of a TPC as utilized in BFE-TPC (Source: [31]).**

BFE-TPC employs Bose-Chaudhuri-Hocquenghem (BCH) [36] codes along rows and columns. A key feature of BFE-TPC is incorporating a bit-flipping technique within its decoding algorithm, which enhances error correction capabilities for uncorrectable submatrices in high error-rate conditions.

The BFE-TPC encoder processes input data row-wise, combining binary matrix multiplication and polynomial division logic for rows and columns, respectively. The BFE-TPC decoder integrates an embedded processor to manage bit-flipping operations, which reduces uncorrectable errors in challenging scenarios. Prototyped using FPGA technology, the BFE-TPC achieves high throughput with significant reductions in hardware area and latency compared to LDPC implementations.

Experimental results demonstrate that BFE-TPC meets industry standards for Uncorrectable Bit Error Rate (UBER) at levels up to $10^{-15}$ in Triple-Level Cell (TLC) NAND flash memory.

## 3.22 Check-Bit Region Exploration in Two-Dimensional Error Correction Codes

Freitas et al. [21] delved into the efficacy of various 2D-ECCs, particularly examining their performance in correcting errors in the check-bit region. The dimensions of the codewords of the 2D-ECCs vary based on the specific 2D-ECC applied, with configurations designed to maximize reliability while minimizing redundancy overhead.

Figure 27 demonstrates how the authors explored the trade-off of increasing the number of check regions to protect the same 16-bit data area ($D_0$ to $D_{15}$). While Figure 27(a) organization represents a Straightforward 2D-ECC, with 4 redundancy bits protecting 4 data bits per row, Figure 27(b) introduces a modified product code by intersecting row and column redundancies. Finally, Figure 27(c) extends this approach by incorporating an additional diagonal redundancy check.

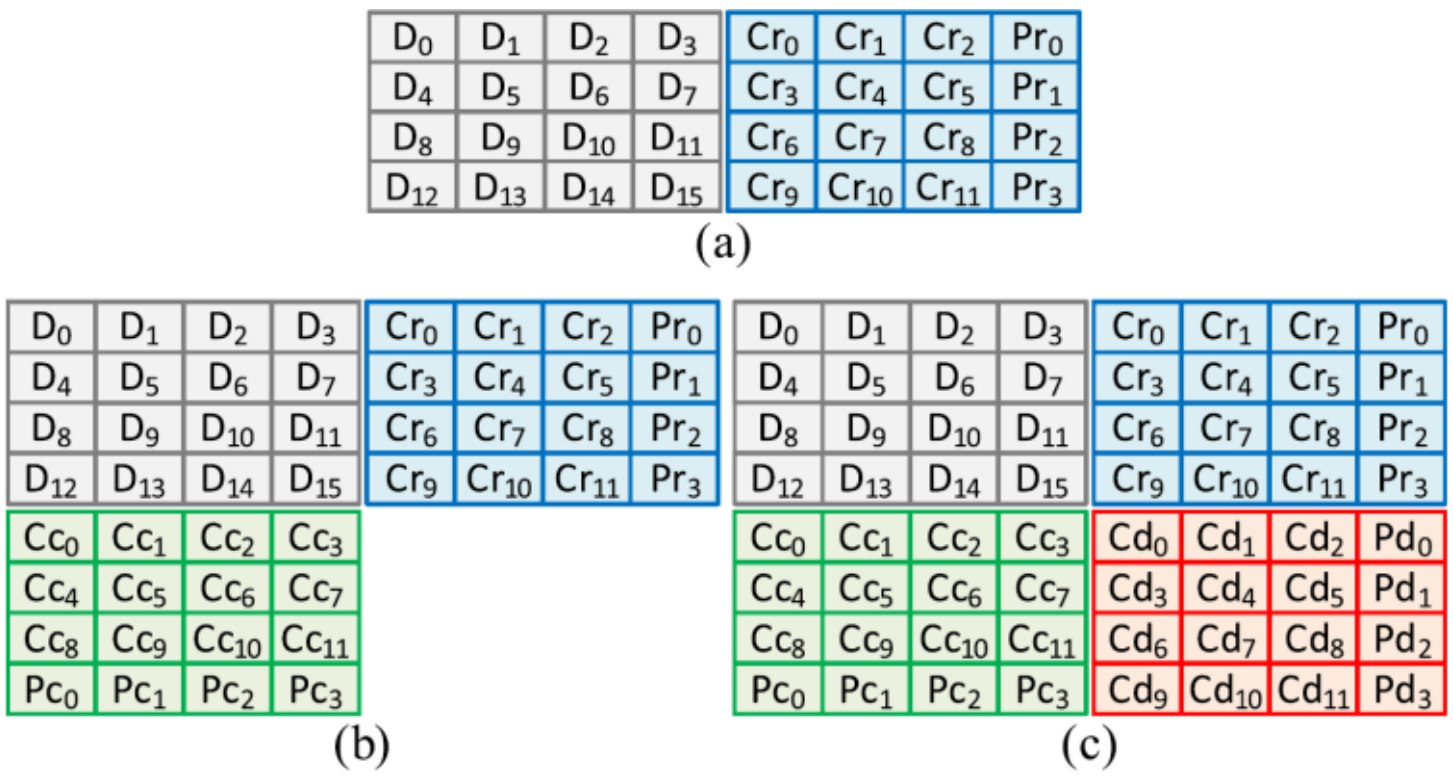

**Figure 27. Three 2D-ECC organizations evaluated in the study; all configurations include 16 data bits and 16 redundancy bits (Source: [21]).**

The study emphasizes the need to optimize 2D-ECC structures to balance correction rates for data and check bit regions, given the growing challenges MBUs pose in modern memory systems. One of the paper's key findings is the differential error correction performance between the data and check bit regions. While codes like Matrix excel in protecting the data region, their effectiveness in the check bit region is relatively lower. On the other hand, certain codes tailored to provide additional redundancy in the check bit region achieve a more balanced error correction rate across both regions, making them suitable for environments with high MBU occurrences.

The paper discusses synthesizing these codes for hardware technologies like SRAM and DRAM. Simulation results demonstrate that these codes provide a good trade-off between error correction capability, area consumption, and power dissipation. The research underscores the importance of optimizing redundancy distribution to enhance the overall efficacy of 2D-ECCs in protecting both data and check bit regions.

## 3.23 Related Work Summary

Table 4 aims to provide the reader with an overview of the main characteristics of the 2D-ECC works presented in this chapter. This allows the reader to understand how the proposed work relates to others and highlights its innovative nature.

**Table 4. Comparative summary of important aspects of the presented works (Source: Author).**

| Work | Year | Classification | Application | Data area (bits) | Redundancy area (bits) | Fault injection method | CMOS technology |
|---|---|---|---|---|---|---|---|
| Argyrides et al. [1] | 2010 | Mixed Code | Generic | 32, 64 | 24, 32 | Adjacent | - |
| Guo et al. [25] | 2014 | Mixed Code | Memory | 32 | 36 | Random | 180 nm |
| Rahman et al. [48] | 2015 | Mixed Code | Memory | 64 | 27 | Exhaustive | - |
| Erozan, Çavuş [11] | 2015 | Product Code | Memory | 32 | 47 | Adjacent | * |
| Raha, Vinodhini, Murty [47] | 2017 | Mixed Code | Memory | 32 | 28 | Random | - |
| Moran et al. [39] | 2018 | Mixed Code | Memory | 32 | 28 | Adjacent | 45 nm |
| Freitas et al. [14] | 2020 | Extended Product Code | Space | 16 | 32 | Exhaustive | 65 nm |
| Silva et al. [59] | 2020 | Mixed Code | Critical | 32, 32 | 64, 57 | Adjacent | 65 nm |
| Freitas et al. [13] | 2020 | Product Code | Space | 32 | 32 | Adjacent, exhaustive | 65 nm |
| Freitas et al. [15] | 2021 | Product Code | Space | 16 | 48 | Adjacent | 65 nm |
| Sen et al. [53] | 2021 | Extended Product Code | Critical | 32 | 32 | Random | - |
| Tejas et al. [66] | 2022 | Mixed Code | Space | 32 | 28 | Random | 180 nm |
| Chen et al. [6] | 2022 | Product Code | Memory | 64 | 16 | - | 40 nm |
| Freitas et al. [16] | 2022 | Product Code | Space | 16 | 32 | Exhaustive | 65 nm |
| Yuqi et al. [69] | 2023 | Product Code | Memory | 64 | 28 | Random | - |
| Magalhães et al. [32] | 2023 | Product Code | Space | 16 | 24 | Specific standards | - |
| Deepanjali and Noor [8][7] | 2024 | Straightforward 2D-ECC | Space | 32 | 19 | Adjacent, random | 40 nm |
| Muniz et al. [40] | 2024 | Straightforward 2D-ECC Mixed Code | | 16 | 16 | Exhaustive | 28 nm |
| Jin et al. [29] | 2024 | Mixed Code | Memory | 16, 16 | 28, 30 | Random | 65 nm |
| Freitas et al. [20] | 2024 | Mixed Code | Generic | 16, 32 | 16, 32 | Exhaustive | 65 nm |
| Luo et al. [31] | 2024 | Product Code | Memory | 65664, 16416, 8192 | 11826, 1980, 1024 | Random | 28 nm, * |
| Freitas et al. [21] | 2024 | Straightforward 2D-ECC Product Code | Generic | 16, 16, 16 | 16, 32, 48 | Adjacent, exhaustive | 65 nm |
| **This work** | 2025 | Overlapping | Memory | 4, 9, 16 | 8, 10, 12 | Exhaustive | 28 nm |

*"–" work does not provide information on the subject.*

*"*" the synthesis was performed for a technology other than CMOS.*

# 4. OVERLAPPING ECC PROPOSAL

This chapter presents the fundamental concepts underlying the overlapping technique, including a mathematical analysis of its error correction potential and the expansion of the proposed approach.

## 4.1 Introduction

A review of the literature has shown that two-Dimensional (2D) approaches exploit the crossing of ECCs, ensuring that each bit in a data or parity matrix is encoded by two or more ECCs, as illustrated in Figure 28. These approaches enable researchers to develop cross-verification algorithms, enhancing both error detection and correction capabilities.

| | | | | | | |
|---|---|---|---|---|---|---|
| $D_{0,0}$ | $D_{0,1}$ | $D_{0,2}$ | $D_{0,3}$ | $C_{0,0}$ | $C_{0,1}$ | $C_{0,2}$ |
| $D_{1,0}$ | $D_{1,1}$ | $D_{1,2}$ | $D_{1,3}$ | $C_{1,0}$ | $C_{1,1}$ | $C_{1,2}$ |
| $D_{2,0}$ | $D_{2,1}$ | $D_{2,2}$ | $D_{2,3}$ | $C_{2,0}$ | $C_{2,1}$ | $C_{2,2}$ |
| $D_{3,0}$ | $D_{3,1}$ | $D_{3,2}$ | $D_{3,3}$ | $C_{3,0}$ | $C_{3,1}$ | $C_{3,2}$ |
| $R_{0,0}$ | $R_{0,1}$ | $R_{0,2}$ | $R_{0,3}$ | | | |
| $R_{1,0}$ | $R_{1,1}$ | $R_{1,2}$ | $R_{1,3}$ | | | |
| $R_{2,0}$ | $R_{2,1}$ | $R_{2,2}$ | $R_{2,3}$ | | | |

**Figure 28. Example traditional 2D-ECCs organization highlighting the intersection of the second-row ECC with the second-column ECC, leading to the shared encoding of the data bit $D_{1,1}$ (Source: Author).**

Figure 28 exemplifies a 4×4 data matrix, where each row and column contain 4 data bits, resulting in 16 data bits ($D_{0,0}$ to $D_{3,3}$). This data region is protected by three parity bits per row and per column. For instance, the column $D_{0,0}$ to $D_{3,0}$ is protected by the parity column $R_{0,0}$ to $R_{2,0}$, while the row $D_{1,0}$ to $D_{1,3}$ is protected by $C_{1,0}$ to $C_{1,2}$.

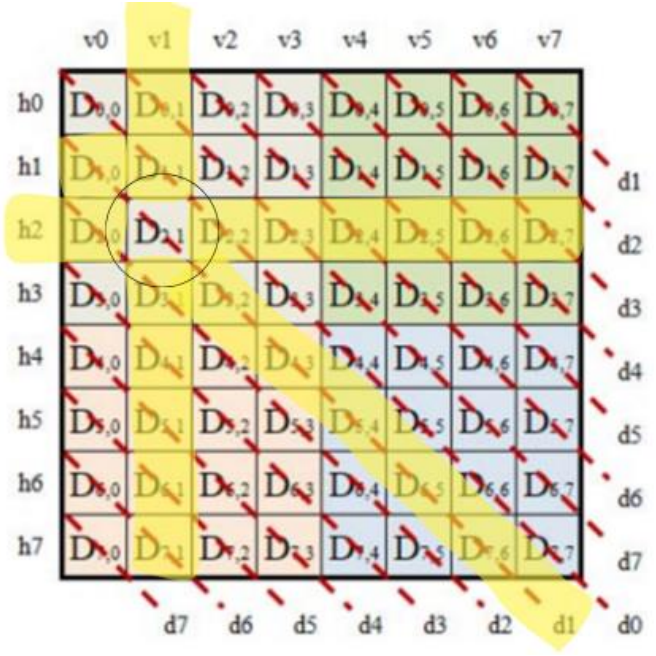

**Figure 29. Encoding structure of the 2D-ECC HVDB, highlighting the three-dimensional intersection of ECCs at bit $D_{2,1}$ (Source: Adapted from [69]).**

At each intersection point, ECC algorithms can leverage error correction opportunities from each intersecting ECC, both independently and jointly, typically through an iterative process. This crossing structure results in a logically multidimensional organization, but for all practical implementations, ECCs are physically arranged in a 2D format. For example, in Figure 21, the data bit $D_{2,1}$ is logically protected by a three-dimensional ECC structure composed of $v_1$, $h_2$, and $d_1$, but it is implemented as a 2D layout.

### 4.1.1 The Proposed Overlapping ECC Approach

The proposed approach extends ECC coverage by overlapping ECCs, ensuring that the same data region is protected by two or more ECCs. Figure 30 illustrates this overlapping structure, where verification areas are rearranged by superimposing an additional row of ECCs over a data row.

| | | | | | | | | | |
|---|---|---|---|---|---|---|---|---|---|
| $R_{0,0}$ | $R_{0,1}$ | $R_{0,2}$ | $D_{0,0}$ | $D_{0,1}$ | $D_{0,2}$ | $D_{0,3}$ | $C_{0,0}$ | $C_{0,1}$ | $C_{0,2}$ |
| $R_{1,0}$ | $R_{1,1}$ | $R_{1,2}$ | $D_{1,0}$ | $D_{1,1}$ | $D_{1,2}$ | $D_{1,3}$ | $C_{1,0}$ | $C_{1,1}$ | $C_{1,2}$ |
| $R_{2,0}$ | $R_{2,1}$ | $R_{2,2}$ | $D_{2,0}$ | $D_{2,1}$ | $D_{2,2}$ | $D_{2,3}$ | $C_{2,0}$ | $C_{2,1}$ | $C_{2,2}$ |
| $R_{3,0}$ | $R_{3,1}$ | $R_{3,2}$ | $D_{3,0}$ | $D_{3,1}$ | $D_{3,2}$ | $D_{3,3}$ | $C_{3,0}$ | $C_{3,1}$ | $C_{3,2}$ |

**Figure 30. Example organization of the overlapped-ECCs. This figure highlights how the entire second row of data ($D_{1,0}$, $D_{1,1}$, $D_{1,2}$, $D_{1,3}$) is protected by two ECCs, with their parity bits represented by R (row-based ECC) and C (column-based ECC) (Source: Author)**

### 4.1.2 Encoding and Decoding Differences Between Crossed and Overlapping ECCs

The encoding of overlapping ECCs is performed independently for each overlapping code, just as in the crossed ECC approach. However, the key difference emerges during decoding, particularly when an error is detected.

The overlapping approach increases the sharing of data bits compared to the crossing approach, creating new opportunities for error detection and correction. However, this comes at the cost of higher decoding complexity, requiring a trade-off analysis between *increasing error detection and correction efficacy* and *reducing efficiency due to higher latency, power consumption, and area overhead*.

### 4.1.3 Trade-offs and Practical Considerations

It is important to note that power dissipation and latency increase significantly in the presence of errors since, upon error detection, the decoder activates the correction circuitry. This conditional impact on performance makes the proposed approach suitable for a broad range of high-speed applications, particularly in scenarios with sporadic error incidence.

## 4.2 Analysis of Hamming Distance

The overlapping of ECCs is characterized by the independent encoding, while decoding can be performed either independently or combined. Because of this model, the total Hamming distance of the codestruct is obtained by summing the distances of each individual codeword. The overlapping codewords evaluated in this study employ extended Hamming coding, which has a minimum distance of 4 [37]. In isolation, these codes would be capable of correcting 1 error and detecting 3 errors, as stated in Equations (1) and (2) of Section 2.4. However, by summing the distances, code overlapping results in a total distance of 8, allowing for the detection of up to 7 errors and the correction of up to 3 errors.

It is important to note that these values represent theoretical limits of correction and detection imposed by the logical combination of all encoding possibilities. However, evaluating all possible cases is not a practical solution, especially as the number of the codeword bits increases significantly.

This work proposes an intermediate solution, employing an algorithm limited to correcting and detecting 2 and 4 errors, respectively. This approach provides a balanced trade-off between efficacy and efficiency. Moreover, in cases where no errors are detected, the decoding latency remains like the encoding latency, enabling high-frequency to read and write operations. The values described here are presented in Chapter 6.

## 4.3 Overlapping ECC Model Explored in This Work

The ECC overlapping model is broad, as there are numerous possibilities for combining error correction methods and organizing data and check bits. For example, Figure 31 shows a 2D codestruct that combines two ECCs (e.g., Hamming) sharing the same data area ($D_{0,0}$-$D_{3,3}$) with a vector-type ECC, where each bit serves as a column parity check.

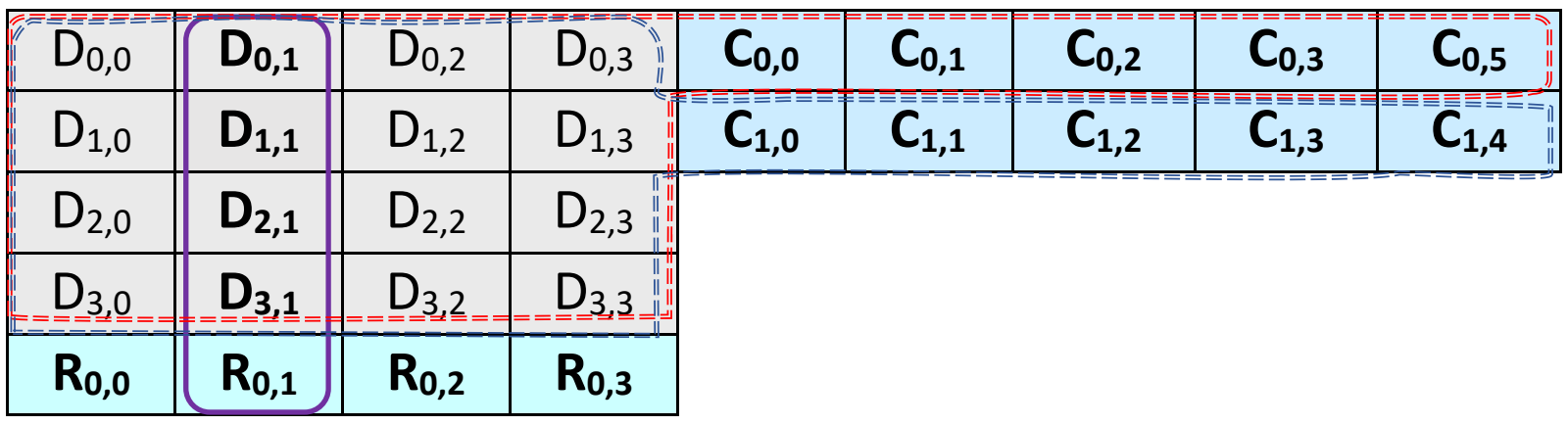

**Figure 31. Example of a 2D-ECC codestruct with overlapping highlighting the entire data sharing ($D_{0,0}$-$D_{3,3}$) by two ECCs, with check bits represented by the letter C, and the crossing of the second column ($D_{0,1}$, $D_{1,1}$, $D_{2,1}$, $D_{3,1}$) with the ECC represented by $R_{0,1}$ – possibly a parity bit (Source: Author).**

The codestruct illustrated in Figure 32 is a modified version of Figure 31, where the ECC that verifies data columns is removed, and an additional check bit is added to the rows. One possible implementation is that each ECC is an extended Hamming code, and both ECCs protect (overlap) the same data area. Note that the codestruct in Figure 32 has two fewer bits than the one in Figure 31, potentially allowing for distinct decoding algorithms that explore the trade-offs between effectiveness and efficiency.

| | | | | | | | | | |
|---|---|---|---|---|---|---|---|---|---|
| $D_{0,0}$ | $D_{0,1}$ | $D_{0,2}$ | $D_{0,3}$ | **$C_{0,0}$** | **$C_{0,1}$** | **$C_{0,2}$** | **$C_{0,3}$** | **$C_{0,5}$** | **$R_0$** |
| $D_{1,0}$ | $D_{1,1}$ | $D_{1,2}$ | $D_{1,3}$ | **$C_{1,0}$** | **$C_{1,1}$** | **$C_{1,2}$** | **$C_{1,3}$** | **$C_{1,4}$** | **$R_1$** |
| $D_{2,0}$ | $D_{2,1}$ | $D_{2,2}$ | $D_{2,3}$ | | | | | | |
| $D_{3,0}$ | $D_{3,1}$ | $D_{3,2}$ | $D_{3,3}$ | | | | | | |

**Figure 32. Example of an overlapped-ECC codestruct. This example highlights the entire data sharing ($D_{0,0}$-$D_{3,3}$) by two ECCs, with check bits represented by the letters C and R – possibly an extended Hamming code (Source: Author).**

This model employs the extended Hamming code to protect the data area $D_{0,0}$-$D_{3,3}$, with double overlapping, as two extended Hamming codes cover the same area. However, it is possible to overlap other codes, such as the Decimal Matrix Code (DMC), as explained in Section 3.2.

Although this work focuses on extended Hamming codes and their double overlapping, where two codes protect the same area, additional code overlaps could be applied, incorporating three or more verification levels with different ECC schemes.

The main advantage of the extended Hamming code lies in its double-error detection capability, as discussed in Section 2.7, enhancing the reliability of data transmission and storage. While it does not increase storage capacity (the number of data bits remains the same), it significantly improves error protection, making the system more trustable. When two extended Hamming codes cover the same data area in a crosswise manner, additional redundancies improve the ability to detect and correct complex errors, such as:

- In a simple error correction system, each extended Hamming code functions independently within its axis (e.g., one for rows and another for columns).
- This approach allows for single-error correction in any direction and enhanced detection of multiple errors, as the crossing enables error identification in multiple positions that would otherwise remain undetected or uncorrected by a single code.
- The cross-checking mechanism prevents single-fault propagation, increasing reliability.
- With global parity, there is additional redundancy across both axes, making it more likely to detect errors that would bypass a single extended Hamming code.

### 4.3.1 Explored Overlapped-ECC Codestructs

Three overlapping ECC models were explored, each utilizing only two overlapping codes. These ECCs were applied to 2×2, 3×3, and 4×4 matrices.

#### 2×Ham_2×2 Structure

The 2×2 matrix, as illustrated in Figure 33, consists of an extended Hamming code, which has three check bits and one global parity bit protecting four data bits – $ExHam(8,4)$. This matrix is overlapped by two $ExHam(8,4)$ and is referred to in this work as 2×Ham_2×2.

| | | | | | |
|---|---|---|---|---|---|
| D | D | Co0 | Co1 | Co2 | Po |
| D | D | Ci0 | Ci1 | Ci2 | Pi |

**Figure 33. An example of an overlapped-ECC - 2×Ham_2×2 codestruct (Source: Author).**

The 2×Ham_2×2 structure includes (i) $C_o0...C_o2$ outer Hamming check bits; (ii) $P_o$ Global parity bit for protecting both data and check bits; (iii) $C_i0...C_i2$ inner Hamming check bits; (iv) $P_i$ internal parity bit; and (v) D data area.

#### 2×Ham_3×3 Structure

Figure 34 displays the second model that explores a 3×3 data area, consisting of nine data bits and four check bits. This matrix is overlapped by two $ExHam(14,9)$ and is referred to in this work as 2×Ham_3×3.

| | | | | | | | |
|---|---|---|---|---|---|---|---|
| D | D | D | Co0 | Co1 | Co2 | Co3 | Po |
| D | D | D | Ci0 | Ci1 | Ci2 | Ci3 | Pi |
| D | D | D | | | | | |

**Figure 34. 2×Ham_3×3 codestruct (Source: Author).**

The 2×Ham_3×3 structure includes (i) Co0…Co3 outer Hamming check bits; (ii) Po Global parity bit for protecting both data and check bits; (iii) Ci0…Ci3 inner Hamming check bits; (iv) Pi internal parity bit; and (v) D data area, containing nine data bits. Reminder that four Hamming check bits can protect up to eleven data bits.

#### 2×Ham_4×4 Structure

The third model contains 16 data bits in a 4×4 data area, protected by five check bits, using an extended Hamming code; i.e., $ExHam(22,16)$, as depicted in Figure 35. This matrix is overlapped by two $ExHam(22,16)$ and is referred to in this work as 2×Ham_4×4. Note that five check bits can potentially cover 26 data bits, performing a $ExHam(32,26)$.

| D | D | D | D | Co0 | Co1 | Co2 | Co3 | Co4 | Po |
|---|---|---|---|---|---|---|---|---|---|
| D | D | D | D | Ci0 | Ci1 | Ci2 | Ci3 | Ci4 | Pi |
| D | D | D | D | | | | | | |
| D | D | D | D | | | | | | |

**Figure 35. 2×Ham_4×4 codestruct (Source: Author).**

The 2×Ham_4×4 structure includes (i) Co0…Co4 outer Hamming check bits; (ii) Po Global parity bit for protecting both data and check bits; (iii) Ci0…Ci4 inner Hamming check bits; (iv) Pi internal parity bit; and (v) D data area, containing 16 data bits. Reminder that five Hamming check bits can protect up to 26 data bits.

## 4.4 Data Addressing in Overlapping ECCs

The overlapping ECC examples used in this work employ Hamming coding, which allows data bits to be addressed by multiple combinations of parity bits. Each combination is designed to identify errors that may occur in both data bits and parity bits.

For instance, the standard $Ham(7,4)$ code uses 3 parity bits to protect 4 data bits ($d_0 \ldots d_4$), in addition to the 3 parity bits ($c_0 \ldots c_2$). As illustrated in Figure 36, the addresses 1, 2, and 4 correspond to the parity bits, while the data bits are assigned addresses 3, 5, 6, and 7 – which are, in fact, combinations of the parity bit addresses. Additionally, when no error is detected, the error address is set to zero.

$$\begin{array}{cccccccc} [\ d_0 & d_1 & d_2 & d_3 & c_0 & c_1 & c_2\ ] \\ 3 & 5 & 6 & 7 & 1 & 2 & 4 \\ & \textit{data bits} & & & \textit{check bits} & & \end{array}$$

**Figure 36. Standard address mapping of a $Ham(7,4)$ (Source: Author).**

Although $Ham(7,4)$ coding assigns the addresses 3, 5, 6, and 7 to the data bits, these bits can be logically arranged in any of the 24 possible permutations of these four addresses (i.e., 4! combinations), as partially illustrated in Figure 37.

| $d_0$ | $d_1$ | $d_2$ | $d_3$ | $d_0$ | $d_1$ | $d_2$ | $d_3$ | $d_0$ | $d_1$ | $d_2$ | $d_3$ | $d_0$ | $d_1$ | $d_2$ | $d_3$ |
|---|---|---|---|---|---|---|---|---|---|---|---|---|---|---|---|
| 3 | 5 | 6 | 7 | 5 | 3 | 6 | 7 | 6 | 5 | 3 | 7 | 7 | 5 | 6 | 3 |
| 3 | 5 | 7 | 6 | 5 | 3 | 7 | 6 | 6 | 5 | 7 | 3 | 7 | 5 | 3 | 6 |
| 3 | 6 | 5 | 7 | 5 | 6 | 3 | 7 | 6 | 3 | 5 | 7 | 7 | 6 | 5 | 3 |
| 3 | 6 | 7 | 5 | 5 | 6 | 7 | 3 | 6 | 3 | 7 | 5 | 7 | 6 | 3 | 5 |
| 3 | 7 | 5 | 6 | 5 | 7 | 3 | 6 | 6 | 7 | 5 | 3 | 7 | 3 | 5 | 6 |
| 3 | 7 | 6 | 5 | 5 | 7 | 6 | 3 | 6 | 7 | 3 | 5 | 7 | 3 | 6 | 5 |

**Figure 37. Possible data address permutations for the $Ham(7,4)$ code (Source: Author).**

This flexibility in mapping physical and logical addresses allows the same error to have different addresses depending on the overlapping ECC used. Figure 38 illustrates an error in bit $d_2$, where two $Ham(7,4)$ codes (ECC_1 and ECC_2) identify the error at different logical addresses.

| | $d_0$ | $d_1$ | $d_2$ | $d_3$ | $c_0$ | $c_1$ | $c_2$ | Error address |
|---|---|---|---|---|---|---|---|---|
| ECC_1 | 3 | 5 | 6 | 7 | 1 | 2 | 4 | 6 |
| ECC_2 | 6 | 3 | 7 | 5 | 1 | 2 | 4 | 7 |

**Figure 38. Two $Ham(7,4)$ codes (ECC_1 and ECC_2) detecting an error in data bit $d_2$, where each ECC assigns a different logical error address (Source: Author).**

As discussed in Section 2.6, Hamming coding employs the exclusive OR (XOR) operation to generate the address of an error within a codeword. Consequently, when multiple errors occur, the error address is computed as the XOR of the addresses of the individual bit errors. This means that multiple errors in overlapping ECCs may be assigned distinct addresses, depending on how the data bits are mapped.

For example, consider the same $Ham(7,4)$ codewords as in Figure 39, but now with errors in bits $d_2$ and $d_3$. For ECC_1, these correspond to errors at addresses 6 and 7, whereas for ECC_2, they correspond to addresses 7 and 5. Applying the XOR operation between 6 and 7 and between 7 and 5 results in addresses 1 and 2, respectively. Assuming no other double-error combination generates the same address pair, this pattern could be used as an identifier for a double-error correction algorithm.

| | | $d_0$ | $d_1$ | $d_2$ | $d_3$ | $c_0$ | $c_1$ | $c_2$ | Error address |
|---|---|---|---|---|---|---|---|---|---|
| ECC_1 | (decimal) | 3 | 5 | 6 | 7 | 1 | 2 | 4 | 1 |
| | (hexa) | 011 | 101 | 110 | 111 | 001 | 010 | 100 | 001 |
| ECC_2 | (decimal) | 6 | 3 | 7 | 5 | 1 | 2 | 4 | 2 |
| | (hexa) | 110 | 011 | 111 | 101 | 001 | 010 | 100 | 010 |

**Figure 39. Two $Ham(7,4)$ codewords (ECC_1 and ECC_2) detecting errors in data bits $d_2$ and $d_3$; the final error address is obtained using the XOR operation on the individual error addresses (Source: Author).**

However, in some cases, different error combinations may result in the same XOR-generated address, making it difficult for a decoder to correctly infer the positions of the errors. For example, both error pairs (3,5) and (5,7) produce the same address 6 when XOR is applied (Figure 40). This ambiguity could lead to incorrect bit modifications, potentially corrupting correctly stored data in a code structure. To address this, a mapping strategy is required to distribute logical addresses across overlapping ECCs in a way that ensures unique addresses for all targeted error patterns.

| | | $d_0$ | $d_1$ | $d_2$ | $d_3$ | $c_0$ | $c_1$ | $c_2$ | Error address |
|---|---|---|---|---|---|---|---|---|---|
| ECC_1 | (decimal) | 3 | 5 | 6 | 7 | 1 | 2 | 4 | 6 |
| | (hexa) | 011 | 101 | 110 | 111 | 001 | 010 | 100 | 110 |
| ECC_2 | (decimal) | 6 | 3 | 7 | 5 | 1 | 2 | 4 | 6 |
| | (hexa) | 110 | 011 | 111 | 101 | 001 | 010 | 100 | 110 |

**Figure 40. A codestruct scenario with 4-bit error. ECC_1 has errors in $d_0$ and $d_1$, while ECC_2 has errors in $d_2$ and $c_0$, but the compound error address for both codes is 6 (Source: Author).**

Figure 40 illustrates a scenario with four errors—an extremely aggressive case in which more than half of the codestruct bits are corrupted. However, less aggressive scenarios can still result in the same address for both codewords, as seen in the three-error case shown in Figure 41.

| | | $d_0$ | $d_1$ | $d_2$ | $d_3$ | $c_0$ | $c_1$ | $c_2$ | Error address |
|---|---|---|---|---|---|---|---|---|---|
| ECC_1 | (decimal) | 3 | 5 | 6 | 7 | 1 | 2 | 4 | 1 |
| | (hexa) | 011 | 101 | 110 | 111 | 001 | 010 | 100 | 001 |
| ECC_2 | (decimal) | 6 | 3 | 7 | 5 | 1 | 2 | 4 | 1 |
| | (hexa) | 110 | 011 | 111 | 101 | 001 | 010 | 100 | 001 |

**Figure 41. A codestruct scenario with 3-bit error. ECC_1 has errors in $d_2$ and $d_2$, while ECC_2 has errors in $d_3$ and $c_2$, the compound error address for both codes is 1 (Source: Author).**

Finally, scenarios with double errors are also susceptible to the generation of non-unique addresses. In the proposed code overlay approach, such scenarios occur only with codes larger than $Ham(7,4)$. However, to help the reader visualize the issue, Figure 42 presents an encoding where the bit order in ECC_2 has been altered—meaning that the parity bits of ECC_1 and ECC_2 are not arranged in the same order.

(a)

| | | $d_0$ | $d_1$ | $d_2$ | $d_3$ | $c_0$ | $c_1$ | $c_2$ | Error address |
|---|---|---|---|---|---|---|---|---|---|
| ECC_1 | (decimal) | 3 | 5 | 6 | 7 | 1 | 2 | 4 | 6 |
| | (hexa) | 011 | 101 | 110 | 111 | 001 | 010 | 100 | 110 |
| ECC_2 | (decimal) | 6 | 3 | 7 | 5 | 1 | 2 | 4 | 5 |
| | (hexa) | 110 | 011 | 111 | 101 | 001 | 010 | 100 | 101 |

(b)

| | | $d_0$ | $d_1$ | $d_2$ | $d_3$ | $c_0$ | $c_1$ | $c_2$ | Error address |
|---|---|---|---|---|---|---|---|---|---|
| ECC_1 | (decimal) | 3 | 5 | 6 | 7 | 1 | 2 | 4 | 6 |
| | (hexa) | 011 | 101 | 110 | 111 | 001 | 010 | 100 | 110 |
| ECC_2 | (decimal) | 6 | 3 | 5 | 7 | 2 | 1 | 4 | 5 |
| | (hexa) | 110 | 011 | 111 | 111 | 010 | 010 | 100 | 101 |

**Figure 42. Two error scenarios, (a) and (b), produce the same combined addressing, preventing the decoder from correctly identifying the actual error (Source: Author).**

Therefore, a double error affecting the bits $d_0$ and $d_1$, as well as another affecting the bits $d_3$ and $c_0$, results in the same composite address (6 and 5) for both scenarios. This ambiguity in addressing prevents the decoder from correctly determining which double error should be corrected.

To prevent scenarios where the decoder cannot accurately determine the number of errors to be corrected, it is crucial to perform a preprocessing step that ensures the generation of unique composite addresses for the overlaid ECCs. In the specific cases of the codes analyzed in this work, this task was carried out for the 2×Ham_3×3 and 2×Ham_4×4 codes.

### 4.4.1 Addition of a Parity Bit in the Overlapping ECCs Explored

Figure 39 illustrates an encoding scheme in which errors in bits $d_2$ and $d_3$ result in the composite addresses 1 and 2, representing a double-bit error. However, a double-bit error in parity bits $c_0$ and $c_1$ of ECC_1 and ECC_2, respectively, would also generate the same addresses (1 and 2). This ambiguity prevents the decoder from correctly inferring the actual location of the double error.

To resolve this issue, a parity bit was added to each overlapping ECC, effectively replacing the standard Hamming codes with extended Hamming codes. In this approach, parity operates independently within each code, allowing for the differentiation of double-bit errors as described above.

Figure 43(a) and (b) illustrate how this differentiation is achieved using even parity to identify the number of errors present. In Figure 43(a), the parity bit remains 0, indicating that no odd number of errors has occurred. Conversely, in Figure 43(b), the parity bit is set to 1, signaling the presence of an odd number of errors, distinguishing between different error cases.

**(a)**

| | $d_0$ | $d_1$ | $d_2$ | $d_3$ | $c_0$ | $c_1$ | $c_2$ | $p$ | Error address |
|---|---|---|---|---|---|---|---|---|---|
| ECC_1* | 3 | 5 | 6 | 7 | 1 | 2 | 4 | 0 | 1 |
| ECC_2* | 6 | 3 | 7 | 5 | 1 | 2 | 4 | 0 | 2 |

**(b)**

| | $d_0$ | $d_1$ | $d_2$ | $d_3$ | $c_0$ | $c_1$ | $c_2$ | $p$ | Error address |
|---|---|---|---|---|---|---|---|---|---|
| ECC_1* | 3 | 5 | 6 | 7 | 1 | 2 | 4 | 1 | 1 |
| ECC_2* | 6 | 3 | 7 | 5 | 1 | 2 | 4 | 1 | 2 |

**Figure 43. Use of extended Hamming codes to improve error correction efficacy. (a) presents a double-bit error in data, identified by the parity bit remaining 0, while (b) presents a double-bit error in parity bits, detected by the parity bits being set to 1 (Source: Author).**

## 4.5 Generalization of the ECC Overlapping Technique

This section discusses the generalization of the ECC overlapping technique proposed in this work. To explore this concept, we examine generalizations regarding the number of overlapping ECCs, their heterogeneity, and the format of the protected data region.

### 4.5.1 Number of Overlapping ECCs

The proposed technique is highly scalable and can be extended to support multiple simultaneous overlaps within the same data area. The resulting structure consists of a single data region protected by multiple distinct sets of parity bits.

Figure 44(a – d) illustrates cases in which the same data area is protected by one to four ECCs. Overlapping occurs when two or more ECCs are applied to the same data region.

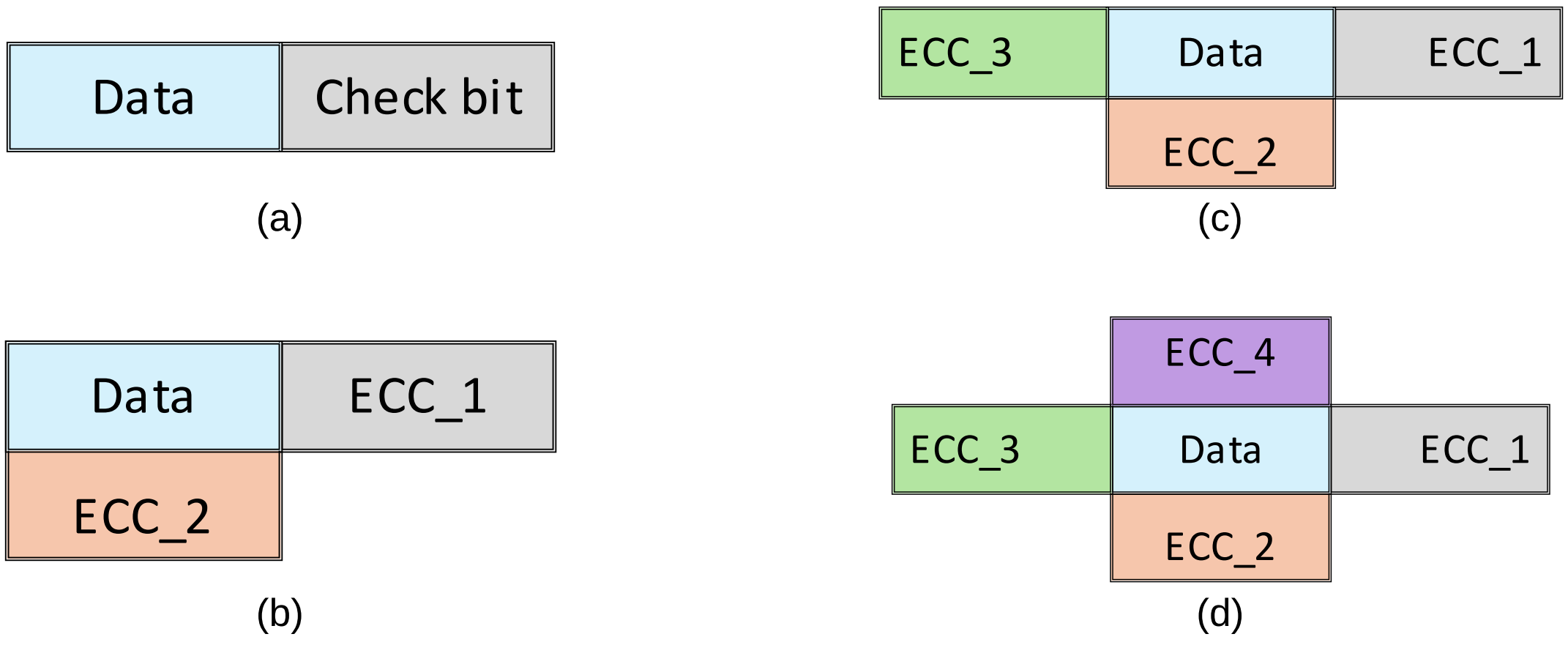

**Figure 44. Generalization of ECC overlapping based on the number of codewords. (a) shows a basic structure with a single data area and a single verification region. (b), (c), and (d) depict the same codeword from (a), but with the addition of two, three, and four overlapping codeword regions, respectively (Fonte: Autor).**

These overlapping codewords can be conceptualized as “layers” covering the data area, each with its own set of control bits based on different coding perspectives.

The application of multiple simultaneous codes not only leverages the scalability of the technique, but also demonstrates its adaptability to practical scenarios where redundancy and reliability are essential requirements. Consequently, multi-overlapping ECCs enhance error correction in critical data regions, offering new possibilities for reliable communication and storage systems.

### 4.5.2 Homogeneity of Overlapping ECCs

In the general case of overlapping ECCs, each code can have a different encoding nature, leading to a heterogeneous overlapping approach. For example, in Figure 44(b), ECC_1 could be an extended Hamming code, while ECC_2 could be a decimal-based error correction code, as illustrated in Section 3.2. This characteristic expands the range of possible combinations, allowing for flexible trade-offs between error correction effectiveness and code efficiency. However, it is important to note that this work focuses exclusively on homogeneous ECCs.

### 4.5.3 Data Area Format

A square matrix structure facilitates the scalability and optimization of the proposed technique, as it allows for efficient data and parity bit distribution, as discussed in Section 4.3. However, the technique does not require data to be organized as a matrix, and if matrices are used, they do not necessarily need to be square. For example, Figure 45 presents a rectangular matrix organization, which maintains the same data capacity as the 2×Ham_4×4 structure.

| | | | | | | | | | | | | | |
|---|---|---|---|---|---|---|---|---|---|---|---|---|---|
| D | D | D | D | D | D | D | D | Co0 | Co1 | Co2 | Co3 | Co4 | Po |
| D | D | D | D | D | D | D | D | Ci0 | Ci1 | Ci2 | Ci3 | Ci4 | Pi |

**Figure 45. Rectangular matrix organization for overlapping ECCs, providing the same data area capacity as the 2×Ham_2×4 structure (Fonte: Autor).**

# 5. DETAILS OF THE 2×HAM_3×3 CODE

This section provides a detailed explanation of the structure of the 2×Ham_3×3 code, including its encoding and decoding functions, as well as the blocks that implement the encoder and decoder. For conciseness, the 2×Ham_2×2 and 2×Ham_4×4 codes are not detailed in this document; however, they follow a similar construction logic. Java and VHDL implementations of these codes are available on GitHub at https://github.com/AndrewARF/OVERLAPPING-ERROR-CORRECTION-CODES-ON-TWO-DIMENSIONAL-STRUCTURES

## 5.1 Codestruct Organization

Figure 46 illustrates the basic elements of the 2×Ham_3×3 code, which consists of a 9-bit data area ($\mathrm{D0} \ldots \mathrm{D8}$) and two 5-bit parity check vectors ($\mathrm{Co0} \ldots \mathrm{Co3}$, $\mathrm{Po}$ and $\mathrm{Ci0} \ldots \mathrm{Ci3}$, $\mathrm{Pi}$), encoded using extended Hamming codes, forming a codestruct of 19 bits. These parity check vectors independently encode the same data area ($\mathrm{D0} \ldots \mathrm{D8}$), characterizing the overlapping of ECCs.

| | | | | | | | |
|---|---|---|---|---|---|---|---|
| D0 | D1 | D2 | Co0 | Co1 | Co2 | Co3 | Po |
| D3 | D4 | D5 | Ci0 | Ci1 | Ci2 | Ci3 | Pi |
| D6 | D7 | D8 | | | | | |

**Figure 46. Organization of the 2×Ham_3×3 code, where $\mathrm{D0} \ldots \mathrm{D8}$ represent the data area, and $\mathrm{Co0} \ldots \mathrm{Co3}$, $\mathrm{Po}$ and $\mathrm{Ci0} \ldots \mathrm{Ci3}$, $\mathrm{Pi}$ are the parity check bits (Source: Author).**

To facilitate the description of the overlapping ECCs, we define:

- **OuterHam** - the red-coded ECC, which contains the parity bits $\mathrm{Co0} \ldots \mathrm{Co3}$ and $\mathrm{Po}$;
- **InnerHam** - the blue-coded ECC, which contains the parity bits $\mathrm{Ci0} \ldots \mathrm{Ci3}$ and $\mathrm{Pi}$.

## 5.2 Data Encoding

Figure 47 illustrates a block diagram implementing the encoding process of the 2×Ham_3×3 code. The encoder receives the data area as input, represented in the figure as the “Data matrix”, and simultaneously generates the OuterHam and InnerHam using two distinct Hamming generators and two parity generators with the same encoding scheme. These Hamming generators differ because their encoding depends on the data addressing analysis described in Section 4.4. This analysis results in multiple unique combinations that enable the decoder to infer the necessary bit corrections for double-error scenarios.

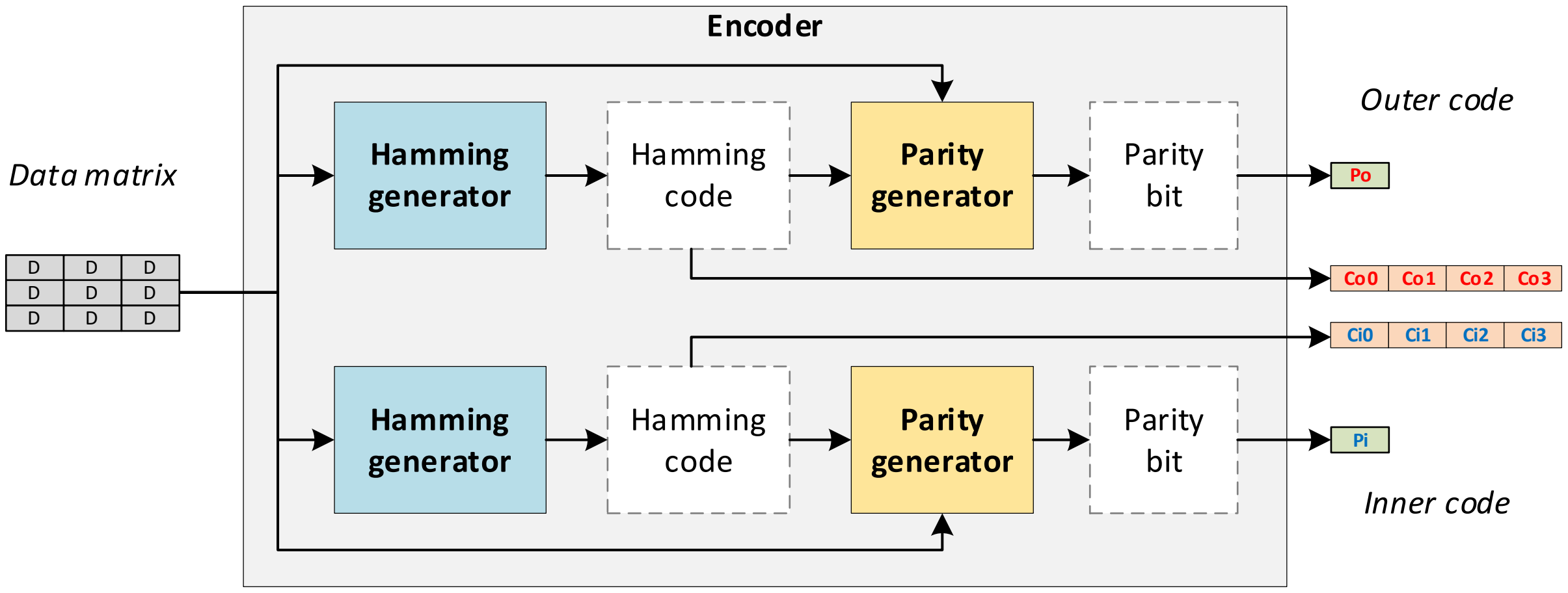

**Figure 47. Block diagram of the 2×Ham_3×3 Encoder (Fonte: Autor).**

As discussed in Section 4.4, we chose not to analyze three-error combinations in this work. However, we highlight that this remains a possibility for future research, as both the 2×Ham_3×3 and 2×Ham_4×4 codes are protected by an extended Hamming scheme that allows encoding a larger amount of data.

For the 2×Ham_3×3 case, four parity check bits are used, allowing for $2^4$ possible combinations, one of which corresponds to the error-free state (zero errors), leaving 15 valid error states. Since each independent ECC in the 2×Ham_3×3 structure has its own 4-bit verification, 11 combinations remain for addressing data errors. Given that the data area contains only 9 bits, two additional combinations remain unused. These extra combinations expand the number of unique solutions for two-error correction and create room for solutions that provide at least partial protection against three-error cases. While these solutions may not cover all possible three-error combinations, they enhance the overall error correction efficacy of the code.

The execution of the data addressing analysis algorithm resulted in the addresses shown in Figure 48(a) and (b) for the OuterHam and InnerHam codes, respectively.

| Dados | | | Co0 | Co1 | Co2 | Co3 | Po |
|---|---|---|---|---|---|---|---|
| 11 | 13 | 3 | 8 | 4 | 2 | 1 | - |
| 10 | 12 | 5 | | | | | |
| 14 | 6 | 15 | | | | | |

(a)

| Dados | | | Ci0 | Ci1 | Ci2 | Ci3 | Pi |
|---|---|---|---|---|---|---|---|
| 9 | 7 | 14 | | | | | |
| 13 | 10 | 12 | 8 | 4 | 2 | 1 | - |
| 5 | 3 | 15 | | | | | |

(b)

**Figure 48. Codestruct addressing for (a) OuterHam and (b) InnerHam (Fonte: Autor).**

Because of this addressing, Equations (30) to (33) and Equations (34) to (37) describe the encoding of the parity check bits for OuterHam and InnerHam, respectively.

$$\mathrm{Co0} = \mathrm{D0} \oplus \mathrm{D1} \oplus \mathrm{D3} \oplus \mathrm{D4} \oplus \mathrm{D6} \oplus \mathrm{D8} \tag{30}$$

$$\mathrm{Co1} = \mathrm{D1} \oplus \mathrm{D4} \oplus \mathrm{D6} \oplus \mathrm{D8} \tag{31}$$

$$\mathrm{Co2} = \mathrm{D2} \oplus \mathrm{D3} \oplus \mathrm{D6} \oplus \mathrm{D8} \tag{32}$$

$$\mathrm{Co3} = \mathrm{D0} \oplus \mathrm{D1} \oplus \mathrm{D2} \oplus \mathrm{D5} \oplus \mathrm{D8} \tag{33}$$

$$\mathrm{Ci0} = \mathrm{D0} \oplus \mathrm{D2} \oplus \mathrm{D3} \oplus \mathrm{D4} \oplus \mathrm{D5} \oplus \mathrm{D8} \tag{34}$$

$$\mathrm{Ci1} = \mathrm{D2} \oplus \mathrm{D3} \oplus \mathrm{D4} \oplus \mathrm{D5} \oplus \mathrm{D8} \tag{35}$$

$$\mathrm{Ci2} = \mathrm{D1} \oplus \mathrm{D2} \oplus \mathrm{D4} \oplus \mathrm{D7} \oplus \mathrm{D8} \tag{36}$$

$$\mathrm{Ci3} = \mathrm{D0} \oplus \mathrm{D1} \oplus \mathrm{D3} \oplus \mathrm{D6} \oplus \mathrm{D7} \oplus \mathrm{D8} \tag{37}$$

To clarify and illustrate the effect of Hamming codes on addressing, note that the parity check bits $\mathrm{Co3}$ and $\mathrm{Ci3}$ have a weight of 1, meaning they serve as parity bits for all odd-numbered addresses. Specifically:

- OuterHam – $\mathrm{Co3}$ verifies $\mathrm{D0}, \mathrm{D1}, \mathrm{D2}, \mathrm{D5}$ and $\mathrm{D8}$;
- InnerHam – $\mathrm{Ci3}$ verifies $\mathrm{D0}, \mathrm{D1}, \mathrm{D3}, \mathrm{D6}, \mathrm{D7}$ and $\mathrm{D8}$.

Another example is address 15, which is obtained through an exclusive OR ($\oplus$) operation on all four parity check bits. For both codes, address 15 corresponds to the logical position $\mathrm{D8}$, meaning $\mathrm{D8}$ is involved in the composition of all parity check bits for both OuterHam and InnerHam.

Finally, both overlapping ECCs use extended Hamming encoding, which requires the inclusion of an additional parity bit, as illustrated in Figure 49(a) and (b). This bit, which is not addressed by the Hamming code, functions as a wrapper encapsulating the Hamming-protected codewords. Consequently, the parity bits $\mathrm{Po}$ and $\mathrm{Pi}$, computed in Equations (39) and (40), respectively, provide protection for all data and check bits.

| | | | | | | | |
|---|---|---|---|---|---|---|---|
| D0 | D1 | D2 | Co0 | Co1 | Co2 | Co3 | Po |
| D3 | D4 | D5 | | | | | |
| D6 | D7 | D8 | | | | | |

(a)

| | | | | | | | |
|---|---|---|---|---|---|---|---|
| D0 | D1 | D2 | | | | | |
| D3 | D4 | D5 | Ci0 | Ci1 | Ci2 | Ci3 | Pi |
| D6 | D7 | D8 | | | | | |

(b)

**Figure 49. Inclusion of a parity bit to protect data and parity check bits in the codes (a) OuterHam and (b) InnerHam (Source: Author).**

$$Po = D0 \oplus D1 \oplus D2 \oplus D3 \oplus D4 \oplus D5 \oplus D6 \oplus D7 \oplus D8 \oplus Co0 \oplus Co1 \oplus Co2 \oplus Co3 \quad (38)$$

$$Pi = D0 \oplus D1 \oplus D2 \oplus D3 \oplus D4 \oplus D5 \oplus D6 \oplus D7 \oplus D8 \oplus Ci0 \oplus Ci1 \oplus Ci2 \oplus Ci3 \quad (39)$$

## 5.3 Decoding of the 2×Ham_3×3 Codestruct

Figure 50 illustrates the block organization for decoding the 2×Ham_3×3 codestruct. The decoder outputs the potentially corrected data area along with signals indicating whether errors were detected and/or corrected.

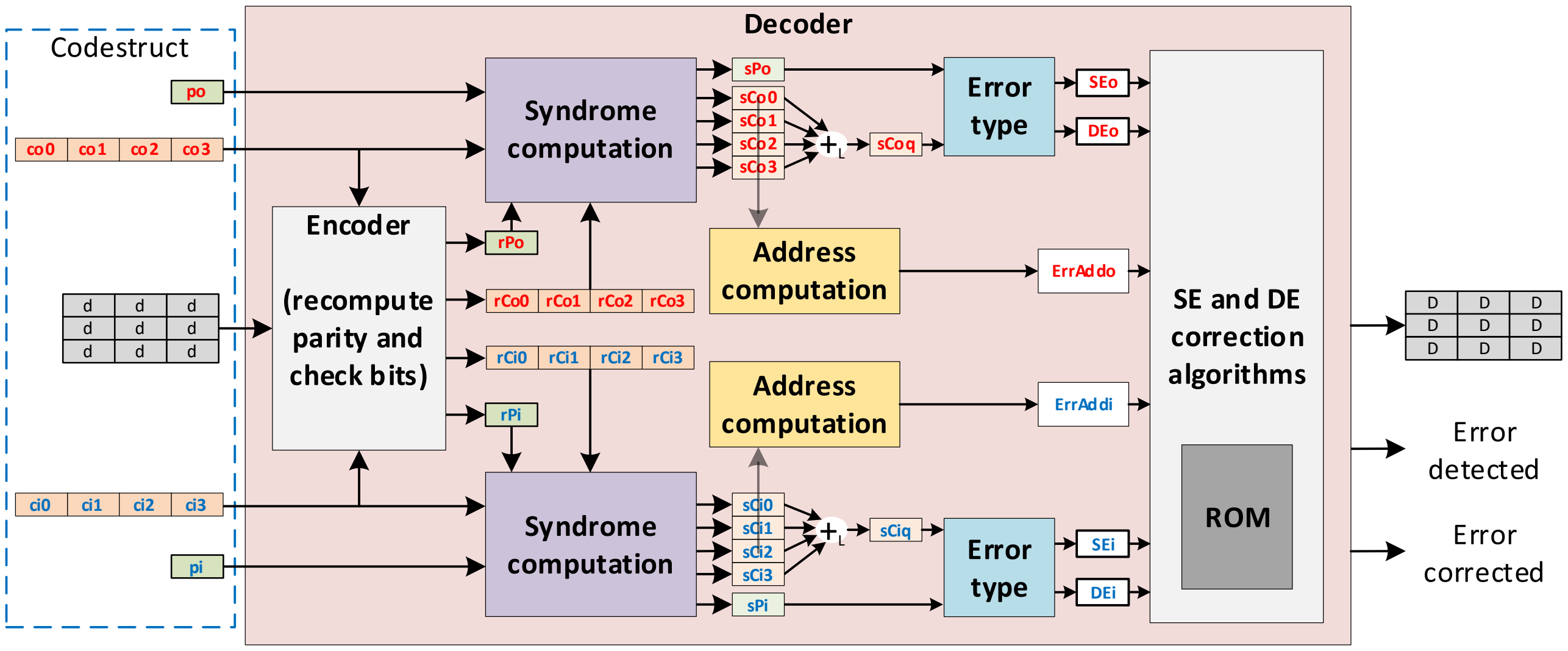

**Figure 50. Block diagram of the 2×Ham_3×3 Decoder (Fonte: Autor).**

The decoding includes an encoding circuit that recalculates the parity bits ($rPo$ and $rPo$) and check bits ($rCo0 \ldots rCo3$ and $rCi0 \ldots rCi3$) based on the data area stored in the codestruct for both OuterHam and InnerHam. Thus, Equations (40) to (49) correspond to (30) to (39) from the encoding. To distinguish decoding from encoding, the codestruct bits are represented using lowercase letters in the decoding equations. These bits are the same as those used in encoding but may have been affected by storage or transmission errors.

$$rCo0 = d0 \oplus d1 \oplus d3 \oplus d4 \oplus d6 \oplus d8 \quad (40)$$

$$rCo1 = d1 \oplus d4 \oplus d6 \oplus d8 \quad (41)$$

$$rCo2 = d2 \oplus d3 \oplus d6 \oplus d8 \quad (42)$$

$$\mathrm{rCo3} = \mathrm{d0} \oplus \mathrm{d1} \oplus \mathrm{d2} \oplus \mathrm{d5} \oplus \mathrm{d8} \tag{43}$$

$$\mathrm{rCi0} = \mathrm{d0} \oplus \mathrm{d2} \oplus \mathrm{d3} \oplus \mathrm{d4} \oplus \mathrm{d5} \oplus \mathrm{d8} \tag{44}$$

$$\mathrm{rCi1} = \mathrm{d2} \oplus \mathrm{d3} \oplus \mathrm{d4} \oplus \mathrm{d5} \oplus \mathrm{d8} \tag{45}$$

$$\mathrm{rCi2} = \mathrm{d1} \oplus \mathrm{d2} \oplus \mathrm{d4} \oplus \mathrm{d7} \oplus \mathrm{d8} \tag{46}$$

$$\mathrm{rCi3} = \mathrm{d0} \oplus \mathrm{d1} \oplus \mathrm{d3} \oplus \mathrm{d6} \oplus \mathrm{d7} \oplus \mathrm{d8} \tag{47}$$

$$\mathrm{rPo} = \mathrm{d0} \oplus \mathrm{d1} \oplus \mathrm{d2} \oplus \mathrm{d3} \oplus \mathrm{d4} \oplus \mathrm{d5} \oplus \mathrm{d6} \oplus \mathrm{d7} \oplus \mathrm{d8} \oplus \mathrm{co0} \oplus \mathrm{co1} \oplus \mathrm{co2} \oplus \mathrm{co3} \tag{48}$$

$$\mathrm{rPi} = \mathrm{d0} \oplus \mathrm{d1} \oplus \mathrm{d2} \oplus \mathrm{d3} \oplus \mathrm{d4} \oplus \mathrm{d5} \oplus \mathrm{d6} \oplus \mathrm{d7} \oplus \mathrm{d8} \oplus \mathrm{ci0} \oplus \mathrm{ci1} \oplus \mathrm{ci2} \oplus \mathrm{ci3} \tag{49}$$

The recalculated parity check and parity bits are compared with their counterparts in the codeword, producing syndrome bits, as defined in (50) to (59). Notably, the syndromes are computed using a single XOR operation ($\oplus$), as a syndrome bit takes the logical value 1 (TRUE) whenever the recalculated bit differs from its corresponding bit in the codestruct.

$$\mathrm{sCo0} = \mathrm{rCo0} \oplus \mathrm{co0} \tag{50}$$

$$\mathrm{sCo1} = \mathrm{rCo1} \oplus \mathrm{co1} \tag{51}$$

$$\mathrm{sCo2} = \mathrm{rCo2} \oplus \mathrm{co2} \tag{52}$$

$$\mathrm{sCo3} = \mathrm{rCo3} \oplus \mathrm{co3} \tag{53}$$

$$\mathrm{sCi0} = \mathrm{rCi0} \oplus \mathrm{ci0} \tag{54}$$

$$\mathrm{sCi1} = \mathrm{rCi1} \oplus \mathrm{ci1} \tag{55}$$

$$\mathrm{sCi2} = \mathrm{rCi2} \oplus \mathrm{ci2} \tag{56}$$

$$\mathrm{sCi3} = \mathrm{rCi3} \oplus \mathrm{ci3} \tag{57}$$

$$\mathrm{sPo} = \mathrm{rPo} \oplus \mathrm{po} \tag{58}$$

$$sPi = rPi \oplus pi \tag{59}$$

The information in Table 3 enables the identification of errors and the classification of whether they are single errors (SE), double errors (DE), or a single parity bit error. Analyzing Table 3 requires combining the Hamming syndrome bits from both OuterHam and InnerHam. This unification is performed using a logical OR ($\vee$) operation on all syndrome bits, generating the $sCoq$ and $sCiq$ bits, as described in (60) and (61).

$$sCoq = sCo0 \vee sCo1 \vee sCo2 \vee sCo3 \tag{60}$$

$$sCiq = sCi0 \vee sCi1 \vee sCi2 \vee sCi3 \tag{61}$$

The signals $DEo$, $SEo$, $DEi$ and $SEi$ are obtained using (64), (65), (68) and (69). It is important to note that the decoding method used in this work does not rely on the parity error information described in Equations (63) and (67). This is because the information from the other fields is already enough to achieve the full double-error correction capability. However, it is worth mentioning that parity error information can be useful in mitigating more aggressive scenarios, such as triple errors. Nevertheless, this work focuses exclusively on single and double errors.

$$\text{No error in outer Hamming} = \overline{sCoq} \wedge \overline{sPo} \tag{62}$$

$$\text{Parity error in outer Hamming} = \overline{sCoq} \wedge sPo \tag{63}$$

$$DEo = sCoq \wedge \overline{sPo} \tag{64}$$

$$SEo = sCoq \wedge sPo \tag{65}$$

$$\text{No error in inner Hamming} = \overline{sCiq} \wedge \overline{sPi} \tag{66}$$

$$\text{Parity error in inner Hamming} = \overline{sCiq} \wedge sPi \tag{67}$$

$$DEi = sCiq \wedge \overline{sPi} \tag{68}$$

$$SEi = sCiq \wedge sPi \tag{69}$$

The error address calculations for OuterHam ($\mathrm{ErrAddo}$) and InnerHam ($\mathrm{ErrAddi}$) are obtained by summing the values of each Hamming syndrome, as described in the arithmetic equations (70) and (71).

$$\mathrm{ErrAddo} = \mathrm{sCo0} \times 8 + \mathrm{sCo1} \times 4 + \mathrm{sCo2} \times 2 + \mathrm{sCo3} \tag{70}$$

$$\mathrm{ErrAddi} = \mathrm{sCi0} \times 8 + \mathrm{sCi1} \times 4 + \mathrm{sCi2} \times 2 + \mathrm{sCi3} \tag{71}$$

Each overlapped ECC obtains the error detection information ($\mathrm{ErrDet}$) by applying a logical OR operation to all syndrome bits, as described in (72). Meanwhile, the error correction information is determined according to the single and double error correction algorithms.

$$\mathrm{ErrDet} = \mathrm{sCoq} \vee \mathrm{sCiq} \vee \mathrm{sPo} \vee \mathrm{sPi} \tag{72}$$

### 5.3.1 Decoding Algorithm

Figure 51 presents a partial pseudocode of the 2×Ham_3×3 decoding algorithm. The algorithm stores whether an error has been detected in the Boolean variable `errorDetected` and then proceeds with the following verification steps:

1. Single-error detection in OuterHam;
2. Single-error detection in InnerHam;
3. Double-error detection, using a procedure that combines both codes.

It is important to note that these procedures are executed only if an error address is different from 0.

```
integer OuterTab[] ← {-1, 12, 11, 2, 10, 5, 7, -1, 9, -1, 3, 0, 4, 2, 6, 8}
integer InneTab[] ← {-1, 12, 11, 7, 10, 6, -1, 1, 9, 0, 4, -1, 5, 3, 2, 8}

boolean decoding() {
      errorDetected ← sPo ∨ sPi ∨ sCoq ∨ SCiq

      if(ErrAddo = 0 ∨ ErrAddi = 0)
            return errorDetected
      if(SEo = TRUE) {
            flipBit(outerTab[ErrAddo])
            return errorDetected
      }
      if(SEi = TRUE) {
            flipBit(innerTab[ErrAddi])
            return errorDetected
      }
```

```
      if(DEo = TRUE ∧ DEi = TRUE) {
            if(ErrAddo = 1) {
                  if(ErrAddi = 4)
                        flipBits(11, 10)
                  else if(ErrAddi = 10)
                        flipBits(14, 15)
                  else if(ErrAddi = 13)
                        flipBits(13, 12)
            }
            else if(ErrAddo = 2) {
                  if(ErrAddi = 8)
                        flipBits(13, 15)
                  else if(ErrAddi = 15)
                        flipBits(12, 14)
            }
                                          ...

            else if(ErrAddo = 15) {
                  if(ErrAddi = 1)
                        flipBits(10, 5)
                  if(ErrAddi = 4)
                        flipBits(3, 12)
            }
      }
      return errorDetected
}
```

**Figure 51. Partial pseudocode of the 2×Ham_3×3 decoding algorithm (Fonte: Autor).**

Single-error correction is performed using the `flipBit` function, illustrated in Figure 52. This function receives the physical address of the erroneous bit and uses `OuterTab` and `InnerTab` to map it to the appropriate location. For double-error correction, the process involves a combination of the error addresses (ErrAddo and ErrAddi) to determine the logical addresses of the two bits that will be passed to the `flipBits` function.

```
flipBit(integer pos) {
      if(pos ≠ -1)
            D[pos] ← D[pos] = 0 ? 1 : 0
}


flipBits(integer addA, integer addB) {
      integer posA ← outerTab[addA]
      if(posA ≠ -1)
            D[posA] ← D[posA] = 0 ? 1 : 0

      integer posB ← outerTab[addB]
      if(posB ≠ -1)
            D[posB] ← D[posB] = 0 ? 1 : 0

}
```

**Figure 52. Pseudocode of the `flipBit` and `flipBits` functions used in the 2×Ham_3×3 decoding algorithm (Fonte: Autor).**

Both `OuterTab` and `InnerTab` represent the mapping between logical and physical addresses of the codeword bits. The algorithm is designed to correct only the data area, meaning that logical addresses 1, 2, 4, and 8 are marked as -1, as they correspond to the physical positions of parity check bits. Additionally:

- *Logical address 0* is reserved to indicate no error detected;
- Logical addresses 7 and 9 in `OuterTab` and 6 and 11 in `InnerTab` are unused. These addresses remain free because the four parity bits in this code can protect up to 11 data bits, while the data area contains only 9 bits (see Section 4.4).

Finally, parity check bits do not need correction, since they can be recomputed from an error-free data area.

The `flipBits` function internally implements `OuterTab`, as the decoding function passes addresses from this vector. However, it is important to note that the function could also be implemented using `InnerTab`, requiring only an address remapping. For example, in the case where `ErrAddo = 1` and `ErrAddi = 4`, the `OuterTab` addresses 11 and 10 are modified. These correspond to `InnerTab` addresses 9 and 13, respectively, as seen in Figure 48(a) and (b).

# 6. EXPERIMENTAL RESULTS

This chapter presents the experiments designed to explore the potential and validate the overlapping ECC technique. We worked with the overlapped ECCs 2×Ham_2×2, 2×Ham_3×3, and 2×Ham_4×4, and used data available in the literature to compare them with other ECCs designed for similar purposes. Specifically, the chapter is divided into six sections: (6.1) exploration of the error correction capability of the proposed codes for various scenarios, as well as a comparison of one of these ECCs with other state-of-the-art ECCs; (6.2 and 6.3) exploration of the error detection capability and reliability of the proposed codes. Additionally, we provide a mathematical description of how to evaluate ECC reliability over time; (6.4) synthesis of the encoders and decoders of the proposed ECCs to obtain data on area consumption, power dissipation, and latency, enabling an understanding of their physical characteristics in a 28nm CMOS technology; and finally, (6.5) scalability analysis of the proposed technique for the specific implementation case chosen (extended Hamming) and comparison with the theoretical scalability of other state-of-the-art ECCs.

## 6.1 Error Correction

Figure 53 illustrates the research methodology adopted in the set of experiments concerning the analysis of the error correction capability of the three explored codes to validate the overlapping ECC technique: 2×Ham_2×2, 2×Ham_3×3, and 2×Ham_4×4.

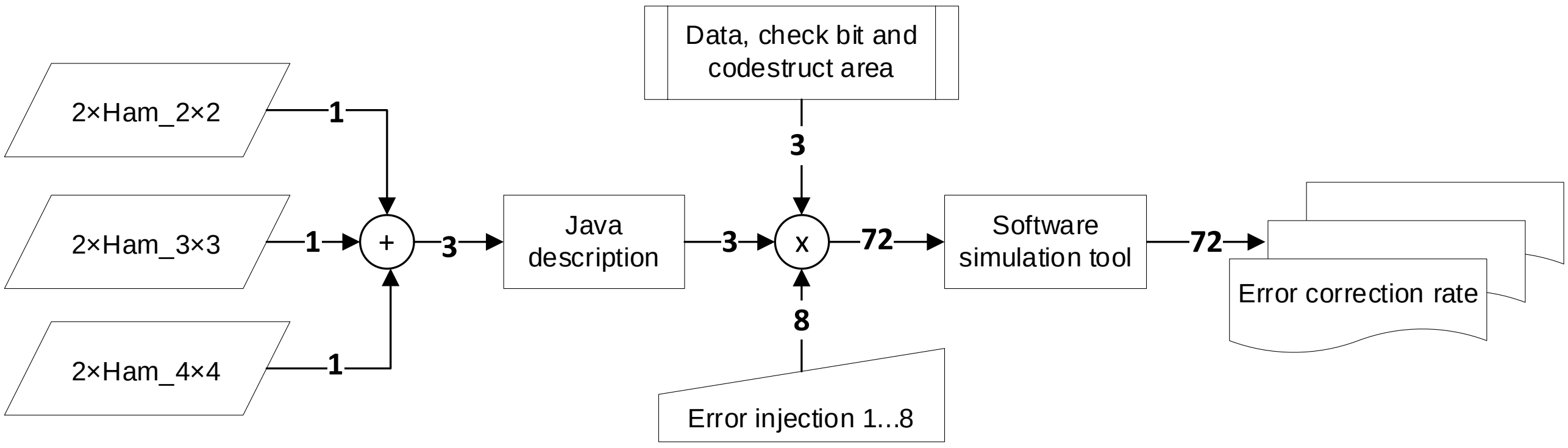

**Figure 53. Methodology for collecting the error correction rates of the overlapped ECCs. The figure shows that 72 simulations were generated, allowing data collection for three regions of each ECC, considering an exhaustive injection of patterns with 1 to 8 simultaneous errors (Source: Author).**

The analysis of the error correction rate employed an exhaustive error injection model ranging from 1 to 8 errors for each region of interest in the codestruct. These regions include isolated data bits, isolated check bits, and all bits within the codestruct. The number of evaluation combinations increases significantly with the number of bits in the evaluated region and the number of errors, as illustrated in Table 5.

**Table 5. Number of combinations analyzed according to the ECC region and number of simultaneous errors (Source: Author).**

| Number of errors | Error correction rate on region (%) | | | | | | | | |
|---|---|---|---|---|---|---|---|---|---|
| | Data | | | Check bits | | | Codestruct | | |
| | 2×2 | 3×3 | 4×4 | 2×2 | 3×3 | 4×4 | 2×2 | 3×3 | 4×4 |
| 1 | 4 | 9 | 16 | 8 | 10 | 12 | 12 | 19 | 28 |
| 2 | 6 | 36 | 120 | 28 | 45 | 66 | 66 | 171 | 378 |
| 3 | 4 | 84 | 560 | 56 | 120 | 220 | 220 | 969 | 3276 |
| 4 | 1 | 126 | 1820 | 70 | 210 | 495 | 495 | 3876 | 20475 |
| 5 | 0 | 126 | 4368 | 56 | 252 | 792 | 792 | 11628 | 98280 |
| 6 | 0 | 84 | 8008 | 28 | 210 | 924 | 924 | 27132 | 376740 |
| 7 | 0 | 36 | 11440 | 8 | 120 | 792 | 792 | 50388 | 1184040 |
| 8 | 0 | 9 | 12870 | 1 | 45 | 495 | 495 | 75582 | 3108105 |

*Legend: **2×2, 3×3** and **4×4** are abbreviations of **2×Ham_2×2, 2×Ham_3×3** and **2×Ham_4×4**, respectively*

We described the encoders and decoders of the overlapped ECCs in Java and implemented a parameterizable simulation environment, allowing the injection of a variable number of errors into controlled regions of each ECC. Table 6 presents the percentage of corrected errors for all evaluated ECCs (2×Ham_2×2, 2×Ham_3×3, and 2×Ham_4×4) across three regions (data, check bits, and codestruct), considering an exhaustive injection of 1 to 8 accumulated errors in the region of interest.

**Table 6. Comparison of error correction rates for 2×Ham_2×2, 2×Ham_3×3, and 2×Ham_4×4, with an exhaustive injection of 1 to 8 errors in data regions, check bits, and codestruct (Source: Author).**

| Number of errors | Error correction rate on region (%) | | | | | | | | |
|---|---|---|---|---|---|---|---|---|---|
| | Data | | | Check bits | | | Codestruct | | |
| | 2×2 | 3×3 | 4×4 | 2×2 | 3×3 | 4×4 | 2×2 | 3×3 | 4×4 |
| 1 | 100.00 | 100.00 | 100.00 | 100.00 | 100.00 | 100.00 | 100.00 | 100.00 | 100.00 |
| 2 | 100.00 | 100.00 | 100.00 | 100.00 | 100.00 | 100.00 | 100.00 | 100.00 | 100.00 |
| 3 | 0.00 | 0.00 | 0.00 | 100.00 | 100.00 | 100.00 | 40.45 | 24.87 | 19.57 |
| 4 | 0.00 | 0.00 | 0.00 | 91.43 | 90.00 | 90.30 | 17.78 | 9.11 | 5.09 |
| 5 | - | 0.00 | 0.00 | 71.43 | 69.84 | 73.11 | 8.84 | 3.56 | 1.99 |
| 6 | - | 0.00 | 0.00 | 53.57 | 56.67 | 64.94 | 3.57 | 1.04 | 0.87 |
| 7 | - | 0.00 | 0.00 | 62.50 | 61.67 | 67.30 | 1.01 | 0.28 | 0.19 |
| 8 | - | 0.00 | 0.00 | 100.00 | 75.56 | 69.49 | 0.20 | 0.11 | 0.03 |

*Legend: **2×2, 3×3** and **4×4** are abbreviations of **2×Ham_2×2, 2×Ham_3×3** and **2×Ham_4×4**, respectively*
*"-" uncollected value*

The results show that the proposed algorithm achieves 100% correction for up to two errors, regardless of the affected region. As described in Chapters 4 and 5, the explored algorithmic proposal does not handle corrections beyond two errors in the data area, justifying the results with zero correction starting from three errors. The "-" representation in the column of the 2×Ham_2×2 code appears because the data area is limited to 4 bits, making it impossible to generate 5 or more errors. Additionally, the explored algorithms can correct up to three errors when these errors are concentrated only in the check bits region. Although this scenario may seem of lesser interest—since the main goal is to protect data—for larger codes where scalability significantly reduces the check bits region, designers may choose to implement this region using more expensive and less radiation-sensitive memory (i.e., radiation-hardened memory [43]), thus achieving greater reliability.

As discussed in the decoding algorithms detailed in the example in Chapter 5, it is possible to increase the correction rate for the data area, but at the cost of larger hardware, higher power dissipation and latency. However, this evaluation is part of future studies.

### 6.1.1 Error Correction Rate Comparison

To fairly evaluate the error correction capability of the overlapped technique against other state-of-the-art ECCs, we used the criterion of maintaining the same number of data bits while exploring different codestructs. Figure 54 illustrates the methodology employed to conduct this experiment.

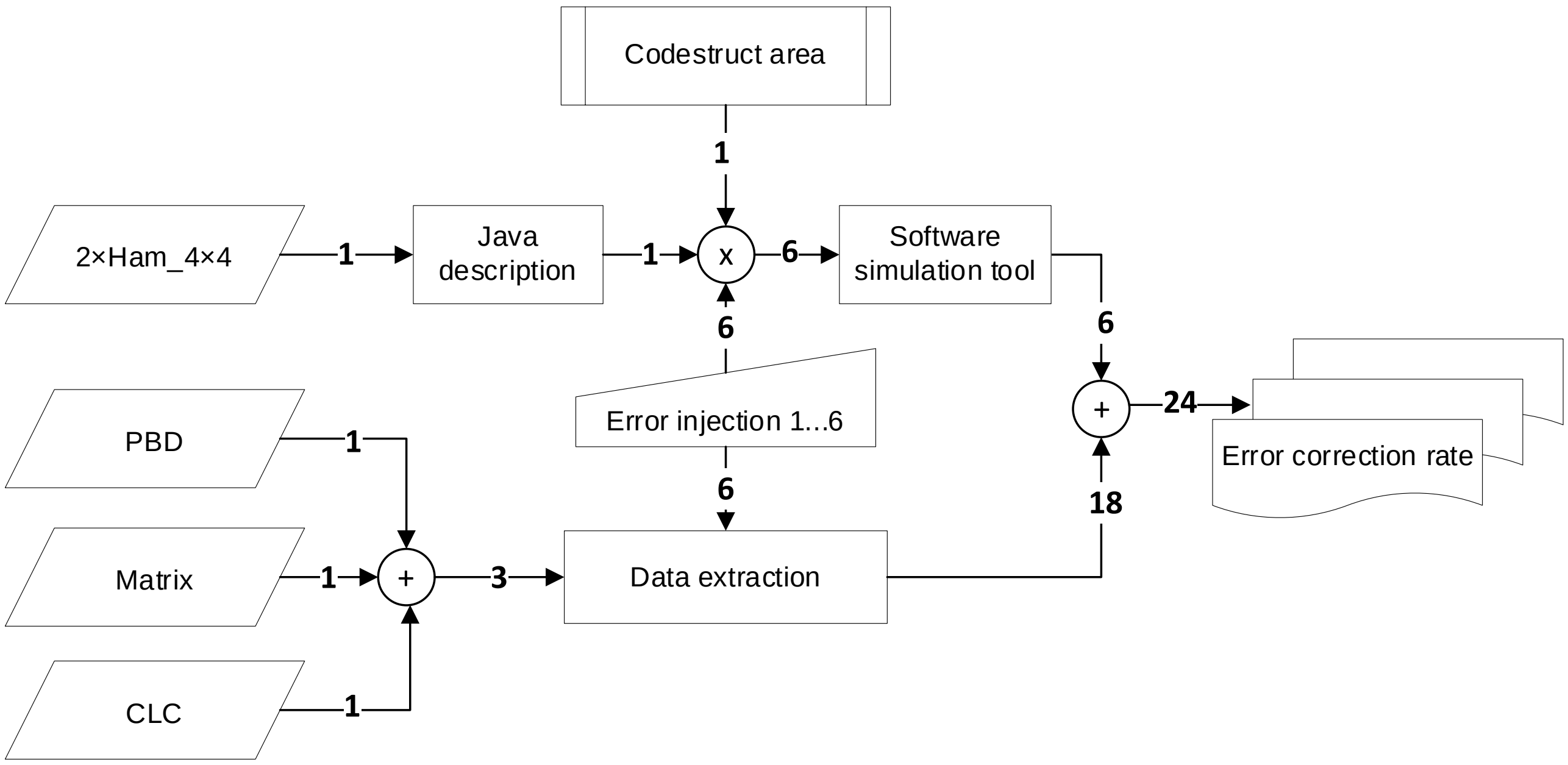

**Figure 54. Methodology used to compare the proposed ECC technology with three state-of-the-art ECCs (Source: Author).**

Initially, we selected recent ECCs from the literature that had similar objectives (i.e., protecting critical system data) and provided experimental results under the same error scenarios (i.e., exhaustive injection). The research identified three ECCs: Matrix [1], CLC [57] and PBD [23]—all 2D-ECCs protecting 16 data bits—which allowed us to compare them with the 2×Ham_4×4 code. Figure 55 illustrates the codestructs of these four ECCs, enabling the reader to better understand the employed approach.

| | | | | | | |
|---|---|---|---|---|---|---|
| $D_{0,0}$ | $D_{0,1}$ | $D_{0,2}$ | $D_{0,3}$ | $C_0$ | $C_1$ | $C_2$ |
| $D_{1,0}$ | $D_{1,1}$ | $D_{1,2}$ | $D_{1,3}$ | $C_3$ | $C_4$ | $C_5$ |
| $D_{2,0}$ | $D_{2,1}$ | $D_{2,2}$ | $D_{2,3}$ | $C_6$ | $C_7$ | $C_8$ |
| $D_{3,0}$ | $D_{3,1}$ | $D_{3,2}$ | $D_{3,3}$ | $C_9$ | $C_{10}$ | $C_{11}$ |
| $P_0$ | $P_1$ | $P_2$ | $P_3$ | | | |

**(a)**

| | | | | | | | | |
|---|---|---|---|---|---|---|---|---|
| $D_{0,0}$ | $D_{0,1}$ | $D_{0,2}$ | $D_{0,3}$ | $R_{0,0}$ | $R_{0,1}$ | $R_{0,2}$ | $R_{0,3}$ | $Pr_0$ |
| $D_{1,0}$ | $D_{1,1}$ | $D_{1,2}$ | $D_{1,3}$ | $R_{1,0}$ | $R_{1,1}$ | $R_{1,2}$ | $R_{1,3}$ | $Pr_1$ |
| $D_{2,0}$ | $D_{2,1}$ | $D_{2,2}$ | $D_{2,3}$ | $R_{2,0}$ | $R_{2,1}$ | $R_{2,2}$ | $R_{2,3}$ | $Pr_2$ |
| $D_{3,0}$ | $D_{3,1}$ | $D_{3,2}$ | $D_{3,3}$ | $R_{3,0}$ | $R_{3,1}$ | $R_{3,2}$ | $R_{3,3}$ | $Pr_3$ |

**(b)**

| | | | | | | | |
|---|---|---|---|---|---|---|---|
| $D_{0,0}$ | $D_{0,1}$ | $D_{0,2}$ | $D_{0,3}$ | $C_0$ | $C_1$ | $C_2$ | $Pr_0$ |
| $D_{1,0}$ | $D_{1,1}$ | $D_{1,2}$ | $D_{1,3}$ | $C_3$ | $C_4$ | $C_5$ | $Pr_1$ |
| $D_{2,0}$ | $D_{2,1}$ | $D_{2,2}$ | $D_{2,3}$ | $C_6$ | $C_7$ | $C_8$ | $Pr_2$ |
| $D_{3,0}$ | $D_{3,1}$ | $D_{3,2}$ | $D_{3,3}$ | $C_9$ | $C_{10}$ | $C_{11}$ | $Pr_3$ |
| $Pc_0$ | $Pc_1$ | $Pc_2$ | $Pc_3$ | $Pc_4$ | $Pc_5$ | $Pc_6$ | $Pc_7$ |

**(c)**

| | | | | | | | | | |
|---|---|---|---|---|---|---|---|---|---|
| $D_{0,0}$ | $D_{0,1}$ | $D_{0,2}$ | $D_{0,3}$ | $Ce_0$ | $Ce_1$ | $Ce_2$ | $Ce_3$ | $Ce_4$ | Pe |
| $D_{1,0}$ | $D_{1,1}$ | $D_{1,2}$ | $D_{1,3}$ | $Ci_0$ | $Ci_1$ | $Ci_2$ | $Ci_3$ | $Ci_4$ | Pi |
| $D_{2,0}$ | $D_{2,1}$ | $D_{2,2}$ | $D_{2,3}$ | | | | | | |
| $D_{3,0}$ | $D_{3,1}$ | $D_{3,2}$ | $D_{3,3}$ | | | | | | |

**(d)**

**Figure 55. Codestructs of (a) Matrix, (b) PBD, (c) CLC, and (d) 2×Ham_4×4 (Source: Author).**

As a common verification scenario for error correction capability, all ECCs presented results from an exhaustive exploration of 1 to 6 errors across the entire codestruct. Therefore, we selected the corresponding data from the 2×Ham_4×4 code, which can be visualized in the last column of Table 6.

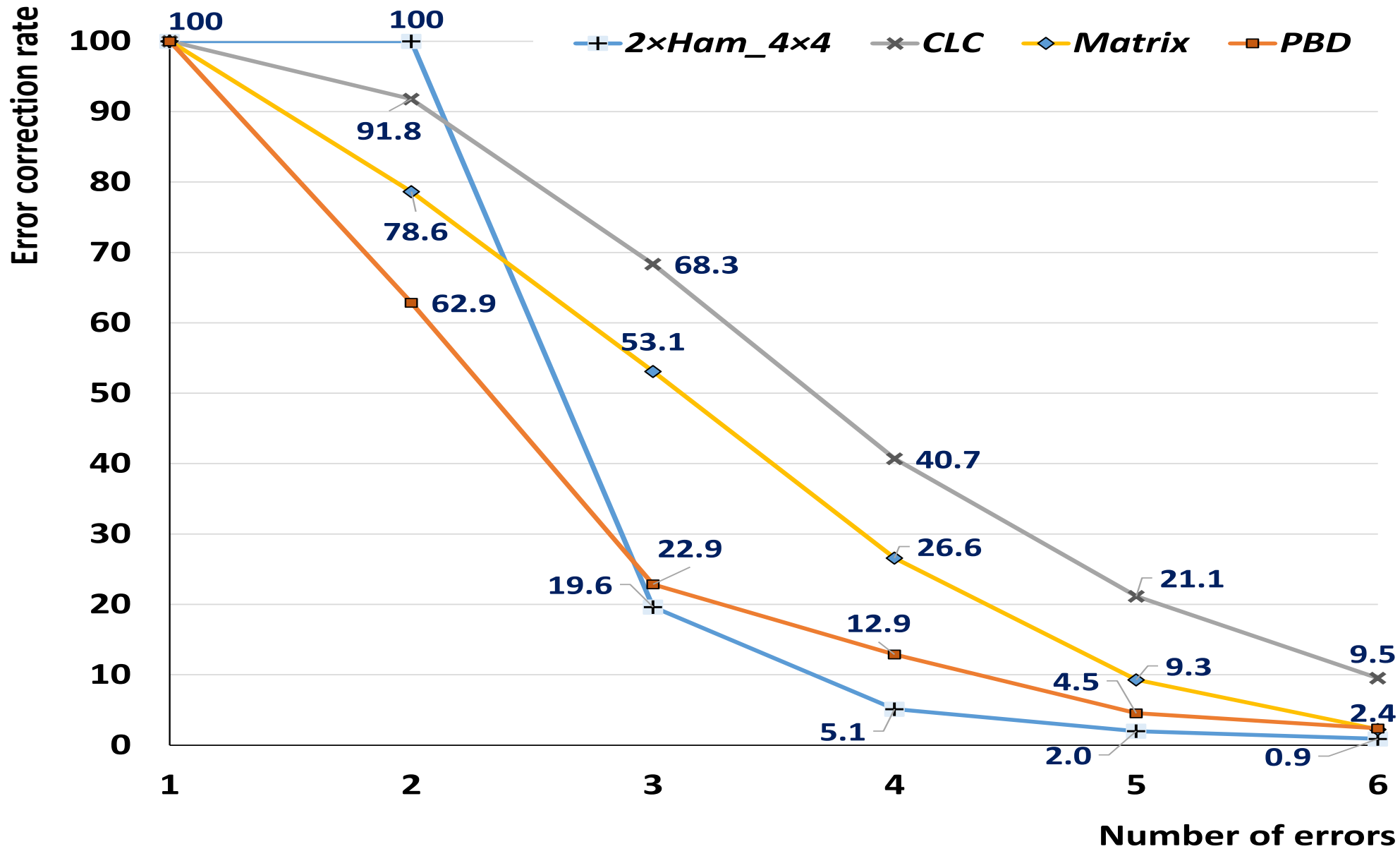

**Figure 56. Error correction rates of the evaluated 2D-ECCs, considering 1 to 6 errors exhaustively distributed across the codestructs (Source: Author).**

Figure 56 illustrates the error correction rates of the four ECCs. The results show that 2×Ham_4×4 meets the goal of recovering from up to two errors anywhere in the codestruct, whereas the other ECCs cannot achieve 100% correction for two errors in any codestruct position. On the other hand, the approach adopted by 2×Ham_4×4 penalizes correction for three or more errors, significantly reducing its correction rate. It is important to note that different ECCs employ distinct encoding and decoding structures and algorithms, resulting in varying costs in terms of area consumption, power dissipation, and latency, which are not evaluated here. However, the structures shown in Figure 55 reveal that 2×Ham_4×4 has the lowest redundancy cost, requiring only 12 bits, while Matrix, PBD, and CLC require 16, 20, and 24 redundancy bits, respectively. This characteristic is further examined in the next sections.

## 6.2 Error Detection

Figure 57 illustrates the set of experiments conducted to analyze the error detection capability of the 2×Ham_2×2, 2×Ham_3×3, and 2×Ham_4×4 codes. Although presented separately, Figure 57 is almost identical to Figure 53, except that the results analyzed pertain to detection rather than correction. In fact, it is the same set of simulations but with new data.

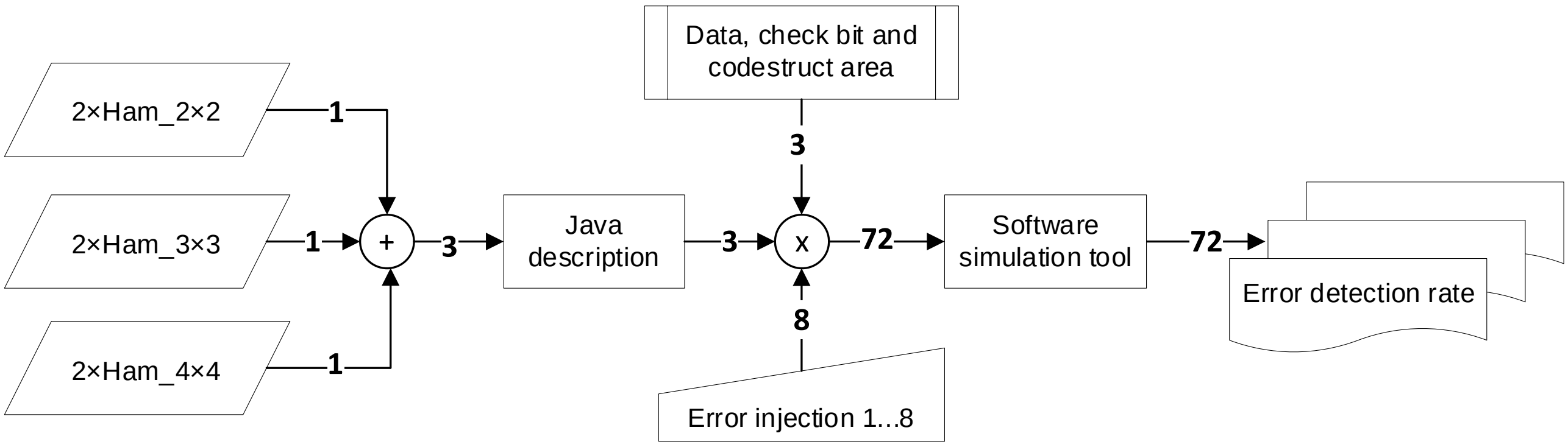

**Figure 57. Methodology employed to collect the error detection rates of the overlapped ECCs. This figure, which is similar to Figure 53, shows that 72 simulations, allowing data collection for 3 regions of each ECC, with exhaustive injection of patterns with 1 to 8 simultaneous errors (Source: Author).**

The results of the experiment set are illustrated in Table 7. The main insight provided by Table 7 is the ability of the 2×Ham_2×2, 2×Ham_3×3, and 2×Ham_4×4 codes to detect at least error patterns with up to four simultaneous errors in any region. Combining this characteristic with correction capability, we can define these codes as belonging to the DECQED (Double Error Correction, Quadruple Error Detection) class, which is uncommon but highlights the potential of the proposed approach for handling critical systems. While error correction may often be more desirable to designers as it allows the system to operate with the received data, error detection is crucial for ensuring the reliability of read information. Additionally, if an error situation is detected, the system can request the information retransmission or re-reading.

**Table 7. Comparison of error detection rates for ECCs 2×Ham_2×2, 2×Ham_3×3, and 2×Ham_4×4, with exhaustive injection of 1 to 8 errors in data, check bits, and codestruct regions (Source: Author).**

| Number of errors | Error detection rate on region (%) | | | | | | | | |
|---|---|---|---|---|---|---|---|---|---|
| | Data | | | Check bits | | | Codestruct | | |
| | 2×2 | 3×3 | 4×4 | 2×2 | 3×3 | 4×4 | 2×2 | 3×3 | 4×4 |
| 1 | 100.00 | 100.00 | 100.00 | 100.00 | 100.00 | 100.00 | 100.00 | 100.00 | 100.00 |
| 2 | 100.00 | 100.00 | 100.00 | 100.00 | 100.00 | 100.00 | 100.00 | 100.00 | 100.00 |
| 3 | 100.00 | 100.00 | 100.00 | 100.00 | 100.00 | 100.00 | 100.00 | 100.00 | 100.00 |
| 4 | 100.00 | 100.00 | 100.00 | 100.00 | 100.00 | 100.00 | 100.00 | 100.00 | 100.00 |
| 5 | - | 100.00 | 100.00 | 100.00 | 100.00 | 100.00 | 99.49 | 99.92 | 99.98 |
| 6 | - | 100.00 | 99.90 | 100.00 | 100.00 | 100.00 | 99.35 | 99.90 | 99.97 |
| 7 | - | 100.00 | 100.00 | 100.00 | 100.00 | 100.00 | 99.49 | 99.91 | 99.98 |
| 8 | - | 100.00 | 99.88 | 100.00 | 100.00 | 100.00 | 99.80 | 99.91 | 99.98 |

*Legend: **2×2**, **3×3** and **4×4** are abbreviations of **2×Ham_2×2**, **2×Ham_3×3** and **2×Ham_4×4**, respectively*
*"-" uncollected value*

The results in Table 7 also suggest that the proposed codes have a high capacity for detecting error patterns in any evaluated region, with the worst-case detection efficiency being 99.35%. Additionally, it is possible to identify 100% error detection for all error patterns occurring exclusively in the check bits region. It is important to note that error patterns with 1 to 8 simultaneous errors do not cover all possible cases, meaning that patterns with a higher number of simultaneous errors could result in lower detection rates. This is a potential area for future research, but if confirmed, it would support the use of radiation-hardened memory [43] exclusively in the data area. Here, the reasoning is the inverse of that presented in Section 6.1, as the goal is to obtain correct data or at least determine the presence of errors in the read information. Furthermore, for smaller codes, such as 2×Ham_2×2, there are only 4 data bits for 8 check bits.

## 6.3 Reliability

Reliability is directly related to the ability to identify, correct, and prevent errors in memory systems, especially in critical scenarios such as space missions or resource-constrained systems. Different methods and metrics can be adopted to assess system reliability. This work is based on the model presented by Freitas et al. [14], which addresses reliability aspects in the context of ECC-protected memories over time. More specifically, the authors explore the Mean Time To Failure (MTTF) metric and how it is used to evaluate the effectiveness of ECC in memory exposed to error occurrences over time. The authors employ probabilistic mathematical models to study fault injection over time and the eventual ability of the code to correct these faults. For better comprehension, we present part of the mathematical modeling below.

Freitas et al. [14] use the *Poisson binomial distribution*, a discrete probability distribution, described in Equation (73), which calculates the probability $P_n^i$ of *i* errors occurring in a memory word with $n$ bits.

$$P_n^i = \binom{n}{i} \times p^i \times (1-p)^{n-i} \tag{73}$$

The fundamental considerations of a binomial distribution are: (i) each trial has only two possible outcomes, success or failure (a binomial situation known as a Bernoulli trial); in this case, whether a failure occurred or not; (ii) the probabilities of success $p^i$ and failure $(1-p)^{n-i}$ in each trial are independent of other trials; (iii) the variable of interest is the number of successes *i* (errors) in $n$ trials (bits where the error can occur). Considering the probabilities of success $p(t)^i$ and failure $(1-p(t))^{n-i}$ over a time interval $t$, these can be represented by Equations (74) and (75), respectively.

$$p(t)^i = (1-e^{-\lambda t})^i \tag{74}$$

$$(1-p(t))^{n-i} = e^{-\lambda t(n-i)} \tag{75}$$

Equations (74) and (75) use the parameter $\lambda$, measured in failures per bit per day, as an indicator of error occurrence severity in the experimental scenario. For instance, considering a satellite $S$ in space orbit or during launch from Earth, $\lambda$ is potentially higher when $S$ enters orbit due to the absence of Earth's magnetic field, which otherwise reduces exposure to various cosmic radiations. Substituting (74) and (75) into (73), now considering the binomial distribution $P(t)_n^i$ over time $t$, we obtain Equation (76).

$$P_n^i(t) = \binom{n}{i} \times (1-e^{-\lambda t})^i \times e^{-\lambda t(n-i)} \tag{76}$$

Additionally, Equation (77) shows the probability $N_n^i(t)$ of errors occurring that do not affect system operation at a given time $t$.

$$N_n^i(t) = \sum_{i=1}^{\sigma} P_n^i(t) \times \mathcal{E}(i) \qquad (77)$$

$N_n^i(t)$ assumes that all errors occurring up to time $t$ can be corrected by some mechanism (e.g., an ECC decoder). Here, $\sigma$ represents the accumulated number of errors, and $\mathcal{E}(i)$ is the error correction rate for $i$ errors. We use $\sigma$ within the range of 1 to 8 accumulated errors since these values are available in the last three columns of Table 6 for each of the three evaluated codes. A more precise probability value for $N_n^i(t)$ tends to be higher, as it includes the sum over the entire range of errors (not just those evaluated in this work). However, it is worth noting that these values are quite small, as ECCs' correction capability significantly reduces for more than 8 errors.

Equation (78) presents the calculation of ECC reliability $r_n(t)$ for a codestruct of $n$ bits over time $t$, considering both the probability of system operation being affected and not being affected. Equation (78) expands into Equation (79) when incorporating Equation (77).

$$r_n(t) = 1 - P_n(t) + N_n^i(t) \qquad (78)$$

$$r_n(t) = 1 - P_n(t) + \sum_{i=1}^{\sigma} P_n^i(t) \times \mathcal{E}(i) \qquad (79)$$

Finally, the MTTF metric considers the sum of all reliability values over time, as described in Equation (80). Note that our experiments illustrate only $r_n(t)$.

$$MTTF = \int_0^{\infty} r_n(t)\,\mathrm{dt} \qquad (80)$$

Figure 58 illustrates the methodology employed to evaluate the reliability of systems protected by the three proposed codes (2×Ham_2×2, 2×Ham_3×3, and 2×Ham_4×4).

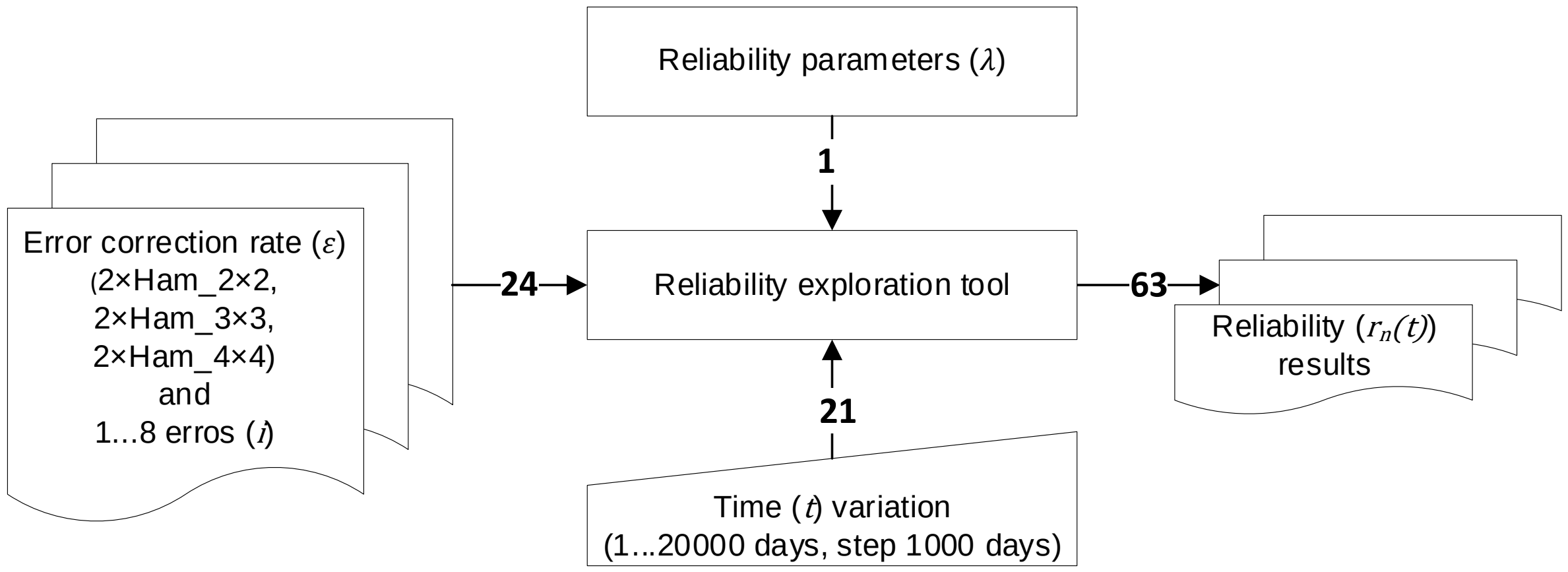

**Figure 58. Methodology used to evaluate the reliability of memory areas protected by the codes 2×Ham_2×2, 2×Ham_3×3, and 2×Ham_4×4 (Source: Author).**

For this experiment, we implemented a tool to calculate reliability values over the first 20,000 days of operation, starting from the first day with a time step of 1,000 days. Additionally, we adopted $\lambda = 10^{-5}$, based on [14], which means a probability of one bit failing every 10,000 days. Note that as $n$ increases, the probability of failures also increases due to the higher number of bits in the codestruct.

The experiment, illustrated in Figure 59, resulted in 63 reliability evaluation points, with 21 points for each explored ECC.

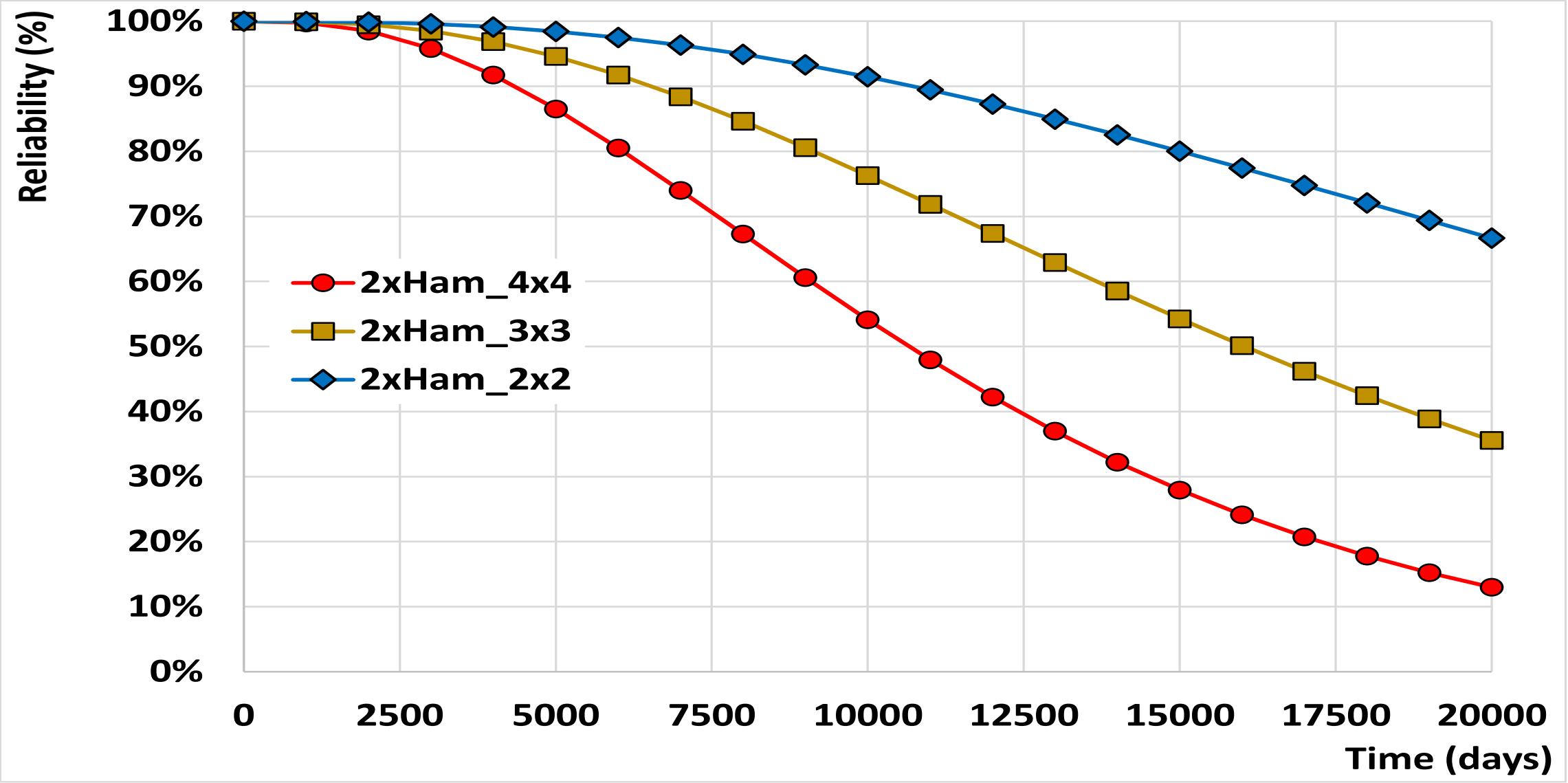

**Figure 59. Reliability of the codes 2×Ham_2×2, 2×Ham_3×3, and 2×Ham_4×4 over a period of 20,000 days. The graph uses $\lambda = 10^{-5}$ failures per bit per day (Source: Author)**

The results show that reliability degradation depends on the number of bits and the error correction rate. Since error correction rates are nearly identical, codes with more bits are more penalized, degrading their reliability more rapidly.

Reliability information can be used to define a scrubbing rate [68] depending on the employed ECC. Scrubbing is a technique that employs periodic reads at predefined intervals to prevent errors from accumulating beyond the point where codestruct correction is no longer possible. For the evaluated codes, the technique considers that ECCs can recover 100% of cases with up to 2 errors. Thus, a system that controls reliability should reduce read intervals for the 2×Ham_4×4 code, while spacing them out for the 2×Ham_2×2 code. This provides insights for balancing ECC efficiency and effectiveness, as using scrubbing increases power consumption. The exploration of scrubbing periods, focusing on integrating the ECC overlapping technique for satellite memory protection, is part of future work.

## 6.4 Power Dissipation, Area Consumption, and Latency

Figure 60 presents the methodology used to evaluate operational and manufacturing costs, i.e., power dissipation, area consumption, and latency, for the 2×Ham_2×2, 2×Ham_3×3, and 2×Ham_4×4 codes.

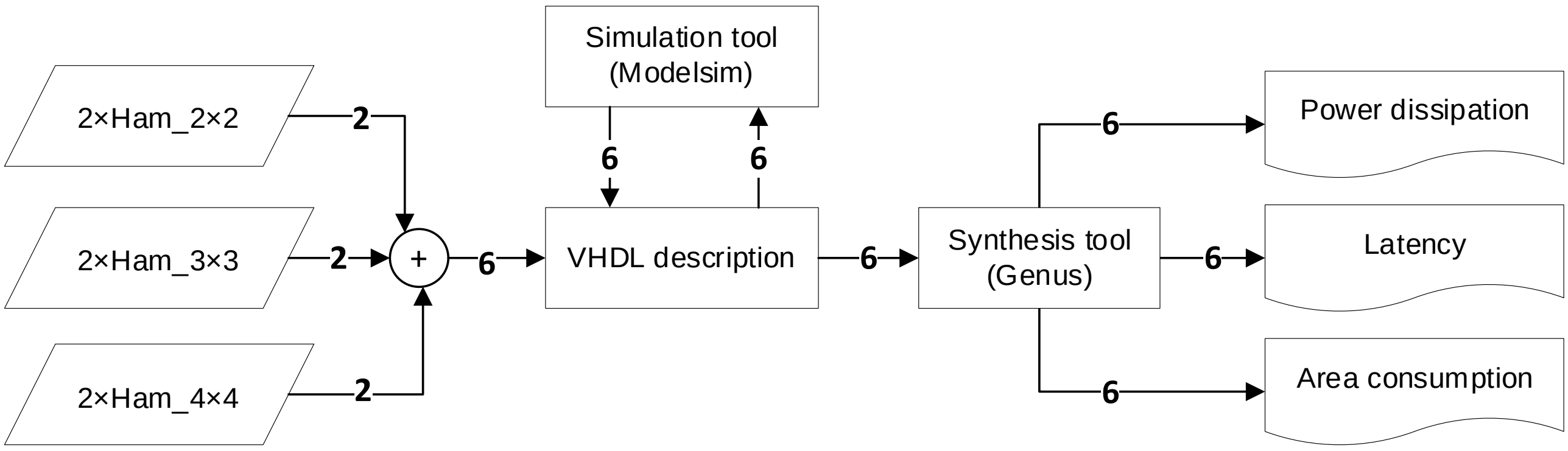

**Figure 60. Methodology applied to obtain data on power dissipation, latency, and area consumption for the overlapped ECCs explored in this work. The encoder and decoder codes were described in VHDL for each ECC, validated through simulation, and logically synthesized (Source: Author).**

We described the encoding and decoding circuits in RTL VHDL[1] for each overlapped ECC using purely combinational circuits. The complete set of VHDL descriptions is available on GitHub (https://github.com/Andrewarf/overlapping-error-correction-codes-on-two-dimensional-structures); additionally, “Appendix B – VHDL Code for 2×Ham3×3 Synthesis” contains the description of the 2×Ham_3×3 code. This low-level description allows synthesis tools to correctly infer the circuits implementing the ECC encoders/decoders.

Since the circuit descriptions were manually created based on their Java implementations, we included a validation step through VHDL simulation. For this validation, we used the ModelSim tool, which allowed us to verify various data encoding patterns, error injection, and subsequent codestruct decoding. Once the six VHDL descriptions were validated (one encoding circuit and one decoding circuit for each of the three ECCs), we proceeded with logical synthesis for a 28nm CMOS technology under normal operating conditions. Table 8, containing the synthesis results, shows that the increased complexity of the decoder significantly impacts area consumption, power dissipation, and latency. One of the main reasons is that the encoder is a subset of the decoder, as observed in Figure 47 and Figure 50, which clearly justifies the increase in consumed area and dissipated power.

[1] RTL VHDL– VHSIC (Very High Speed Integrated Circuits) Hardware Description Language, Register-Transfer Level is a language for describing digital circuits at the register transfer level.

**Table 8. Comparative table of values obtained from the logical synthesis of the 2×Ham_2×2, 2×Ham_3×3, and 2×Ham_4×4 codes. For each code, data on power dissipation, area consumption, and latency of encoders and decoders were obtained (Source: Author).**

| Metric | | 2×Ham_2×2 | | 2×Ham_3×3 | | 2×Ham_4×4 | |
|---|---|---|---|---|---|---|---|
| | | Encoder | Decoder | Encoder | Decoder | Encoder | Decoder |
| Area ($um^2$) | | 35.7 | 140.5 | 80.4 | 371.8 | 146.5 | 778.0 |
| Power (uW) | | 2.1 | 9.3 | 4.7 | 24.0 | 9.7 | 47.9 |
| Latency (ps) | Error not detected | 333.0 | 419.7 | 522.0 | 683.4 | 751.0 | 961.3 |
| | Error detected | | 841.0 | | 1508.0 | | 2073.0 |

Table 9 displays that, on average, the areas of the decoders are 4.6 times larger than that of the encoders, while the decoders dissipate more than 4.8 times the power of the encoders. Latency varies depending on the decoding result. If the decoder detects an error, it must execute the error correction algorithm, which has an average latency 2.7 times higher than that of the encoder. However, if no error is detected, the decoder can abort the process much earlier, reducing operational latency. Notably, these values were obtained through logical synthesis, which does not have the same accuracy and precision as physical synthesis. However, physical synthesis is part of a set of experiments planned for future work, along with the analysis of encoders and decoders for other state-of-the-art ECCs, such as Matrix [1], CLC [57] and PBD [23], which were discussed in Section 6.1.1.

**Table 9. Comparative table with the same values as Table 8, but including columns with partial and total average information (Source: Author).**

| Metric | | 2×Ham_2×2 | | | 2×Ham_3×3 | | | 2×Ham_4×4 | | | Total average |
|---|---|---|---|---|---|---|---|---|---|---|---|
| | | Encoder | Decoder | Average | Encoder | Decoder | Average | Encoder | Decoder | Average | |
| Area ($um^2$) | | 35.7 | 140.5 | 3.94 | 80.4 | 371.8 | 4.62 | 146.5 | 778 | 5.31 | 4.62 |
| Power (uW) | | 2.1 | 9.3 | 4.43 | 4.7 | 24 | 5.11 | 9.7 | 47.9 | 4.94 | 4.82 |
| Latency (ps) | N. det. | 333.0 | 419.7 | 1.26 | 522.0 | 683.4 | 1.31 | 751.0 | 961.3 | 1.28 | 1.28 |
| | Det. | | 841.0 | 2.53 | | 1508.0 | 2.89 | | 2073.0 | 2.76 | 2.72 |

Another aspect analyzed is the increase in synthesis costs with the expansion of the data area, both for encoders and decoders. For this case, we generated Table 10, which normalizes the values in Table 8 based on the number of data bits in each ECC, i.e., 4, 9, and 16 for the 2×Ham_2×2, 2×Ham_3×3, and 2×Ham_4×4 ECCs, respectively.

The values in Table 10 present two trends of interest to designers. The first trend is the increase in area consumption and power dissipation as the code scales, both for the encoder and the decoder. The only observed exception is the power dissipation of the 2×Ham_2×2 decoder compared to the 2×Ham_3×3 decoder, which shows a slight reduction from 0.53 µW to 0.52 µW. This trend suggests an increase in costs when scaling the codes.

**Table 10. Comparative table with the values from Table 8 normalized by the number of data area bits in each ECC (Source: Author).**

| Metric | | 2×Ham_2×2 | | 2×Ham_3×3 | | 2×Ham_4×4 | |
|---|---|---|---|---|---|---|---|
| | | Encoder | Decoder | Encoder | Decoder | Encoder | Decoder |
| Area ($um^2$) | | 8.93 | 35.13 | 8.93 | 41.31 | 9.16 | 48.63 |
| Power (uW) | | 0.53 | 2.33 | 0.52 | 2.67 | 0.61 | 2.99 |
| Latency (ps) | Error not detected | 83.25 | 104.93 | 58.00 | 75.93 | 46.94 | 60.08 |
| | Error detected | | 210.25 | | 167.56 | | 129.56 |

On the other hand, the second trend is a reduced latency, which increases the operating frequency of read and write circuits, opening the possibility for developing high-speed ECCs. It is important to note that these two verified trends should be explored with a greater variation in scale to obtain values with higher statistical significance. This analysis can be conducted in future studies.

## 6.5 Scalability of the Overlapped-ECCs

The scalability of an ECC reflects variations in effectiveness and efficiency due to changes in the data protection area. A scalable ECC maintains its effectiveness (e.g., error correction and detection rates) while preserving or improving its efficiency (e.g., energy consumption or area usage). Figure 61 shows the methodology used to evaluate scalability.

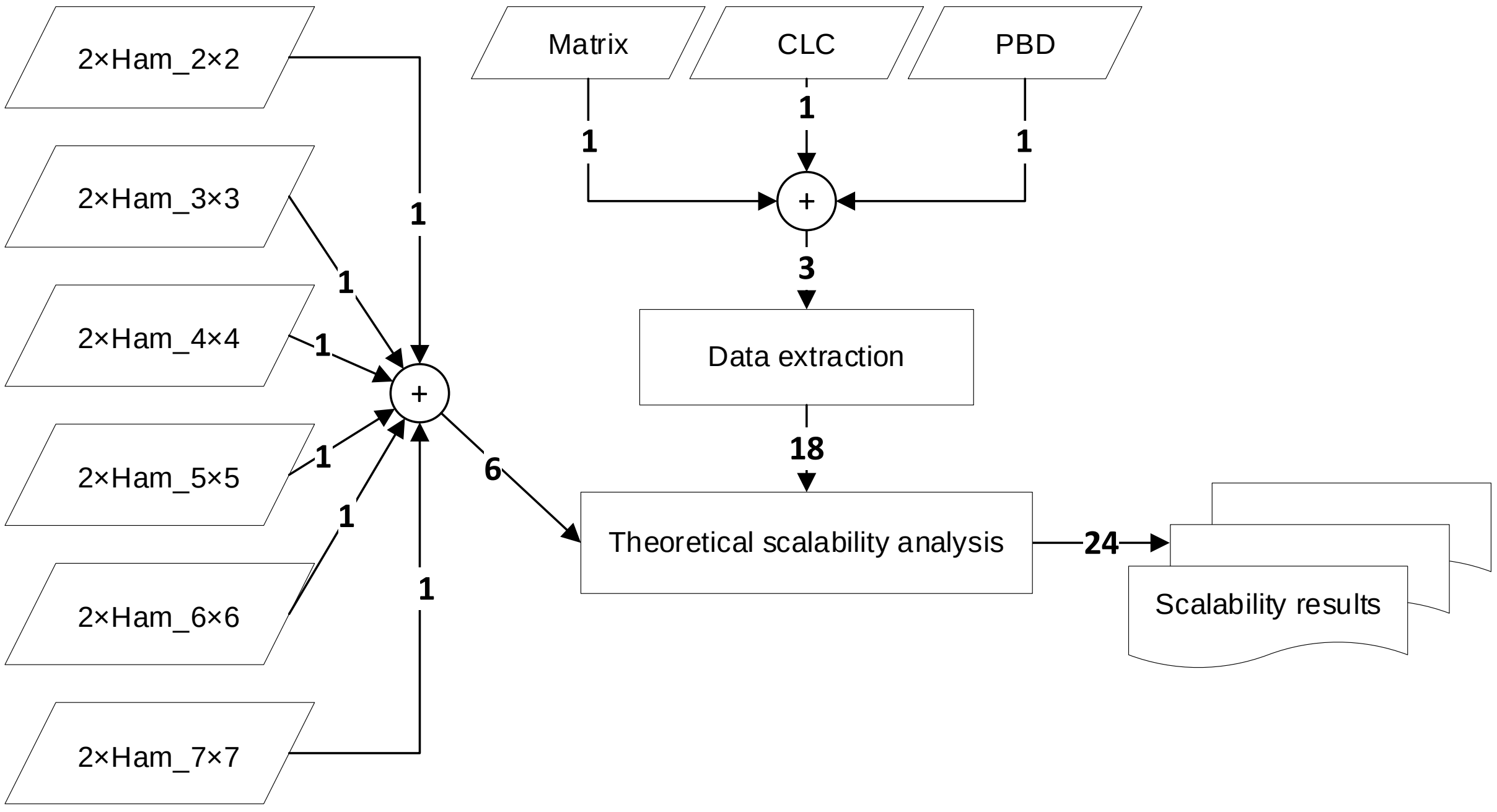

**Figure 61. Methodology used to assess the scalability of the overlapped ECC case study adopted in this work, along with the scalability of three state-of-the-art ECCs. The analysis is entirely theoretical, based on the extrapolation of data obtained from the literature (Source: Author).**

This section employs the evaluated overlapped ECC model (i.e., using two extended Hamming codes to protect the same data region), conducting a theoretical analysis of square data matrices ranging from 4 to 49 bits. Additionally, we use the same 2D-ECCs presented in Section 6.1.1 (i.e., Matrix [1], CLC [57] and PBD [23]) to compare with state-of-the-art ECCs with similar requirements. It is essential to highlight that this section is purely theoretical and does not involve the implementation of all the codes described here.

Beyond the three overlapped ECCs extensively discussed throughout this work (i.e., 2×Ham_2×2, 2×Ham_3×3, and 2×Ham_4×4), we analyzed the codestruct designs of three additional ECCs: 2×Ham_5×5, 2×Ham_6×6, and 2×Ham_7×7, aiming to extend the theoretical scalability analysis. The design of these three new ECCs follows the same model as the previous ones, expanding the Hamming protection areas to accommodate the growth of data areas. The choice of square matrices was made to maintain a consistent form factor. Regarding the Matrix, CLC, and PBD ECCs, we first analyzed and understood the formation rules of each ECC before scaling them to sizes equivalent to the overlapped ECCs. Notably, the choice of square data matrices allowed for perfect scaling of these 2D-ECCs available in the literature, ensuring a fair comparison between all ECCs.

To evaluate scalability, we used redundancy cost ($rc$) as a metric that divides the number of check bits ($\#cb$) by the total number of bits in the codestruct ($\#cs$), as expressed in Equation (81). The scalability results are illustrated in Table 11.

$$rc = \frac{\#cb}{\#cs} \tag{81}$$

**Table 11. Redundancy costs for square data matrices ranging from 2×2 to 7×7; green and red highlight the lowest (better) and highest (worst) scalability results, respectively (Source: Author).**

| Data matrix | N | Overlapped ECC | | | Matrix | | | PBD | | | CLC | | |
|---|---|---|---|---|---|---|---|---|---|---|---|---|---|
| | | #cb | #cs | rc | #cb | #cs | rc | #cb | #cs | rc | #cb | #cs | rc |
| **2×2** | 4 | 8 | 12 | 0.67 | 8 | 12 | 0.67 | 5 | 9 | **0.56** | 14 | 18 | **0.78** |
| **3×3** | 9 | 10 | 19 | **0.53** | 12 | 21 | 0.57 | 12 | 21 | 0.57 | 19 | 28 | **0.68** |
| **4×4** | 16 | 12 | 28 | **0.43** | 16 | 32 | 0.50 | 20 | 36 | 0.56 | 24 | 40 | **0.60** |
| **5×5** | 25 | 12 | 37 | **0.32** | 25 | 50 | 0.50 | 32 | 57 | 0.56 | 35 | 60 | **0.58** |
| **6×6** | 36 | 14 | 50 | **0.28** | 30 | 66 | 0.45 | 45 | 81 | **0.56** | 41 | 77 | **0.53** |
| **7x7** | 49 | 14 | 63 | **0.22** | 35 | 84 | 0.42 | 62 | 111 | **0.56** | 47 | 96 | **0.49** |

The results reveal that the proposed overlapped technique is the ECC with the highest potential, scaling efficiently while significantly reducing redundancy costs. Even for very small data matrices, overlapped ECCs ranked second, alongside the Matrix code. Despite the lower cost of 2×2 PBD matrix, its poor scalability results in worse performance than CLC for size 6×6 or larger square matrices.

Isolated scalability results only indicate the cost of each ECC. However, when combined with effectiveness results, such as the error correction rates presented in Section 6.1.1, they provide designers with significant insights into the benefits and trade-offs of each choice, helping evaluate whether the project requirements are met. Specifically for the evaluated ECCs, we highlight that the scaling of overlapped ECCs results in a smaller storage area than other 2D-ECCs, directly impacting memory area consumption and energy efficiency. Additionally, the error correction rates presented in Section 6.1.1, combined with the smaller number of bits in the codestructs, suggest that our proposed method may achieve higher reliability levels over time. Confirming this hypothesis requires further studies, which are planned for future research.

# 7. CONCLUSIONS

The study on overlapped ECCs allowed us to evaluate their effectiveness in error detection and correction across matrices of different sizes, considering factors such as correction and detection rates, long-term reliability, area consumption, power dissipation, and latency.

The experimental results demonstrated that, for single and double errors, the proposed code overlapping technique achieves a 100% correction rate in both the parity check area and the data area, ensuring high reliability. However, as the number of errors increases, the correction effectiveness drops significantly. For triple and quadruple errors, the correction rate remains high in the parity check area but becomes zero in the data area. When five or more errors occur, the correction rate drops drastically, highlighting the limitations of the proposed approach.

It is important to emphasize that the technique is not mathematically limited to correcting only double errors; this is a limitation of the error correction algorithm applied to the technique. Research into more efficient correction algorithms is part of future work.

Additionally, error detection is significantly higher, reaching 100% or nearly 100% in all analyzed scenarios, ensuring that all cases up to four errors are correctly identified.

The long-term reliability was analyzed for data areas of 4 (2×Ham_2×2), 9 (2×Ham_3×3), and 16 (2×Ham_4×4) bits. As expected, larger matrices exhibited greater degradation due to an increased number of failure points. In this regard, smaller matrices performed better. For example, the 2×Ham_2×2 code maintained a reliability above 60% after 20,000 days, whereas the 2×Ham_4×4 code suffered sharper degradation, with reliability dropping to around 20% in the same period.

The analysis of area consumption, power dissipation, and latency revealed that matrix size directly impacts these factors. The encoder circuit grew by a factor of four, while the decoder circuit increased by more than five times when transitioning from 2×Ham_2×2 to 2×Ham_4×4. Power consumption followed the same trend, being approximately five times higher for the larger matrix. Latency also increased significantly, particularly in the decoder, where the processing time in error-free scenarios was 2.5 times longer for 2×Ham_4×4 compared to 2×Ham_2×2.

Based on these results, we conclude that the matrix size selection depends on system constraints and requirements:

2×Ham_2×2 proved to be the most efficient option in terms of hardware consumption and longevity, making it ideal for applications requiring long-term reliability with area and power constraints.

2×Ham_4×4, while offering stronger short-term protection, comes with a higher computational cost and greater long-term degradation.

2×Ham_3×3 represents a balance between these two extremes, making it a viable choice for systems requiring a trade-off between performance, reliability, and efficiency.

Additionally, it is crucial to emphasize that the latency and power dissipation of the decoders are very similar to those of the encoders in scenarios without detected errors. This means that decoder complexity is only noticeable in the presence of errors, potentially reducing operational time.

Finally, comparisons with other ECCs of similar capacity demonstrate that the proposed technique is highly promising, featuring a lower relative bit consumption cost while achieving 100% correction for up to two random errors.

# 8. PUBLICATIONS

Throughout this master's dissertation, two scientific papers related to the proposed work were written and accepted for publication. These papers are in the references [40] and [22] and listed below:

- A. Muniz, L. Mazzoco, W. Savaris, E. Pissolatto, T. Beneditto, A. Fritsch, J. Silveira, C. Marcon, "Comparing Structures of Two-Dimensional Error Correction Codes", Microelectronics Reliability, vol. 161, pp. 115481:1-9, Oct. 2024.
- A. Fritsch, W. Savaris, A. Muniz, G. Borba, R. Girardi, J. Silveira, C. Marcon "Overlapped Error Correction Codes in Two-Dimensional Structures", Proceedings of the IEEE CASS Latin American Symposium on Circuits and Systems (LASCAS), pp. 1-6, 2025.

Additionally, a third paper detailing this work is currently being prepared, with plans for submission to *IEEE Transactions on VLSI (TVLSI)*.

The tools used to obtain the results presented in this dissertation are publicly available on GitHub at https://github.com/andrewarf/overlapping-error-correction-codes-on-two-dimensional-structures .For completeness and ease of understanding, the Java code used for evaluating error detection and correction effectiveness is provided in "*Appendix A – Java Code for Simulating 2×Ham3×3 Code*", while the VHDL code used for synthesis data collection is included in "*Appendix B – VHDL Code for 2×Ham3×3 Synthesis*".

# 10. APPENDIX A – JAVA CODE FOR SIMULATING 2×HAM3×3 CODE

This appendix presents four Java files used to validate the 2×Ham3×3 code and to extract data on error detection and correction effectiveness:

- **CodeStruct.java** – Defines the class that implements the encoding for both the outer Hamming and inner Hamming codes. We emphasize that, to simplify the textual description of Chapter 5, we chose to present the data as a one-dimensional vector. However, the Java implementation uses a two-dimensional matrix, resulting in a slight difference between the textual description and the code implementation;
- **CodeStructWithError.java** – Contains methods and data structures used in the decoder, as well as a routine for inserting errors at specific positions;
- **Decoder.java** – Has the decoding algorithm, partially described in 5.3.1; and
- **MainSystem.java** – Includes routines for exploring error positioning, generating a specified number of errors, and performing exhaustive error generation.

**CodeStruct.java**

```
public class CodeStruct {
      public int D[][] = new int[3][3];
      public int Co[] = new int[4];
      public int Ci[] = new int[4];
      public int Po, Pi;

      public CodeStruct(int D[][]) throws Exception {
            for(int r=0; r<this.D.length; r++)
                  for(int c=0; c<this.D[0].length; c++)
                        this.D[r][c] = D[r][c];

            encodeOuterHamming(Co);
            Po = encodeHammingParity(Co);
            encodeInnerHamming(Ci);
            Pi = encodeHammingParity(Ci);
      }

      protected void encodeInnerHamming(int ham[]) {
            ham[0] = D[0][0] ^ D[0][2] ^ D[1][0] ^ D[1][1] ^ D[1][2]^ D[2][2];
            ham[1] = D[0][1] ^ D[0][2] ^ D[1][0] ^ D[1][2] ^ D[2][0] ^ D[2][2];
            ham[2] = D[0][1] ^ D[0][2] ^ D[1][1] ^ D[2][1] ^ D[2][2];
            ham[3] = D[0][0] ^ D[0][1] ^ D[1][0] ^ D[2][0] ^ D[2][1] ^ D[2][2];
      }

      protected void encodeOuterHamming(int ham[]) {
            ham[0] = D[0][0] ^ D[0][1] ^ D[1][0] ^ D[1][1] ^ D[2][0] ^ D[2][2];
            ham[1] = D[0][1] ^ D[1][1] ^ D[1][2] ^ D[2][0] ^ D[2][1] ^ D[2][2];
            ham[2] = D[0][0] ^ D[0][2] ^ D[1][0] ^ D[2][0] ^ D[2][1] ^ D[2][2];
            ham[3] = D[0][0] ^ D[0][1] ^ D[0][2] ^ D[1][2] ^ D[2][2];
      }
```

```
	protected int encodeHammingParity(int ham[]) {
		int parity = ham[0];
		for(int k=1; k<ham.length; k++)
			parity = parity ^ ham[k];
		for(int j=0; j<this.D.length; j++) {
			for(int k=0; k<this.D[0].length; k++)
				parity = parity ^ D[j][k];
		}
		return parity;
	}

	public boolean isEqual(CodeStruct ecc) {
		for(int k=0; k<3; k++) {
			for(int j=0; j<3; j++)
				if(D[k][j] != ecc.D[k][j])
					return false;
		return true;
	}
}
```

## CodeStructWithError.java

```
public class CodeStructWithErrror extends CodeStruct {
	public int recCo[] = new int[4];
	public int recCi[] = new int[4];
	public int recPo, recPi;
	public int sCo[] = new int[4];
	public int sCi[] = new int[4];
	public int sPo, sPi;
	public int sCoq, sCiq;
	public int EArO, EArI;
	public boolean DErO, DErI;
	public boolean SErO, SErI;

	public CodeStructWithErrror(CodeStruct iCode, int err[]) throws Exception {
		super(iCode.D);
		setErrorPattern(err);
		recomputeControlVariables();
	}

	public void recomputeControlVariables() {
		recomputeCheckBitsAndParity();
		computeSyndromes();
		computeErrorAddress();
		computeSE_DE();
	}

	public void recomputeCheckBitsAndParity() {
		encodeOuterHamming(recCo);
		recPo = encodeHammingParity(Co);
		encodeInnerHamming(recCi);
		recPi = encodeHammingParity(Ci);
	}

	public void computeSyndromes() {
		for(int k=0; k<sCi.length; k++)
			sCi[k] = Ci[k] == recCi[k] ? 0 : 1;
		sPi = Pi == recPi ? 0 : 1;
		for(int k=0; k<sCo.length; k++)
			sCo[k] = Co[k] == recCo[k] ? 0 : 1;
		sPo = Po == recPo ? 0 : 1;
		sCiq = (sCi[0]==1 || sCi[1]==1 || sCi[2]==1 || sCi[3]==1) ? 1 : 0;
		sCoq = (sCo[0]==1 || sCo[1]==1 || sCo[2]==1 || sCo[3]==1) ? 1 : 0;
	}
```

```
	public void computeErrorAddress() {
		EArO = sCo[0] * 8 + sCo[1] * 4 + sCo[2] * 2 + sCo[3];
		EArI = sCi[0] * 8 + sCi[1] * 4 + sCi[2] * 2 + sCi[3];
	}

	public void computeSE_DE() {
		SErI = sCiq==1 && sPi==1;
		DErI = sCiq==1 && sPi==0;
		SErO = sCoq==1 && sPo==1;
		DErO = sCoq==1 && sPo==0;
	}

	private void setErrorPattern(int errorPattern[]) throws Exception {
		for(int k = 0; k < errorPattern.length; k++)
			setError(errorPattern[k]);
	}

	private static int invertBit(int value) {
		return value == 0 ? 1 : 0;
	}

	private void setError(int errorPosition) throws Exception {
		int row, column;
		if(errorPosition < 9) {
			row = errorPosition / 3;
			column = errorPosition % 3;
			D[row][column] = invertBit(D[row][column]);
		}
		else if(errorPosition >= 9 && errorPosition < 13) {
			errorPosition -= 9;
			Co[errorPosition] = invertBit(Co[errorPosition]);
		}
		else if(errorPosition == 13) {
			Po = invertBit(Po);
		}
		else if(errorPosition >= 14 && errorPosition < 18) {
			errorPosition -= 14;
			Ci[errorPosition] = invertBit(Co[errorPosition]);
		}
		else if(errorPosition == 18) {
			Pi = invertBit(Pi);
		}
		else
			throw new Exception("Error position = " + errorPosition);
	}
}
```

## Decoder.java

```
public class Decoder {
	static int outerRowTab[] = {-1,-1,-1,0,-1,1,2,-1,-1,-1,1,0,1,0,2,2};
	static int outerColumnTab[] = {-1,-1,-1,2,-1,2,1,-1,-1,-1,0,0,1,1,0,2};
	static int innerRowTab[] = {-1,-1,-1,2,-1,2,-1,0,-1,0,1,-1,1,1,0,2};
	static int innerColumnTab[] = {-1,-1,-1,1,-1,0,-1,1,-1,0,1,-1,2,0,2,2};

	static void flitDoubleError(CodeStructWithErrror ecc, int posA, int posB) {
		int rowA = outerRowTab[posA];
		int columnA = outerColumnTab[posA];
		int rowB = outerRowTab[posB];
		int columnB = outerColumnTab[posB];

		ecc.D[rowA][columnA] = ecc.D[rowA][columnA]==0 ? 1 : 0;
		ecc.D[rowB][columnB] = ecc.D[rowB][columnB]==0 ? 1 : 0;
	}
```

```
    public static boolean decoding(CodeStructWithErrror ecc) {
        boolean detErr;
        detErr = ecc.sPo!=0||ecc.sPi!=0||ecc.EArO!=0||ecc.EArI!=0 ? true:false;
        if(ecc.EArO == 0 || ecc.EArI == 0)
            return detErr;
        if(ecc.SErO) {
            int row = outerRowTab[ecc.EArO];
            int column = outerColumnTab[ecc.EArO];
            if(row!=-1 && column!=-1)
                ecc.D[row][column] = ecc.D[row][column]==0 ? 1 : 0;
            ecc.recomputeControlVariables();
            return detErr;
        }
        if(ecc.SErI) {
            int row = innerRowTab[ecc.EArI];
            int column = innerColumnTab[ecc.EArI];
            if(row!=-1 && column!=-1)
                ecc.D[row][column] = ecc.D[row][column]==0 ? 1 : 0;
            ecc.recomputeControlVariables();
            return detErr;
        }
        if(ecc.DErO && ecc.DErI) {
            if(ecc.EArO==1) {
                if(ecc.EArI==4)
                    flitDoubleError(ecc, 11, 10);
                else if(ecc.EArI==10)
                    flitDoubleError(ecc, 14, 15);
                else if(ecc.EArI==13)
                    flitDoubleError(ecc, 13, 12);
            }
            else if(ecc.EArO==2) {
                if(ecc.EArI==8)
                    flitDoubleError(ecc, 13, 15);
                else if(ecc.EArI==15)
                    flitDoubleError(ecc, 12, 14);
            }
            else if(ecc.EArO==3) {
                if(ecc.EArI==2)
                    flitDoubleError(ecc, 13, 14);
                else if(ecc.EArI==5)
                    flitDoubleError(ecc, 12, 15);
                else if(ecc.EArI==15)
                    flitDoubleError(ecc, 5, 6);
            }
            else if(ecc.EArO==4) {
                if(ecc.EArI==6)
                    flitDoubleError(ecc, 11, 15);
                else if(ecc.EArI==8)
                    flitDoubleError(ecc, 10, 14);
            }
            else if(ecc.EArO==5) {
                if(ecc.EArI==2)
                    flitDoubleError(ecc, 10, 15);
                else if(ecc.EArI==12)
                    flitDoubleError(ecc, 11, 14);
                else if(ecc.EArI==13)
                    flitDoubleError(ecc, 3, 6);
            }
            else if(ecc.EArO==6) {
                if(ecc.EArI==2)
                    flitDoubleError(ecc, 3, 5);
                else if(ecc.EArI==7)
                    flitDoubleError(ecc, 10, 12);
                else if(ecc.EArI==14)
                    flitDoubleError(ecc, 11, 13);
            }
```

```
                else if(ecc.EArO==7) {
                      if(ecc.EArI==3)
                            flitDoubleError(ecc, 11, 12);
                      else if(ecc.EArI==10)
                            flitDoubleError(ecc, 13, 10);
                }
                else if(ecc.EArO==8) {
                      if(ecc.EArI==6)
                            flitDoubleError(ecc, 14, 6);
                      else if(ecc.EArI==7)
                            flitDoubleError(ecc, 11, 3);
                      else if(ecc.EArI==11)
                            flitDoubleError(ecc, 13, 5);
                }
                else if(ecc.EArO==9) {
                      if(ecc.EArI==3)
                            flitDoubleError(ecc, 3, 10);
                      else if(ecc.EArI==6)
                            flitDoubleError(ecc, 12, 5);
                      else if(ecc.EArI==12)
                            flitDoubleError(ecc, 6, 15);
                }
                else if(ecc.EArO==10) {
                      if(ecc.EArI==3)
                            flitDoubleError(ecc, 5, 15);
                      else if(ecc.EArI==9)
                            flitDoubleError(ecc, 12, 6);
                      }
                else if(ecc.EArO==11) {
                      if(ecc.EArI==4)
                            flitDoubleError(ecc, 13, 6);
                      else if(ecc.EArI==9)
                            flitDoubleError(ecc, 5, 14);
                }
                else if(ecc.EArO==12) {
                      if(ecc.EArI==1)
                            flitDoubleError(ecc, 3, 15);
                      else if(ecc.EArI==14)
                            flitDoubleError(ecc, 10, 6);
                }
                else if(ecc.EArO==13) {
                      if(ecc.EArI==10)
                            flitDoubleError(ecc, 11, 6);
                      else if(ecc.EArI==11)
                            flitDoubleError(ecc, 3, 14);
                }
                else if(ecc.EArO==14) {
                      if(ecc.EArI==5)
                            flitDoubleError(ecc, 11, 5);
                      else if(ecc.EArI==9)
                            flitDoubleError(ecc, 13, 3);
                }
                else if(ecc.EArO==15) {
                      if(ecc.EArI==1)
                            flitDoubleError(ecc, 10, 5);
                      if(ecc.EArI==4)
                            flitDoubleError(ecc, 3, 12);
                }
                ecc.recomputeControlVariables();
          }
          return detErr;
    }
}
```

**MainSystem.java**

```
public class MainSystem {
      private static int numErrors, initialElement = 0, numElements = 19;
      private static long numberOfDecodigns = 0, numberOfErrorsDetected = 0;
      private static long errorsAfterDecoding = 0;
      private static CodeStruct initialCode;
      private static CodeStructWithErrror eccWithErrors;

      static void errGen(int errIdx, int errPat[], int elemIdx) throws Exception {
            if(errIdx == numErrors) {
                  eccWithErrors = new CodeStructWithErrror(initialCode, errPat);
                  if(Decoder.decoding(eccWithErrors)==true)
                        numberOfErrorsDetected++;
                  numberOfDecodigns++;
                  if(!initialCode.isEqual(eccWithErrors))
                        errorsAfterDecoding++;
                  return;
            }
            if(elemIdx >= numElements)
                  return;
            errPat[errIdx] = elemIdx;
            errGen(errIdx+1, errPat, elemIdx+1);
            errGen(errIdx, errPat, elemIdx+1);
      }
      private static void setInitialCode() throws Exception {
            int D[][] = {      {0, 0, 0},
                               {0, 0, 0},
                               {0, 0, 0},};
            initialCode = new CodeStruct(D);
      }
      private static void printTestIdentification() {
            System.out.println("#Errors=" + numErrors);
      }
      private static void printResults() {
            System.out.println("NumberOfDecodigns = " + numberOfDecodigns);
            System.out.println("\tErrorsDetected = " + numberOfErrorsDetected);
            System.out.println("\tErrorsAfterDecoding = " + errorsAfterDecoding);
      }
      private static void setNumberOfErrors(int nE) {
            numErrors = nE;
      }
      private static void resetSimulationData() {
            numberOfDecodigns = 0;
            numberOfErrorsDetected = 0;
            errorsAfterDecoding = 0;
      }
      private static void setErrorInterval(int inicio, int fim) {
            initialElement = inicio;
            numElements = fim;
      }
      public static void main(String[] args) throws Exception {
            setInitialCode();
            setErrorInterval(0, 19);
            for(int numberOfErrors=1; numberOfErrors<=8; numberOfErrors++) {
                  setNumberOfErrors(numberOfErrors);
                  printTestIdentification();
                  resetSimulationData();
                  errGen(0, new int[numErrors], initialElement);
                  printResults();
            }
      }
}
```

The `setInitialCode` method initializes the data area with an arbitrary bit sequence and properly encodes the ECCs.

The `setErrorInterval` method defines the range [start, end) where errors may occur. To analyze errors in specific areas, the method should be set as follows: [0, 9) for errors in the data area; [9, 19) for errors in the parity check area of both Hamming codes; [0, 19) for evaluating errors across the entire codestruct.

The number of errors to be evaluated is controlled by the loop variable `numberOfErrors`. In each iteration, the algorithm sets an internal variable `numErrors` using the `setNumberOfErrors` method. Then, the `resetSimulationData` method resets some decoding control variables to ensure that changes from one iteration do not affect subsequent results.

Finally, the algorithm calls the recursive method `errGen`, which explores all possible error patterns of size `numErrors` within the address range defined by `setErrorInterval`.

The recursiveness of `errGen` allows for an exhaustive exploration of all possible error patterns applied to the initial codestruct (`initialCode`). When recursion reaches a pattern with the same number of errors as `numErrors`, the `CodeStructWithError` function receives pattern and `initialCode` to generate a new codestruct with errors, referred to as `eccWithErrors`.

Subsequently, `eccWithErrors` is decoded using the approach proposed in this work via the decoding method from the Decoder class, producing a potentially corrected `eccWithErrors`. Additionally, the `numberOfDecodings` counter is incremented.

The final step of the method, once recursion ends, is to verify the decoding effectiveness by comparing `eccWithErrors` with `initialCode`. If the two structures differ, the `errorsAfterDecoding` variable is incremented, indicating that decoding has failed.

# 11. APPENDIX B – VHDL CODE FOR 2×HAM3×3 SYNTHESIS

This appendix contains three VHDL files with the code used to obtain area, latency, and power data:

- **Encoder.vhd** – Defines the entity-architecture pair for the encoder circuit, as illustrated in Figure 47;
- **Decoder.vhd** – Defines the entity-architecture pair for the decoder circuit, as illustrated in Figure 50;
- **Matrix_package.vhd** – Contains the single and double addressing vectors stored in ROM, used in the decoder illustrated in Figure 50. Note that the implementation described in Java as a sequence of *if-else* statements (Section 5.3.1 and Decoder.java in "*Appendix A – Java Code for Simulating 2×Ham3×3 Code*" has been replaced by a sparse matrix).

**Encoder.vhd**

```
===============================================================================
-- Computes Hamming for the data bit positioning defined in the outer ECC
===============================================================================
library IEEE;
use IEEE.STD_LOGIC_1164.ALL;

entity OuterHamming is
      Port (      D   : in std_logic_vector(0 to 8);   -- 9-bit data array
                  ECC : out std_logic_vector(0 to 3)   -- 4-bit outer ECC array
            );
end OuterHamming;

architecture OutHam of OuterHamming is
begin
      ECC(0) <= D(0) xor D(1) xor D(3) xor D(4) xor D(6) xor D(8);
      ECC(1) <= D(1) xor D(4) xor D(5) xor D(6) xor D(7) xor D(8);
      ECC(2) <= D(0) xor D(2) xor D(3) xor D(6) xor D(7) xor D(8);
      ECC(3) <= D(0) xor D(1) xor D(2) xor D(5) xor D(8);
end OutHam;


===============================================================================
-- Computes Hamming for the data bit positioning defined in the inner ECC
===============================================================================
library IEEE;
use IEEE.STD_LOGIC_1164.ALL;

entity InnerHamming is
      Port (      D   : in std_logic_vector(0 to 8);   -- 9-bit data array
                  ECC : out std_logic_vector(0 to 3)   -- 4-bit outer ECC array
            );
end InnerHamming;
```

```
architecture InnHam of InnerHamming is
begin
      ECC(0) <= D(0) xor D(2) xor D(3) xor D(4) xor D(5) xor D(8);
      ECC(1) <= D(1) xor D(2) xor D(3) xor D(5) xor D(6) xor D(8);
      ECC(2) <= D(1) xor D(2) xor D(4) xor D(7) xor D(8);
      ECC(3) <= D(0) xor D(1) xor D(3) xor D(6) xor D(7) xor D(8);
end InnHam;

================================================================================
-- Calculates the parity bit of all data bits along with an ECC
================================================================================
library IEEE;
use IEEE.STD_LOGIC_1164.ALL;

entity Parity is
      Port (      D  : in std_logic_vector(0 to 8);     -- 9-bit data array
                  ECC: in std_logic_vector(0 to 3);     -- 4-bit ECC array
                  par: out std_logic                    -- parity bit
             );
end Parity;

architecture Parity of Parity is
begin
      par <= ECC(0) xor ECC(1) xor ECC(2) xor ECC(3) xor D(0) xor D(1) xor
             D(2) xor D(3)  xor D(4) xor D(5) xor D(6) xor D(7) xor D(8);
end Parity;

================================================================================
-- Calculates the parity bit of all data bits along with an ECC
================================================================================
library IEEE;
use IEEE.STD_LOGIC_1164.ALL;

entity Encoder is
      Port (      D : in std_logic_vector(0 to 8);     -- 9-bit data array
                  Co: out std_logic_vector(0 to 3);    -- 4-bit outer ECC array
                  Ci: out std_logic_vector(0 to 3);    -- 4-bit inner ECC array
                  Po: out std_logic;               -- parity bit of outer codeword
                  Pi: out std_logic                -- parity bit of inner codeword
             );
end Encoder;

architecture Encoder of Encoder is
      signal CoTmp: std_logic_vector(0 to 3);
      signal CiTmp: std_logic_vector(0 to 3);
begin
    OutH: entity work.OuterHamming port map(D => D, ECC => CoTmp);
    OutP: entity work.Parity port map(D => D, ECC => CoTmp, par => Po);
    InH:  entity work.InnerHamming port map(D => D, ECC => CiTmp);
    InP:  entity work.Parity port map(D => D, ECC => CiTmp, par => Pi);
    Co <= CoTmp;
    Ci <= CiTmp;
end Encoder;
```

## Decoder.vhd

```
================================================================================
-- Compute Syndromes
================================================================================
library IEEE;
use IEEE.STD_LOGIC_1164.ALL;
```

```
entity Syndromes is
      Port  (
            C, RC: in std_logic_vector(0 to 3);      -- Read and recomputed Hamming
            P, RP: in std_logic;                     -- Read and recomputed Parity
            sC: out std_logic_vector(0 to 3);        -- Hamming Syndromes
            sP: out std_logic                        -- Parity Syndromes
      );
end Syndromes;

architecture Syndromes of Syndromes is
begin
      sC(0) <= '0' when C(0) = RC(0) else '1';
      sC(1) <= '0' when C(1) = RC(1) else '1';
      sC(2) <= '0' when C(2) = RC(2) else '1';
      sC(3) <= '0' when C(3) = RC(3) else '1';
      sP <= '0' when RP = P else '1';
end Syndromes;

================================================================================
-- Compute SE_DE
================================================================================
library IEEE;
use IEEE.STD_LOGIC_1164.ALL;

entity SE_DE is
      Port (
            sCq, sP: in std_logic;
            SEr, DEr: out std_logic
      );
end SE_DE;

architecture SE_DE of SE_DE is
begin
      SEr <= '1' when (sCq = '1' and sP = '1') else '0';
      DEr <= '1' when (sCq = '1' and sP = '0') else '0';
end SE_DE;

================================================================================
-- Compute scalar of Hamming syndromes
================================================================================
library IEEE;
use IEEE.STD_LOGIC_1164.ALL;

entity ScalarHamSyn is
      Port  (
            sC: in std_logic_vector(0 to 3);
            sCq: out std_logic
      );
end ScalarHamSyn;

architecture ScalarHamSyn of ScalarHamSyn is
begin
      sCq <= '1' when sC(0) = '1' or sC(1) = '1' or sC(2) = '1' or sC(3) = '1';
end ScalarHamSyn;

================================================================================
-- Compute Error Address
================================================================================
library IEEE;
use IEEE.STD_LOGIC_1164.ALL;
use IEEE.NUMERIC_STD.ALL;
```

```
entity ErrorAddress is
      Port  (
            sC: in std_logic_vector(0 to 3);
            EAr: out integer
      );
end ErrorAddress;

architecture ErrorAddress of ErrorAddress is
begin
      EAr <= to_integer(unsigned(sC));
            -- sC(0) * 16 + sC(1) * 8 + sC(2) * 4 + sC(3) * 2 + sC(4);
end ErrorAddress;

================================================================================
-- Decoder
library IEEE;
use IEEE.STD_LOGIC_1164.ALL;
use IEEE.NUMERIC_STD.ALL;
use work.matrix_package.ALL;

entity Decoder is
      Port  (
            D: in std_logic_vector(0 to 8);         -- 16-bit data array input
            Co: in std_logic_vector(0 to 3);        -- 4-bit outer ECC array
            Ci: in std_logic_vector(0 to 3);        -- 4-bit inner ECC array
            Po: in std_logic;                      -- parity bit of outer codeword
            Pi: in std_logic;                      -- parity bit of inner codeword
            Dout: out std_logic_vector(0 to 8);     -- 16-bit data array output
            ErrorDet : out std_logic
      );
end Decoder;

architecture Decoder of Decoder is
      signal RCi, RCo: std_logic_vector(0 to 3); --Hamming/Parity recomputed signals
      signal RPi, RPo: std_logic;

      signal sPi, sPo: std_logic;               --Syndrome signals
      signal sCi, sCo: std_logic_vector(0 to 3);
      signal sCiq, sCoq: std_logic;

      signal SErI, SErO: std_logic;            --Single and Double Error flags
      signal DErI, DErO: std_logic;
      signal EArO, EArI: integer;              --Inner and Outer Error Addresses

      signal Dtemp, DtempSEI, DtempSEO, DtempDE: std_logic_vector(0 to 8);
      signal addISE, addOSE, addIDE, addODE: integer;
      signal errPos: error_positions;

begin
      OutHamRec: entity work.OuterHamming port map(D => D, ECC => RCo);
      InnHamRec: entity work.InnerHamming port map(D => D, ECC => RCi);

      ErrorDet <=  (RCo(0) or RCo(1) or RCo(2) or RCo(3)) or
                   (RCi(0) or RCi(1) or RCi(2) or RCi(3));

      OutParRec: entity work.Parity port map(D => D, ECC => Co, par => RPo);
      InnParRec: entity work.Parity port map(D => D, ECC => Ci, par => RPi);

      entity work.Syndromes port map(C=>Ci,RC=>RCi,P=>Pi,RP=>RPi,sC=>sCi,sP=>sPi);
      entity work.Syndromes port map(C=>Co,RC=>RCo,P=>Po,RP=>RPo,sC=>sCo,sP=>sPo);

      InnScalHamSyn: entity work.ScalarHamSyn port map(sC => sCi, sCq => sCiq);
      OutScalHamSyn: entity work.ScalarHamSyn port map(sC => sCo, sCq => sCoq);

      InnSE_DE: entity SE_DE work.port map(sCq=>sCiq,sP=>sPi,SEr=>SErI,DEr=>DErI);
      OutSE_DE: entity SE_DE work.port map(sCq=>sCoq,sP=>sPo,SEr=>SErO,DEr=>DErO);
```

```
      InnErrAdd: entity work.ErrorAddress port map(sC => sCi, EAr => EArI);
      OutErrAdd: entity work.ErrorAddress port map(sC => sCo, EAr => EArO);

      errPos <=   doubleErrorMap(EArO, EArI) when DErI = '1' and DErO = '1'
                  else (x"00", x"00");

      addIDE <= to_integer(errPos(0)) when DErI = '1' and DErO = '1' else -1;
      addODE <= to_integer(errPos(1)) when DErI = '1' and DErO = '1' else -1;
      addISE <= to_integer(innerAddTab(EArI)) when SErI = '1' else -1;
      addOSE <= to_integer(outerAddTab(EArO)) when SErO = '1' else -1;

      GenerateSEI: for k in 0 to 8 generate
            DtempSEI(k) <= not D(k) when k = addISE else D(k);
            DtempSEO(k) <= not D(k) when k = addOSE else D(k);
            DtempDE(k)  <= not D(k) when k = addIDE or k = addODE else D(k);
      end generate;

      Dtemp <=    DtempSEI when SErI = '1' else
                  DtempSEO when SErO = '1' else
                  DtempDE when DErI = '1' and DErO = '1' else
                  D;

      Dout <= Dtemp when EArI /= 0 and EArO /= 0 else D;

end Decoder;
```

## Matrix_package.vhd

```
================================================================================
library IEEE;
use IEEE.STD_LOGIC_1164.ALL;
use IEEE.NUMERIC_STD.ALL;

package matrix_package is

      type error_positions is array (0 to 1) of signed(7 downto 0);
      type signed_array is array (0 to 15) of signed(7 downto 0);
      type error_map is array (0 to 15, 0 to 15) of error_positions;

      constant outerAddTab : signed_array :=
      (
            to_signed(-1, 8), to_signed(-1, 8), to_signed(-1, 8), to_signed(2, 8),
            to_signed(-1, 8), to_signed(5, 8), to_signed(7, 8), to_signed(-1, 8),
            to_signed(-1, 8), to_signed(-1, 8), to_signed(3, 8), to_signed(0, 8),
            to_signed(4, 8), to_signed(1, 8), to_signed(6, 8), to_signed(8, 8)
      );

      constant innerAddTab : signed_array :=
      (
            to_signed(-1, 8), to_signed(-1, 8), to_signed(-1, 8), to_signed(7, 8),
            to_signed(-1, 8), to_signed(6, 8), to_signed(-1, 8), to_signed(1, 8),
            to_signed(-1, 8), to_signed(0, 8), to_signed(4, 8), to_signed(-1, 8),
            to_signed(5, 8), to_signed(3, 8), to_signed(2, 8), to_signed(8, 8)
      );

      constant doubleErrorMap : error_map :=
      (
            -- Row 0
            (
                  (to_signed(-1, 8), to_signed(-1, 8)), -- Column 0
                  (to_signed(-1, 8), to_signed(-1, 8)), -- Column 1
                  (to_signed(-1, 8), to_signed(-1, 8)), -- Column 2
                  (to_signed(-1, 8), to_signed(-1, 8)), -- Column 3
                  (to_signed(-1, 8), to_signed(-1, 8)), -- Column 4
                  (to_signed(-1, 8), to_signed(-1, 8)), -- Column 5
```

```
            (to_signed(-1, 8), to_signed(-1, 8)), -- Column 6
            (to_signed(-1, 8), to_signed(-1, 8)), -- Column 7
            (to_signed(-1, 8), to_signed(-1, 8)), -- Column 8
            (to_signed(-1, 8), to_signed(-1, 8)), -- Column 9
            (to_signed(-1, 8), to_signed(-1, 8)), -- Column 10
            (to_signed(-1, 8), to_signed(-1, 8)), -- Column 11
            (to_signed(-1, 8), to_signed(-1, 8)), -- Column 12
            (to_signed(-1, 8), to_signed(-1, 8)), -- Column 13
            (to_signed(-1, 8), to_signed(-1, 8)), -- Column 14
            (to_signed(-1, 8), to_signed(-1, 8)) -- Column 15
      ),

      -- Row 1
      (
            (to_signed(-1, 8), to_signed(-1, 8)), -- Column 0
            (to_signed(-1, 8), to_signed(-1, 8)), -- Column 1
            (to_signed(-1, 8), to_signed(-1, 8)), -- Column 2
            (to_signed(-1, 8), to_signed(-1, 8)), -- Column 3
            (to_signed(0, 8), to_signed(3, 8)), -- Column 4
            (to_signed(-1, 8), to_signed(-1, 8)), -- Column 5
            (to_signed(-1, 8), to_signed(-1, 8)), -- Column 6
            (to_signed(-1, 8), to_signed(-1, 8)), -- Column 7
            (to_signed(-1, 8), to_signed(-1, 8)), -- Column 8
            (to_signed(-1, 8), to_signed(-1, 8)), -- Column 9
            (to_signed(6, 8), to_signed(8, 8)), -- Column 10
            (to_signed(-1, 8), to_signed(-1, 8)), -- Column 11
            (to_signed(-1, 8), to_signed(-1, 8)), -- Column 12
            (to_signed(1, 8), to_signed(4, 8)), -- Column 13
            (to_signed(-1, 8), to_signed(-1, 8)), -- Column 14
            (to_signed(-1, 8), to_signed(-1, 8)) -- Column 15
      ),

      -- Row 2
      (
            (to_signed(-1, 8), to_signed(-1, 8)), -- Column 0
            (to_signed(-1, 8), to_signed(-1, 8)), -- Column 1
            (to_signed(-1, 8), to_signed(-1, 8)), -- Column 2
            (to_signed(-1, 8), to_signed(-1, 8)), -- Column 3
            (to_signed(-1, 8), to_signed(-1, 8)), -- Column 4
            (to_signed(-1, 8), to_signed(-1, 8)), -- Column 5
            (to_signed(-1, 8), to_signed(-1, 8)), -- Column 6
            (to_signed(-1, 8), to_signed(-1, 8)), -- Column 7
            (to_signed(1, 8), to_signed(8, 8)), -- Column 8
            (to_signed(-1, 8), to_signed(-1, 8)), -- Column 9
            (to_signed(-1, 8), to_signed(-1, 8)), -- Column 10
            (to_signed(-1, 8), to_signed(-1, 8)), -- Column 11
            (to_signed(-1, 8), to_signed(-1, 8)), -- Column 12
            (to_signed(-1, 8), to_signed(-1, 8)), -- Column 13
            (to_signed(-1, 8), to_signed(-1, 8)), -- Column 14
            (to_signed(4, 8), to_signed(6, 8)) -- Column 15
      ),

      -- Row 3
      (
            (to_signed(-1, 8), to_signed(-1, 8)), -- Column 0
            (to_signed(-1, 8), to_signed(-1, 8)), -- Column 1
            (to_signed(1, 8), to_signed(6, 8)), -- Column 2
            (to_signed(-1, 8), to_signed(-1, 8)), -- Column 3
            (to_signed(-1, 8), to_signed(-1, 8)), -- Column 4
            (to_signed(4, 8), to_signed(8, 8)), -- Column 5
            (to_signed(-1, 8), to_signed(-1, 8)), -- Column 6
            (to_signed(-1, 8), to_signed(-1, 8)), -- Column 7
            (to_signed(-1, 8), to_signed(-1, 8)), -- Column 8
            (to_signed(-1, 8), to_signed(-1, 8)), -- Column 9
            (to_signed(-1, 8), to_signed(-1, 8)), -- Column 10
            (to_signed(-1, 8), to_signed(-1, 8)), -- Column 11
            (to_signed(-1, 8), to_signed(-1, 8)), -- Column 12
```

```
            (to_signed(-1, 8), to_signed(-1, 8)), -- Column 13
            (to_signed(-1, 8), to_signed(-1, 8)), -- Column 14
            (to_signed(5, 8), to_signed(7, 8)) -- Column 15
      ),

      -- Row 4
      (
            (to_signed(-1, 8), to_signed(-1, 8)), -- Column 0
            (to_signed(-1, 8), to_signed(-1, 8)), -- Column 1
            (to_signed(-1, 8), to_signed(-1, 8)), -- Column 2
            (to_signed(-1, 8), to_signed(-1, 8)), -- Column 3
            (to_signed(-1, 8), to_signed(-1, 8)), -- Column 4
            (to_signed(-1, 8), to_signed(-1, 8)), -- Column 5
            (to_signed(0, 8), to_signed(8, 8)), -- Column 6
            (to_signed(-1, 8), to_signed(-1, 8)), -- Column 7
            (to_signed(3, 8), to_signed(6, 8)), -- Column 8
            (to_signed(-1, 8), to_signed(-1, 8)), -- Column 9
            (to_signed(-1, 8), to_signed(-1, 8)), -- Column 10
            (to_signed(-1, 8), to_signed(-1, 8)), -- Column 11
            (to_signed(-1, 8), to_signed(-1, 8)), -- Column 12
            (to_signed(-1, 8), to_signed(-1, 8)), -- Column 13
            (to_signed(-1, 8), to_signed(-1, 8)), -- Column 14
            (to_signed(-1, 8), to_signed(-1, 8)) -- Column 15
      ),

      -- Row 5
      (
            (to_signed(-1, 8), to_signed(-1, 8)), -- Column 0
            (to_signed(-1, 8), to_signed(-1, 8)), -- Column 1
            (to_signed(3, 8), to_signed(8, 8)), -- Column 2
            (to_signed(-1, 8), to_signed(-1, 8)), -- Column 3
            (to_signed(-1, 8), to_signed(-1, 8)), -- Column 4
            (to_signed(-1, 8), to_signed(-1, 8)), -- Column 5
            (to_signed(-1, 8), to_signed(-1, 8)), -- Column 6
            (to_signed(-1, 8), to_signed(-1, 8)), -- Column 7
            (to_signed(-1, 8), to_signed(-1, 8)), -- Column 8
            (to_signed(-1, 8), to_signed(-1, 8)), -- Column 9
            (to_signed(-1, 8), to_signed(-1, 8)), -- Column 10
            (to_signed(-1, 8), to_signed(-1, 8)), -- Column 11
            (to_signed(0, 8), to_signed(6, 8)), -- Column 12
            (to_signed(2, 8), to_signed(7, 8)), -- Column 13
            (to_signed(-1, 8), to_signed(-1, 8)), -- Column 14
            (to_signed(-1, 8), to_signed(-1, 8)) -- Column 15
      ),

      -- Row 6
      (
            (to_signed(-1, 8), to_signed(-1, 8)), -- Column 0
            (to_signed(-1, 8), to_signed(-1, 8)), -- Column 1
            (to_signed(2, 8), to_signed(5, 8)), -- Column 2
            (to_signed(-1, 8), to_signed(-1, 8)), -- Column 3
            (to_signed(-1, 8), to_signed(-1, 8)), -- Column 4
            (to_signed(-1, 8), to_signed(-1, 8)), -- Column 5
            (to_signed(-1, 8), to_signed(-1, 8)), -- Column 6
            (to_signed(3, 8), to_signed(4, 8)), -- Column 7
            (to_signed(-1, 8), to_signed(-1, 8)), -- Column 8
            (to_signed(-1, 8), to_signed(-1, 8)), -- Column 9
            (to_signed(-1, 8), to_signed(-1, 8)), -- Column 10
            (to_signed(-1, 8), to_signed(-1, 8)), -- Column 11
            (to_signed(-1, 8), to_signed(-1, 8)), -- Column 12
            (to_signed(-1, 8), to_signed(-1, 8)), -- Column 13
            (to_signed(0, 8), to_signed(1, 8)), -- Column 14
            (to_signed(-1, 8), to_signed(-1, 8)) -- Column 15
      ),

      -- Row 7
      (
```

```
            (to_signed(-1, 8), to_signed(-1, 8)), -- Column 0
            (to_signed(-1, 8), to_signed(-1, 8)), -- Column 1
            (to_signed(-1, 8), to_signed(-1, 8)), -- Column 2
            (to_signed(0, 8), to_signed(4, 8)), -- Column 3
            (to_signed(-1, 8), to_signed(-1, 8)), -- Column 4
            (to_signed(-1, 8), to_signed(-1, 8)), -- Column 5
            (to_signed(-1, 8), to_signed(-1, 8)), -- Column 6
            (to_signed(-1, 8), to_signed(-1, 8)), -- Column 7
            (to_signed(-1, 8), to_signed(-1, 8)), -- Column 8
            (to_signed(-1, 8), to_signed(-1, 8)), -- Column 9
            (to_signed(1, 8), to_signed(3, 8)), -- Column 10
            (to_signed(-1, 8), to_signed(-1, 8)), -- Column 11
            (to_signed(-1, 8), to_signed(-1, 8)), -- Column 12
            (to_signed(-1, 8), to_signed(-1, 8)), -- Column 13
            (to_signed(-1, 8), to_signed(-1, 8)), -- Column 14
            (to_signed(-1, 8), to_signed(-1, 8)) -- Column 15
      ),

      -- Row 8
      (
            (to_signed(-1, 8), to_signed(-1, 8)), -- Column 0
            (to_signed(-1, 8), to_signed(-1, 8)), -- Column 1
            (to_signed(-1, 8), to_signed(-1, 8)), -- Column 2
            (to_signed(-1, 8), to_signed(-1, 8)), -- Column 3
            (to_signed(-1, 8), to_signed(-1, 8)), -- Column 4
            (to_signed(-1, 8), to_signed(-1, 8)), -- Column 5
            (to_signed(6, 8), to_signed(7, 8)), -- Column 6
            (to_signed(0, 8), to_signed(2, 8)), -- Column 7
            (to_signed(-1, 8), to_signed(-1, 8)), -- Column 8
            (to_signed(-1, 8), to_signed(-1, 8)), -- Column 9
            (to_signed(-1, 8), to_signed(-1, 8)), -- Column 10
            (to_signed(1, 8), to_signed(5, 8)), -- Column 11
            (to_signed(-1, 8), to_signed(-1, 8)), -- Column 12
            (to_signed(-1, 8), to_signed(-1, 8)), -- Column 13
            (to_signed(-1, 8), to_signed(-1, 8)), -- Column 14
            (to_signed(-1, 8), to_signed(-1, 8)) -- Column 15
      ),

      -- Row 9
      (
            (to_signed(-1, 8), to_signed(-1, 8)), -- Column 0
            (to_signed(-1, 8), to_signed(-1, 8)), -- Column 1
            (to_signed(-1, 8), to_signed(-1, 8)), -- Column 2
            (to_signed(2, 8), to_signed(3, 8)), -- Column 3
            (to_signed(-1, 8), to_signed(-1, 8)), -- Column 4
            (to_signed(-1, 8), to_signed(-1, 8)), -- Column 5
            (to_signed(4, 8), to_signed(5, 8)), -- Column 6
            (to_signed(-1, 8), to_signed(-1, 8)), -- Column 7
            (to_signed(-1, 8), to_signed(-1, 8)), -- Column 8
            (to_signed(-1, 8), to_signed(-1, 8)), -- Column 9
            (to_signed(-1, 8), to_signed(-1, 8)), -- Column 10
            (to_signed(-1, 8), to_signed(-1, 8)), -- Column 11
            (to_signed(7, 8), to_signed(8, 8)), -- Column 12
            (to_signed(-1, 8), to_signed(-1, 8)), -- Column 13
            (to_signed(-1, 8), to_signed(-1, 8)), -- Column 14
            (to_signed(-1, 8), to_signed(-1, 8)) -- Column 15
      ),

      -- Row 10
      (
            (to_signed(-1, 8), to_signed(-1, 8)), -- Column 0
            (to_signed(-1, 8), to_signed(-1, 8)), -- Column 1
            (to_signed(-1, 8), to_signed(-1, 8)), -- Column 2
            (to_signed(5, 8), to_signed(8, 8)), -- Column 3
            (to_signed(-1, 8), to_signed(-1, 8)), -- Column 4
            (to_signed(-1, 8), to_signed(-1, 8)), -- Column 5
            (to_signed(-1, 8), to_signed(-1, 8)), -- Column 6
```

```
            (to_signed(-1, 8), to_signed(-1, 8)), -- Column 7
            (to_signed(-1, 8), to_signed(-1, 8)), -- Column 8
            (to_signed(4, 8), to_signed(7, 8)), -- Column 9
            (to_signed(-1, 8), to_signed(-1, 8)), -- Column 10
            (to_signed(-1, 8), to_signed(-1, 8)), -- Column 11
            (to_signed(-1, 8), to_signed(-1, 8)), -- Column 12
            (to_signed(-1, 8), to_signed(-1, 8)), -- Column 13
            (to_signed(-1, 8), to_signed(-1, 8)), -- Column 14
            (to_signed(-1, 8), to_signed(-1, 8)) -- Column 15
      ),

      -- Row 11
      (
            (to_signed(-1, 8), to_signed(-1, 8)), -- Column 0
            (to_signed(-1, 8), to_signed(-1, 8)), -- Column 1
            (to_signed(-1, 8), to_signed(-1, 8)), -- Column 2
            (to_signed(-1, 8), to_signed(-1, 8)), -- Column 3
            (to_signed(1, 8), to_signed(7, 8)), -- Column 4
            (to_signed(-1, 8), to_signed(-1, 8)), -- Column 5
            (to_signed(-1, 8), to_signed(-1, 8)), -- Column 6
            (to_signed(-1, 8), to_signed(-1, 8)), -- Column 7
            (to_signed(-1, 8), to_signed(-1, 8)), -- Column 8
            (to_signed(5, 8), to_signed(6, 8)), -- Column 9
            (to_signed(-1, 8), to_signed(-1, 8)), -- Column 10
            (to_signed(-1, 8), to_signed(-1, 8)), -- Column 11
            (to_signed(-1, 8), to_signed(-1, 8)), -- Column 12
            (to_signed(-1, 8), to_signed(-1, 8)), -- Column 13
            (to_signed(-1, 8), to_signed(-1, 8)), -- Column 14
            (to_signed(-1, 8), to_signed(-1, 8)) -- Column 15
      ),

      -- Row 12
      (
            (to_signed(-1, 8), to_signed(-1, 8)), -- Column 0
            (to_signed(2, 8), to_signed(8, 8)), -- Column 1
            (to_signed(-1, 8), to_signed(-1, 8)), -- Column 2
            (to_signed(-1, 8), to_signed(-1, 8)), -- Column 3
            (to_signed(-1, 8), to_signed(-1, 8)), -- Column 4
            (to_signed(-1, 8), to_signed(-1, 8)), -- Column 5
            (to_signed(-1, 8), to_signed(-1, 8)), -- Column 6
            (to_signed(-1, 8), to_signed(-1, 8)), -- Column 7
            (to_signed(-1, 8), to_signed(-1, 8)), -- Column 8
            (to_signed(-1, 8), to_signed(-1, 8)), -- Column 9
            (to_signed(-1, 8), to_signed(-1, 8)), -- Column 10
            (to_signed(-1, 8), to_signed(-1, 8)), -- Column 11
            (to_signed(-1, 8), to_signed(-1, 8)), -- Column 12
            (to_signed(-1, 8), to_signed(-1, 8)), -- Column 13
            (to_signed(3, 8), to_signed(7, 8)), -- Column 14
            (to_signed(-1, 8), to_signed(-1, 8)) -- Column 15
      ),

      -- Row 13
      (
            (to_signed(-1, 8), to_signed(-1, 8)), -- Column 0
            (to_signed(-1, 8), to_signed(-1, 8)), -- Column 1
            (to_signed(-1, 8), to_signed(-1, 8)), -- Column 2
            (to_signed(-1, 8), to_signed(-1, 8)), -- Column 3
            (to_signed(-1, 8), to_signed(-1, 8)), -- Column 4
            (to_signed(-1, 8), to_signed(-1, 8)), -- Column 5
            (to_signed(-1, 8), to_signed(-1, 8)), -- Column 6
            (to_signed(-1, 8), to_signed(-1, 8)), -- Column 7
            (to_signed(-1, 8), to_signed(-1, 8)), -- Column 8
            (to_signed(-1, 8), to_signed(-1, 8)), -- Column 9
            (to_signed(0, 8), to_signed(7, 8)), -- Column 10
            (to_signed(2, 8), to_signed(6, 8)), -- Column 11
            (to_signed(-1, 8), to_signed(-1, 8)), -- Column 12
            (to_signed(-1, 8), to_signed(-1, 8)), -- Column 13
```

```
                  (to_signed(-1, 8), to_signed(-1, 8)), -- Column 14
                  (to_signed(-1, 8), to_signed(-1, 8)) -- Column 15
            ),

            -- Row 14
            (
                  (to_signed(-1, 8), to_signed(-1, 8)), -- Column 0
                  (to_signed(-1, 8), to_signed(-1, 8)), -- Column 1
                  (to_signed(-1, 8), to_signed(-1, 8)), -- Column 2
                  (to_signed(-1, 8), to_signed(-1, 8)), -- Column 3
                  (to_signed(-1, 8), to_signed(-1, 8)), -- Column 4
                  (to_signed(0, 8), to_signed(5, 8)), -- Column 5
                  (to_signed(-1, 8), to_signed(-1, 8)), -- Column 6
                  (to_signed(-1, 8), to_signed(-1, 8)), -- Column 7
                  (to_signed(-1, 8), to_signed(-1, 8)), -- Column 8
                  (to_signed(1, 8), to_signed(2, 8)), -- Column 9
                  (to_signed(-1, 8), to_signed(-1, 8)), -- Column 10
                  (to_signed(-1, 8), to_signed(-1, 8)), -- Column 11
                  (to_signed(-1, 8), to_signed(-1, 8)), -- Column 12
                  (to_signed(-1, 8), to_signed(-1, 8)), -- Column 13
                  (to_signed(-1, 8), to_signed(-1, 8)), -- Column 14
                  (to_signed(-1, 8), to_signed(-1, 8)) -- Column 15
            ),

            -- Row 15
            (
                  (to_signed(-1, 8), to_signed(-1, 8)), -- Column 0
                  (to_signed(3, 8), to_signed(5, 8)), -- Column 1
                  (to_signed(-1, 8), to_signed(-1, 8)), -- Column 2
                  (to_signed(-1, 8), to_signed(-1, 8)), -- Column 3
                  (to_signed(2, 8), to_signed(4, 8)), -- Column 4
                  (to_signed(-1, 8), to_signed(-1, 8)), -- Column 5
                  (to_signed(-1, 8), to_signed(-1, 8)), -- Column 6
                  (to_signed(-1, 8), to_signed(-1, 8)), -- Column 7
                  (to_signed(-1, 8), to_signed(-1, 8)), -- Column 8
                  (to_signed(-1, 8), to_signed(-1, 8)), -- Column 9
                  (to_signed(-1, 8), to_signed(-1, 8)), -- Column 10
                  (to_signed(-1, 8), to_signed(-1, 8)), -- Column 11
                  (to_signed(-1, 8), to_signed(-1, 8)), -- Column 12
                  (to_signed(-1, 8), to_signed(-1, 8)), -- Column 13
                  (to_signed(-1, 8), to_signed(-1, 8)), -- Column 14
                  (to_signed(-1, 8), to_signed(-1, 8)) -- Column 15
            )
      );

end matrix_package;
```

Pontifícia Universidade Católica do Rio Grande do Sul
Pró-Reitoria de Pesquisa e Pós-Graduação
Av. Ipiranga, 6681 – Prédio 1 – Térreo
Porto Alegre – RS – Brasil
Fone: (51) 3320-3513
E-mail: propesq@pucrs.br
Site: www.pucrs.br